\documentclass{ws-rv975x65}
\usepackage{subfigure}     
\usepackage{ws-rv-van}     
\makeindex
\usepackage{amsmath}
\usepackage{amsfonts}

\usepackage{graphicx}
\usepackage{amssymb,mathtools}
\usepackage{bm}
\usepackage[colorlinks=true, citecolor=blue,linkcolor=blue]{hyperref}
\renewcommand{\theequation}{\thesection.\arabic{equation}}

\long\def\comment#1{ }

\newcommand{\beq}{\begin{eqnarray}}
\newcommand{\eeq}{\end{eqnarray}}
\newcommand{\nn}{\nonumber\\}
\newcommand{\bal}{\begin{align}}
\newcommand{\eal}{\begin{align}}
\def\cG{{\cal G}}
\def\cP{{\cal P}}
\def\cK{{\cal K}}

\def\cD{{\cal D}}
\def\cT{{\cal T}}
\def\cU{{\cal U}}

\def\cH{{\cal H}}

\def\lg{{\langle}}
\def\rg{{\rangle}}

\def\br{\text{br}}
\def\BH{\text{BH}}
\def\el{\text{el}}

\def\Tr{\text{Tr}}
\def\Tjet{\Theta_\text{jet}}

\def\Sii{{S^{(2)}}}
\def\Siio{{S_0^{(2)}}}
\def\Siii{{S^{(3)}}}

\def\tSiii{{\tilde S^{(3)}}}
\def\Siv{{S^{(4)}}}
\def\bdel{\boldsymbol \del}

\newcommand{\rmd}{{\rm d}}

\newcommand{\rme}{{\rm e}}

\newcommand{\rmR}{{\rm Re}}

\def\bp{{p}}
\def\bk{{ k}}

\def\q{{\bm q}}
\def\0{{\boldsymbol 0}}
\def\p{{\boldsymbol p}}
\def\l{{\boldsymbol l}}
\def\k{{\boldsymbol k}}

\def\x{{\boldsymbol x}}
\def\y{{\boldsymbol y}}

\def\r{{\boldsymbol r}}

\def\u{{\boldsymbol u}}
\def\v{{\boldsymbol v}}

\def\Q{{\boldsymbol Q}}

\def\btheta{{\boldsymbol \theta}}

\newcommand{\del}{\partial}

\begin{document}

\begin{center}
{\large \bf Jet Structure in Heavy Ion Collisions}
\end{center}
\author[J. -P. Blaizot  and Y. Mehtar-Tani]{J. -P. Blaizot$^a$  and Y. Mehtar-Tani$^b$}
\address{$^a$ Institut de Physique Th\'{e}orique de Saclay, CNRS/URA2306, CEA Saclay,
F-91191 Gif-sur-Yvette, France\\
$^b$ Institute for Nuclear Theory, University of Washington, Seattle, WA 98195-1550, USA}

\begin{abstract}
We review recent theoretical developments in the study of the structure of jets that are produced in ultra relativistic heavy ion collisions. The core of the review focusses on the dynamics of the parton cascade  that is induced by the interactions of a fast parton crossing a quark-gluon plasma. We recall the basic mechanisms responsible for medium induced radiation, underline the rapid disappearance of coherence effects, and the ensuing probabilistic nature of the medium induced cascade. We discuss how large radiative corrections modify the classical picture of the gluon cascade, and how these can be absorbed in a renormalization of the jet quenching parameter $\hat q $. Then, we analyze the (wave)-turbulent transport of energy along the medium induced cascade, and point out the main characteristics of the angular structure of such a cascade.  Finally, color decoherence of the in-cone jet structure is discussed. Modest contact with phenomenology is presented towards the end of the review. 

\end{abstract}

 \begin{flushright}
INT-PUB-15-006
 \end{flushright}
\body

\section{Introduction}\label{intro}
A \emph{jet} is a cone-shaped beam of energetic particles which originate from a hard parton produced in a high energy collision. Jets have played an important role in the development of Qauntum Chromodynamics (QCD), with three jet events in   $e^+e^-$ annihilation providing a direct experimental evidence for the existence of the gluon, and the quark-gluon vertex\cite{Ellis:1976uc,Wiik:1979cq}.  Moreover, perturbative QCD correctly describes the soft hadronic activity surrounding jets.  For instance, the inclusive intrajet hadron distribution, the so-called fragmentation function, exhibits a humpbacked shape that reflects the suppression of soft gluon radiation at large angle as a result of destructive interferences, as predicted by QCD \cite{Azimov:1985by,	Dokshitzer:1991wu,Khoze:1996dn}. 
Today, jet technology has become a precision tool embodied in powerful Monte Carlo event generators  \cite{Sjostrand:2006za,Bahr:2008qs,Gleisberg:2008ta}, and involve elaborate perturbative QCD calculations (culminating at the NNLO level), that are not only relevant for further testing strong and weak interactions at high energy, but also in searching for physics beyond the Standard Model at the LHC (see e.g.  Refs.~\cite{Currie:2013dwa,Pires:2014rxa}  recent reviews).

In the context of heavy ion collisions, jet occupy a very special place. They are produced in hard collisions over tiny space-time regions, and the partons that originate form such hard collisions propagate through the hot matter produced in the bulk of the heavy ion collisions. As such they constitute ideal ``test particles'' whose study could reveal the properties of the medium they traverse before hadronizing. It was recognized long ago   \cite{Bjorken:1982tu},  for instance, that quark and gluon jets may lose energy while crossing a quark-gluon plasma  produced in heavy ion collisions, resulting in a strong suppression of high-$p_t$ hadrons. This phenomenon, commonly referred to as \emph{ jet-quenching}, was in fact observed at RHIC \cite{Arsene:2004fa,Back:2004je,Adams:2005dq,Adcox:2004mh} and at the LHC \cite{Aad:2010bu,Aamodt:2010jd,Chatrchyan:2011sx}, providing direct evidence for the creation of dense QCD mater in heavy ion collisions. The high energy of the LHC make jets  a very powerful diagnostic tool: they are produced in large number, have large energies (with $p_t$ ranging from $100$ to $300$ GeV, an order of magnitude larger than at RHIC).  Furthermore, the detector capabilities of ALICE,  ATLAS and CMS, allow for the exploration of a variety of  new and more detailed observables,  among which, the jet nuclear modification factor, the dijet asymmetry, the jet profiles \cite{Aad:2010bu,Aad:2011sc,Aad:2012vca,Aad:2014wha,Aad:2014bxa,Chatrchyan:2011sx,Chatrchyan:2012gw,CMS:2012rba,Chatrchyan:2013kwa,CMS:2014uca,Adam:2015ewa}. All this is opening new perspectives in the field and invites new theoretical efforts.

The theory of jet quenching in heavy ion collisions started developing in the early 90's with the first studies of medium induced gluon bremsstrahlung by Gyulassy and Wang \cite{Gyulassy:1993hr,Wang:1994fx}, accompanied by those of Baier, Dokshitzer, Peigné, Mueller, Schiff \cite{ Baier:1994bd,Baier:1996kr,Baier:1996sk} and Zakharov \cite{Zakharov:1996fv,Zakharov:1997uu},   where it was realized that radiative parton energy loss is the dominant mechanism for jet quenching in large QCD media. This pioneering works were followed by many others emphasizing  various aspects of the phenomena and using different formalisms.  For instance, assuming the medium to be dilute, Gyulassy, Levai  and Vitev \cite{Gyulassy:1999zd,Gyulassy:2000er} (see also Ref.~\cite{Wiedemann:2000za}) developed a formalism in which the interactions with the plasma can be calculated order by order in opacity. In a similar approach, due to  Guo and Wang, the so-called High-Twist (HT) \cite{Guo:2000nz,Wang:2001ifa}, the first medium correction to the the leading order splitting functions is computed and embedded in a medium-modified version of the DGLAP equation. Considering the plasma to be dense, an approximation of multiple soft collisions was applied by Armesto, Salgado and Wiedemann \cite{Salgado:2003gb,Armesto:2003jh} where multiple gluon radiations are treated as independent and described with a Poisson distribution \cite{Baier:2001yt}. An analogous approach, accounting for thermal effects, is due to Arnold, Moore and Yaffe \cite{Arnold:2002ja,Arnold:2002zm,Jeon:2003gi}. 

While based on similar assumptions (see Refs.~\cite{CasalderreySolana:2007pr,d'Enterria:2009am,Majumder:2010qh} for reviews), the various approaches fail to provide a consensual quantitative description of the RHIC data \cite{Armesto:2011ht}.  There are perhaps two reasons for that. First, the relatively low momentum range accessible at RHIC, of the order of 10 GeV, may question the applicability of weak coupling techniques. Second, each of the models has limitations related to approximations done regarding  the nature of the medium and the kinematics. For instance, none of the  models addresses from first principles the issue of in-medium multiple parton branchings and resulting interferences, and thus rely on strong assumptions on the space-time structure of the  in-medium parton shower. 
Moreover, at RHIC,  scenarii based on radiative energy loss proved to be inconsistent with the strong suppression of the single non-photonic electrons resulting from semi-leptonic decays of hadrons carrying heavy quarks \cite{Adler:2005xv}, reviving the interest for collisional energy loss \cite{Mustafa:2004dr,Peigne:2005rk,Wicks:2005gt,Djordjevic:2006tw}. At the same time, evidence for almost perfect fluid behavior of the quark-gluon plasma \cite{Kolb:2003dz,Romatschke:2007mq,Heinz:2013th}, diverted the attention towards strong coupling techniques, providing an alternative perspective on jet quenching based on the AdS/CFT correspondence (see Ref.~\cite{CasalderreySolana:2011us} for a recent review, and references therein). \\

In parallel to the aforementioned developments,  the theory of jet-quenching based on essentially weak techniques of QCD,  has progressed within the last few years. The set  goals are the  understanding of the structure of jets in matter, and how this structure  is modified with respect  to that of well understood jets in vacuum \cite{MehtarTani:2010ma,MehtarTani:2011tz,MehtarTani:2011jw,CasalderreySolana:2011rz,MehtarTani:2011gf,MehtarTani:2012cy,Blaizot:2012fh,Blaizot:2013hx,Blaizot:2014ula,Kurkela:2014tla,Apolinario:2014csa,Blaizot:2014rla,Fister:2014zxa,Blaizot:2015jea} (see Ref.~\cite{Mehtar-Tani:2013pia} for a recent review), and the calculation of higher-order corrections  \cite{Wu:2011kc,Liou:2013qya,Fickinger:2013xwa,Kang:2013raa,Blaizot:2013vha,Blaizot:2014bha,Arnold:2015qya}, addressing some of the limitations of the standard approaches of energy loss. Among the major results of these efforts, it was shown that in the limit of asymptotically large or dense media successive medium-induce branchings are quasi-local and independent from one another \cite{Blaizot:2012fh,Blaizot:2013vha}, providing a theoretical basis for the approximations made in some models of jet quenching \cite{Baier:2001yt,Jeon:2003gi,Salgado:2003gb}. Color coherence, that characterizes jet structure in vacuum, has been shown to be significantly altered in a QCD medium \cite{MehtarTani:2010ma,MehtarTani:2011tz,MehtarTani:2011jw}. As a result, medium-induced color decoherence suppresses the interferences between successive gluon radiations. There is also an effort to systematize the calculations of multiple parton branching in the framework of Soft-Collinear Effective Theory (SCET) \cite{Ovanesyan:2011kn,Fickinger:2013xwa}. Last but not least, radiative corrections to some jet-quenching observables were recently carried out\cite{Wu:2011kc,Liou:2013qya,Kang:2013raa,Blaizot:2013vha,Blaizot:2014bha,Ghiglieri:2013gia}, in addition to calculations of the jet quenching parameter on the lattice \cite{Majumder:2012sh,Panero:2013pla}. The present review will present a summary of some of these developments. 

We should also mention that  the theoretical developments mentioned above are accompanied by ongoing efforts to develop Monte-Carlo event generators, that encompass some of the recent advances in the microscopic physics, as well as a realistic modeling of the geometry of the plasma and its hydrodynamical evolution, in order to carry out systematic studies of jet observables with all sorts of experimental biases, such as, $p_t$ cuts, triggers, etc. The most notable ones are:  JEWEL \cite{Zapp:2008af,Zapp:2011ya}, Q- PYTHIA \cite{Armesto:2009fj} and MARTINI \cite{Schenke:2009gb,Young:2011ug}, that account for the BDMPS-Z  radiation mechanism, in addition to PYQUEN/HYDJET\cite{Lokhtin:2008xi}, YAJEM \cite{Renk:2010zx} and MATTER \cite{Majumder:2013re}. These developments will not be covered in the review. Also, the phenomenology of  jet the quenching  is beyond the scope of this review and will be covered in another contribution to the present volume \cite{Qin2015}. \\

Before we embark on the theoretical discussion, a word of warning. For the sake of simplicity we shall most of the time deal with the  ideal scenario in which the parton cascade is   purely gluonic and interacts with a uniform and static plasma. We also assume the medium to be dense enough so that we can apply the multiple soft scattering approximation.  Corrections to this idealized picture can be calculated, and this will be mentioned at proper places, but those corrections do not change the qualitative picture.  The review is organized as follows. In Section~\ref{jets}, the physics of parton branching in vacuum and the BDMPS-Z  mechanism for medium-induced gluon radiation are introduced.  In Section~\ref{prob-pic}, we generalize the BDMPS-Z  spectrum to multiple branchings and develop a probabilistic picture of in-medium gluon cascades. Then, in Section~\ref{RG-qhat}, we analyze radiative corrections to momentum broadening and parton energy loss and show that they can be reabsorbed in a renormalization of the quenching parameter. A complementary discussion on NLO corrections to energy loss and $p_\perp$-broadening in high temperature plasma is presented in this volume Ref.~\cite{Ghiglieri2015} : note however that these corrections are related to soft plasma modes and  are of a  different nature than the radiative corrections involved in the renormalization of $\hat q $. In Section~\ref{energy-angle}, we examine the time evolution of the angular and energy distribution along the medium-induced cascade,  away from the jet core. In Section~\ref{decoherence}, we discuss the physics of in-medium jet decoherence that affects the inner jet structure. Finally,  Section~\ref{outlook} presents a brief summary and outlook.

\section{Jets and parton cascades of two kinds}\label{jets}
Perturbative QCD successfully predicts the emergence of jets in the final state of high energy collisions such as $e^+e^-$ annihilation and hadronic collisions. 
As a matter of fact, the collimation of jet particles relates directly to the collinear nature of QCD parton splittings. The underlying mechanism is akin to QED bremsstrahlung: an electric charge that suffers an acceleration subsequently radiates photons collinearly to its direction of motion. Likewise, quarks (and gluons) carry a color charge and thus can act as sources of collinear gluon bremsstrahlung.  Besides the confining forces of strong interactions that eventually turn patrons into observable hadrons,  the difference with QED (at the partonic level) lies in the fact that, contrary to photons, gluons may also radiate affecting the general pattern of the collimated cascade. 
The nascent parton possesses typically a large virtuality $Q^2$, and therefore radiates gluon, which themselves radiate other gluons, forming a so-called \emph{parton cascade}. Because this cascade generically propagates in vacuum, we shall refer to it as a \emph{vacuum cascade}. 
Such a cascade is accompanied by a  degradation of the virtuality $Q$, which can be used to define an evolution ``time'', $t\equiv \ln Q^2/Q_0^2$, where $Q_0$ is the scale that marks the end point of the parton cascade where hadronization takes place. Typically,  $Q_0\sim \Lambda_\text{QCD}$. Hadronization is a non perturbative phenomenon and one may fear that it erases all memory of the detailed structure of the preceding partonic phase. However, this is not so. It turns out that  inclusive observables are to a large extent determined by the parton dynamics prior to hadronization. This remarkable property, known as the local parton-hadron duality \cite{Dokshitzer:1991wu,Khoze:1996dn},  is largely responsible for the success of perturbative QCD in describing jet observables to a high degree of precision. 

When the energetic parton  propagates in a medium, another mechanism may generate radiation, beyond that due to the possible virtuality of the parton: during its propagation, the parton collides with the constituents of the medium, and each of these collisions may in principle trigger radiation. In fact, because of this mechanism, even an on-shell parton entering a medium will be able to radiate and generate a cascade. We shall refer to such a cascade as to a \emph{medium-induced cascade}. 

We therefore encounter two types of parton cascades: 
\begin{enumerate}
\item The vacuum cascade that forms after a hard process that creates a parton with a large virtuality.
\item The medium-induced parton cascade that forms due to multiple interactions of the jet with a QCD medium. 
\end{enumerate} 
This review will be mostly devoted to an analysis of the medium induced cascade. But in our discussion we shall often refer to the vacuum cascade in order to emphasize some of the distinctive properties of the medium induced cascade. 

In both cascades, coherence and interferences play an important role by suppressing radiation in particular regions of the spectrum. For the vacuum cascade, this is the case of gluons with large transverse wavelengths which do not resolve the individual color charges carried by the jet, and thus are radiated coherently off the total color charge of the jet. In the case of medium induced radiation, we have gluons with long formation times that do not ``see'' the individual constituents on which they scatter, and appear to be radiated coherently off many scattering centers. In both cases the gluon bremsstrahlung spectrum is suppressed with respect to the incoherent case. We now discuss these phenomena in some more detail. 

\subsection{ Jet structure in vacuum: transverse color coherence}\label{vacuum}
Experimentally, a jet is defined by a collection of particles that are contained in a cone of given opening angle $\Tjet$. The characteristic hard scale for the perturbative evolution is then $Q\equiv E\Tjet \gg \Lambda_\text{QCD}$. It corresponds to the largest transverse scale (with respect to the jet axis) that can be carried by one of the jet constituents and serves as an upper bound for the jet invariant transverse mass.  The phase space between the hard scale $Q$ and the non-perturbative scale $Q_0\sim \Lambda_\text{QCD}$ is filled by the vacuum cascade.

The microscopic mechanism generating the cascade is bremssthalung. It is common to both the vacuum and in-medium shower. 
In the collinear limit, the branching probability into two offsprings, with energy fractions $z$ and $1-z$ respectively, reads 
\beq\label{P-br-vac}
dP \equiv \frac{\alpha_s}{4\pi}\frac{\rmd k_\perp^2}{k_\perp^2} P(z) \rmd z,
\eeq 
where $\alpha_s$ is the strong coupling constant, and $P(z)$ the Altarelli-Parisi splitting function \cite{Altarelli:1977zs}. The expression (\ref{P-br-vac}) of the branching probability exhibits a collinear divergence that corresponds to a mass singularity, and a soft divergence, when $k_\perp\to 0$ and $z\to 0$ ($z\to 1$), respectively. The latter is cured in inclusive quantities by real-virtual cancellations, 
while the former is regularized by cutting off the transverse momentum integration at the scale  $Q_0$. 
The collinear divergence provides a large logarithm $\ln Q^2/Q_0^2$  that may compensate the small coupling constant in the expression (\ref{P-br-vac}) of the emission probability. When $\bar\alpha \ln Q^2/Q_0^2\sim 1$, multiple branchings need to be resummed. In this resummation, successive branchings  are strongly ordered in their transverse momenta, with the largest momentum emitted first.  \cite{Gribov:1972ri,Altarelli:1977zs,Dokshitzer:1977sg}. 

It  was soon realized that this picture is not complete and fails to describe the soft sector of jet observables. This is because it does not fully take into account the effects of color coherence: The radiation pattern of a collection of color charges that form a jet, as they originate from a single parton, does not simply reduce to the incoherent sum of the radiation spectra of the individual charges.  If the transverse wave length $\lambda_\perp\sim 1/k_\perp$ of the radiated gluon is larger than the transverse size of the jet $r_\perp \sim \Tjet t_f  $, at the time $t_f = \omega/k_\perp^2$ when the gluon forms, the latter does not resolve the fine structure of the jet. The radiation is then triggered instead by the total charge of the jet, i.e.,  the color charge of the original parton. This leads to a suppression of the radiation as compared to that expected form the various color charges contained in the jet.

Such considerations, including the fact that  gluons carry a color charge and thus can radiate as well \cite{Mueller:1981ex,Ermolaev:1981cm,Bassetto:1982ma,Bassetto:1984ik}, and accounting also for  for energy conservation, constitute the basis of the jet algorithms in vacuum that were first introduced in the Marchesini-Webber model \cite{Marchesini:1987cf}.
 Along a QCD shower a strict ordering in angles accounts for the color coherence and the evolution equation for the inclusive parton distribution reads, 
\beq\label{MLLA}
\frac{\del D(x,Q)}{\del \ln (Q^2/Q_0^2)} = \int_0^1 \frac{\rmd z}{z}  \frac{\alpha_s(k_\perp)}{2\pi}\hat P(z) D(x/z, zQ).  
\eeq 
where $k_\perp\equiv z(1-z)Q$ and $\hat P(z)$ stands for the regularized Altarelli-Parisi splitting function\cite{Altarelli:1977zs} (the dependence on different parton species is omitted). Eq.~(\ref{MLLA}) is referred to as the Modified-Leading-Log-Approximation (MLLA)\cite{Bassetto:1982ma,Bassetto:1984ik,Dokshitzer:1991wu} as it differs from DGLAP\cite{Gribov:1972ri,Altarelli:1977zs,Dokshitzer:1977sg} by the  fact that in the r.h.s. the parton distribution scales with $zQ$ rather than $Q$ to account for angular ordering. Eq.~(\ref{MLLA}) describes in reality the evolution of the parton shower as a function of the jet-opening angle $\Tjet$ with $Q\equiv \Tjet E$ and $E$ the energy of the parent parton: The infinitesimal variation of the parton distribution is given by an initial splitting of the parent parton generating a subjet characterized by the energy $E'=zE$  and the corresponding scale is $Q' = E' \Tjet\equiv zQ$. 

\subsection{Medium-induced gluon radiation: longitudinal coherence}\label{BDMPS-Z-spect}

In a medium of size $L$, an energetic parton experiences multiple collisions with the medium constituents. In principle each of these collisions puts the gluon off-shell, so that it may subsequently emit radiation,  after a typical time, commonly referred to as the \emph{formation time}, $t_f\sim \omega/k_\perp^2$, where $\omega$ and $k_\perp$ are the energy and the transverse momentum of the radiated gluon. However, this simple picture may be strongly affected by coherence effects in multiple scattering.

Such effects were studied long ago  in the context of QED,  by Landau, Pomeranchuk and Migdal (LPM) \cite{Landau:1953gr,Migdal:1956tc}, and are generically referred to as the \emph{Landau-Pomeranchuk-Migdal (LPM) effect}. The QCD analog of the LPM effect\footnote{See Ref.~\cite{Peigne:2008wu} for a recent review on the LPM effect in QED and QCD.} was investigated by Gyulassy and Wang (GW)\cite{Gyulassy:1993hr,Wang:1994fx} first, then by Baier, Dokshitzer, Mueller, Peigné, Schiff and Zakharov (BDMPS-Z) \cite{Baier:1994bd,Baier:1996kr,Baier:1996sk,Zakharov:1997uu,Zakharov:1996fv}. The main focus of these early studies was the energy loss  caused by induced gluon radiation\cite{Baier:2001yt}.  For large media, radiative energy loss scales as $L^2$ where $L$ is the length of the matter traversed, as opposed to  collisional energy loss that scales as $L$ \cite{Braaten:1991we}.

We shall return extensively in this review to the BDMPS-Z mechanism, but it is useful at this stage to have an heuristic understanding of the basic physics involved.
Consider then a hard quark or gluon of energy $E$ propagating in a dense QCD medium. The collisions of the parton with the medium constituents induce radiation, with the typical formation time of a radiated gluon given by $t_f\sim \omega / k_\perp^2$. Now,  during that time scale,  the gluon fluctuation experiences multiple kicks, which may cause its transverse momentum to increase by an amount $k_\perp^2 \sim \hat q t_f$ where $\hat q$ is a diffusion coefficient to be specified shortly.  Assuming that these multiple kicks are the only source of transverse momentum, and combining the two relations above one finds that the typical  time for induced emission, to which we shall refer here as the \emph{branching time} is given by 
\beq\label{tbranching}
\tau_\br=\sqrt{\frac{\omega}{\hat q}}.
\eeq

Multiple collisions cause the broadening of the transverse momentum. In the regime where the collisions are soft, and where a large number of collisions are needed to change the transverse momentum by a significant amount, a diffusion approximation becomes appropriate. In this case, it is customary to define a transport coefficient $\hat q$ that describes the diffusion in momentum space. Assuming that the dominant color field fluctuations of the medium are characterized by a scale $m_D$, to which we refer as the Debye screening mass, the typical transverse momentum exchanged with the medium in a collision is $\sim m_D$. Then the transverse momentum squared acquired by the hard parton traveling a distance $L$ is, by definition of $\hat q$, $\hat q L$. 
Equivalently, if ${\lambda_\el}$ denotes the elastic mean free path, we have $m_D^2\simeq \hat q\lambda_\el$. The coefficient $\hat q$ is referred to as the \emph{jet quenching parameter. }

Now, when $\tau_\br\lesssim \lambda_\text{el}$, each of the $N=L/\lambda_\text{el}$ collisions act incoherently and produce a maximal radiation intensity $\propto N$. This is the so-called Bethe-Heitler regime which yields a spectrum
\beq\label{BHspec}
\omega\frac{\rmd I}{\rmd \omega}\simeq \bar\alpha \frac{L}{\lambda_{el}}, \qquad \left( \bar\alpha\equiv \frac{\alpha_s N_c}{\pi}  \right).
\eeq
On the other hand, when $\tau_\br \gg \lambda_\text{el}$,  a number $N_\text{coh}=\tau_\br/\lambda_\text{el}$ of the collisions act coherently as a single collision to radiate one gluon. As a result the radiation spectrum  scales as $N/N_\text{coh}$ and reads\cite{Zakharov:1996fv,Baier:1996kr,Baier:1996sk,Zakharov:1997uu}.  \footnote{$I$ stands for ``Intensity'', a dimensionless quantity proportional to the number of gluons.}\footnote{In QED, the spectrum scales as $\alpha_s  \sqrt{\omega \omega_c}/E$. It depends explicitly on the energy of the projectile. At asymptotically high energies, the photon radiation dies out, while according to Eq.~(\ref{BDMPS-Z }) gluon radiation does not. }
\beq\label{BDMPS-Z }
\omega\frac{\rmd I}{\rmd \omega}\simeq \bar\alpha\sqrt{\frac{\omega_c}{\omega}},\qquad \omega_c\equiv \hat q L^2, 
\eeq
where $\omega_c$  is such that   $\tau_\br(\omega_c)=L$. When $\omega=\omega_c$, $\omega \rmd I/\rmd \omega\sim \bar\alpha$, as if we  had a single collision. The full Bethe-Heitler result $\omega \rmd I/\rmd \omega\sim \bar\alpha N$ is recovered when $\omega=\hat q\lambda_\el^2\equiv \omega_{BH}$.

Note that for $\omega\gtrsim \omega_c$ the spectrum is strongly suppressed.  In this case, the radiated gluon cannot distinguish the hard event that produced its radiator from the multiple scattering of the latter with the medium. As a result, the suppression $1/\sqrt{\omega}$  turns into a stronger one, that is $1/\omega^2$. A complete calculation, accounting for finite size effects, with the two limiting cases above, yields \cite{Baier:1996kr}
\beq\label{BDMPS-Z -L}
\omega\frac{\rmd I}{\rmd \omega}\simeq 2\bar \alpha \ln\left|\cos\left(\Omega L\right)\right|,
\eeq
where $\bar\alpha\equiv \alpha_sC_R/\pi$, with $C_R\equiv C_{F,A}$ is the color charge of the source of radiation, and 
\beq\label{Omega}
\Omega=\frac{1+i}{2}\sqrt{\frac{\hat q}{ \omega}} \sim \frac{1}{\tau_\br(\omega)}.
\eeq
By radiating gluons, an energetic parton, with energy $E \gg \omega_c$,  that traverses a dense QCD medium, loses a mean energy, \footnote{For parton energies $E\ll \omega_c$ we find that $\Delta E \sim \sqrt{E}$}
\beq\label{MEloss}
\Delta E\equiv \int_0^\infty \rmd\omega \omega \,\frac{\rmd I}{\rmd \omega}= \frac{\pi}{4}\bar \alpha\hat q L^2,
\eeq
Indeed, since the spectrum Eq.~(\ref{BDMPS-Z -L}) falls as $1/\sqrt{\omega}$  when $\omega \ll \omega_c$ (cf. Eq.~(\ref{BDMPS-Z })), the $\omega$ integral in (\ref{MEloss}) is dominated by gluon frequencies of the order of $\omega_c$. This formula relates the mean-energy loss to the typical transverse momentum squared accumulated by the gluons during the maximal formation time $t_f\sim L$, $\lg k_\perp^2\rg \sim \hat q L$, that is, $\Delta E\sim \lg k_\perp^2 \rg L$. 

Consider now  the number of radiated gluons with frequency larger than $\omega$, 
\beq\label{number-g}
N(\omega)=\int^{\omega_c}_\omega \rmd \omega'\frac{\rmd I}{\rmd\omega}.
\eeq
 When $\omega \ll \omega_c$, the number of radiated gluons is dominated by the lower bound, $\omega$, of the integral in Eq.~(\ref{number-g}), for the spectrum rises as $1/\omega^{3/2}$. Thus, we obtain  
\beq\label{number-g-2}
N(\omega)\simeq \sqrt{\frac{\omega_s(L)}{\omega} }.
\eeq
where 
\beq\label{omega_s}
\omega_s (L) \equiv \bar\alpha^2\,\omega_c(L).
\eeq
As long as $N(\omega)$ is smaller than one it can be  identified with a radiation probability, i.e., $N(\omega)\sim P(\omega) <1$.
When the average number of emitted gluons exceeds unity, higher orders in perturbation theory, that involve multiple gluon radiations, become relevant processes that need to be dealt with. According to Eq.~(\ref{number-g-2}), this regime is achieved when $\omega \lesssim \omega_s(L)$.  
This  criterion can be looked at from a different perspective. As we shall see later, the branching rate for a gluon with energy $\omega$ is given by 
\beq
\frac{\bar\alpha}{\tau_\br (\omega)}\equiv \frac{1}{t_\ast(\omega)}, 
\eeq
where, accordingly (see Eq.~(\ref{tbranching}))
\beq\label{stop-time-0}
t_\ast (\omega) = \frac{1}{\bar\alpha}\sqrt{\frac{\omega}{\hat q}}
\eeq
is the time it takes a gluon with energy $\omega$ to branch, a quantity that one could also refer to as an  ``inelastic mean-free path''. Then, the condition $\omega \lesssim \omega_s(L)$ is equivalent to the condition $t_\ast(\omega)  \lesssim L$, that is, during the time $t_\ast(\omega)$ a gluon with energy $\omega$ will branch at least once during the time it spends in the matter. When applied to the leading particle, the quantity $t_\ast(E)$ will  be referred to as the stopping time in Section~\ref{energy-angle}.
\section{Probabilistic picture of medium-induced gluon cascade}\label{prob-pic}

We start now examining more closely the properties of the medium-induced cascade and begin by putting  in place various elements of the formalism that will be used throughout this review.

\subsection{Non relativistic dynamics in the transverse plane}\label{elements}

A high energy jet originates from an ultrarelativistic parton, whose direction of propagation is  chosen as the $z$ axis.  We choose light-cone variables\footnote{With this choice, the 4-coordinate vector is $x\equiv(x^+,x^-,\x)$, where $x^+=(t+z)/2$, $x^-=t-z$, and $\x\equiv (x,y).$} such that  the 4-momentum $p$ of a parton  reads 
\beq
p^+=\frac{1}{2}(E+p_z),\qquad p^-=E-p_z,  \qquad\text{and} \qquad \p\equiv (p_x,p_y),
\eeq 
where $E$ is the energy, $p_z$ the longitudinal momentum, and $\p$ the transverse momentum. 
The angle $\theta$ formed by the 3-momentum $\p$ and the jet axis is given by  $p_\perp \equiv |\p|\sin \theta $. Considering that jets are collimated objects, and that their angular broadening is small, we assume in all our calculations that  $\theta\ll 1$, and  accordingly, 
\beq
p^+\simeq E, \qquad\text{and} \qquad |\p|\equiv p_\perp\simeq  E\theta, 
\eeq
while the light-cone time can be identified with the real time,
\beq
x^+ \simeq t\simeq z.
\eeq
We work in the light-cone gauge $A^+=0$, a convenient choice in the present context. This is because  the current $J^-$, which describes the target through which the jet propagates, is coupled to the $+$ component of the gauge field, which vanishes in this gauge. In turn, gluon radiation off particles propagating in the $-z$ direction is suppressed,  and only the jet constituents, propagating in the $+z$ direction, radiate.  Furthermore, the light-cone gauge allows for a partonic picture of the radiation dynamics.

The hard gluons that compose a collimated jet travel along  nearly straight-line trajectories, and interact with the color field of the medium. Only the $-$ component of this field couples to the high energy gluon with the helicity conserving vertex, $V^{ij+}A^-\sim p^+ A^- \delta^{ij}$, where $i,j = 1,2$  depict the two transverse components of the physical polarization vector $\epsilon_\lambda^i$. The transverse component of the field can be removed by making use of the additional degree of freedom of the light-cone gauge \cite{Blaizot:2008yb}. Moreover, the gluon with very large longitudinal momentum $p^+$, propagates near the light-cone branch, $x^- = 0$, so that, finally, the only component of the medium field that matters is $A(t,\x)\equiv \left.A^-(x)\right|_{x^-=0}$.

The choice of the light-cone gauge along with the eikonal approximation for the vertex imply that the gluon propagator in the medium background field is diagonal in helicity space, i.e., $G^{ij}\sim \delta^{ij} G$, with  $G$ obeying the following equation, 
\beq\label{backprop}
\left[\Box-ig\del^+ A^-(x)\cdot T\right] G(x, x')=\delta(x-x'),
\eeq
where $g$ is the gauge  coupling constant. 
We have already mentioned that the jet probes the medium gauge field only near the light cone, i.e.  $x^-\approx 0$. We may then assume that over such small distance, the medium gauge field is constant, and perform a Fourier transform with respect to $x^-$.\footnote{Since $x^-$ is the conjugate variable to $p^+\sim E$, assuming that the field $A$ does not depend on $x^-$ leads to the conservation of $p^+$ at each vertex coupling the field to the parton. Alternatively, this conservation law could be obtained by observing that the transfer of $+$ momentum from the plasma to the parton is negligible compared to the large energy $E$ of the parton.} We define
\beq
(\x|  \cG(t,t')|\x') =\frac{i}{2E} \int \frac{dx^-}{2\pi }  \rme^{-iE (x-x')^-} G(x-x'), 
\eeq
and from Eq.~(\ref{backprop}) above, one finds that $\cG(t,t')$ obeys the following
 2-dimensional Schr\"odinger equation
\beq\label{SchroG}
\left[i\frac{\del}{\del t} + \frac{\del^2_\perp}{2 E}+gA(t,\x) \right] (\x|\cG(t-t')|\x')=i\delta(t-t')\delta(\x-\x'),
\eeq
where the background field, $A\equiv A_a T^a$, with $T^a$ the generators of $SU$(3) in the adjoint representation, is a matrix in color space, and so is the propagator $\cG$. In the high energy limit, the interesting dynamics is therefore localized in the transverse plane. It is the 2-D equivalent   of a non-relativistic particle in a background potential, where the invariant $p^+= E$ component of the momentum plays the role of the mass.

The propagation of the gluon from $(t_0,\x_0)$ to $(t,\x)$ is described by the Green's function $(\x|\cG(t,t_0)|\x_0)$ that  solves the  2-dimensional Schr\"odinger equation (\ref{SchroG}) (see e.g. Ref.~\cite{Kovner:2003zj}).
Its solution can be formally represented by a path integral in transverse coordinate space:
\beq\label{prop-G}
(\x|\cG(t,t_0)|\x_0)=\Theta(t-t_0)\, \int^{\x}_{\x_0} \cD\r \exp\left[i\frac{E}{2}\int_{t_0}^t \rmd t' \dot \r^2(t') \right]\, \cU_\r(t,t_0),
\eeq
where retarded conditions are used and
\beq
\cU_\r(t,t_0)=\cT\exp\left[ig \int_{t_0}^t \rmd t' A(t',\r(t'))\cdot T\right]
\eeq
is a Wilson-line along the path $\r(t')$, with $\r(t_0)\equiv \x_0$ and $\r(t)\equiv \x$. 
Between two scatterings, the gluon propagation is free, 
\beq\label{free-prop-G}
(\x|\cG_0(t,t_0)|\x_0)=\Theta(t-t_0)\,  \frac{E}{2\pi i (t-t_0)} \exp\left[\frac{iE(\x-\x_0)^2}{2(t-t_0)}\right].
\eeq
The Green's function $\cG(t,t')$ plays the role of a unitary time evolution operator for a free particle evolving from the position state $|\x)$ to the state $|\x')$. In momentum representation,
\beq
|\p) = \int \rmd^2 \x\, \rme^{-i\p\cdot\x}\,|\x), \qquad \int \frac{\rmd^2\p}{(2\pi)^2}\, |\p)(\p| = 1,
\eeq
we define the  amplitude for a gluon of (large) momentum $p_0\equiv(p^+_0,\p_0=0)$, 
present in the system at time $t_0$, to evolve in the medium into a gluon with momentum $p_1=(p^+_1,\p_1)$, where $p^+_0=p^+_1\equiv E$ (to within an irrelevant phase factor),
\beq\label{Mone}
{\cal M}(p_1|p_0)=  
(\p_1|\,{\cal
G}(t_1,t_0)\,|\p_0).
\eeq
 The amplitude ${\cal M}\equiv {\cal M}_{ab}$ is a matrix in color space, propagating a gluon with color $b$ to color $a$. It is also a (diagonal) matrix in the helicity space. We do not write explicitly color and helicity indices to alleviate the notation. 

The free Hamiltonian in 2-dimensional space, reads
\beq
\cH_0=\frac{\p^2}{2E}.
\eeq
The interaction part of the hamiltonian is the sum of two components, $\delta \cH=\delta \cH_1+\delta \cH_2$,
\beq\label{int-H}
(\q|\delta \cH_1|\p)=g A(\q-\p,t), \quad\text{and}\quad (\q; \q'|\delta \cH_2|\p)=g(\q; \q'|V|\p).
\eeq
where $\delta \cH_1$ represents the elastic scattering of a particle off the potential $A(\q,t)$, while  $\delta \cH_2$ represents the splitting of a particle into two. Note that the latter process does not conserve the non relativistic energy $\p^2/2E$, but it conserves the ``mass''  $E$. When using the propagator (\ref{prop-G}) we effectively treat $\delta \cH_1$ to all orders. Later, we shall calculate the branching probability of a gluon to first order in   $\delta \cH_2$. Multiple branchings will be treated via probabilistic cascades.

\subsection{Independent multiple scatterings approximation}\label{independent-scat}

The color field of the plasma, through which the jet propagates, fluctuates from time to time and place to place. The typical scale characterizing such fluctuations is the Debye screening mass $m_D$, which is much smaller than the energy $E$ of the jet. 
As we have seen earlier, the energy $E$ plays the role of the mass in the 2-dimensional non relativistic dynamics in the transverse plane. 
Because this mass is large, the motion in the transverse plane is a slow motion, compared to which the dynamics of the plasma appears to be very fast. It follows that 
the interactions mediated by exchange of plasma fluctuations are thus nearly instantaneous\footnote{Said differently, because of Lorentz time dilation, the high energy parton has a very poor resolution in $x^+$, so that the medium correlations appear to him as effectively local in time.}. In performing the average over the plasma fluctuations, we may then use the following correlator
\beq\label{field-cor}
\lg A^a(t,\x)A^{\ast b}(t',\y) \rg = n \delta^{ab} \delta(t'-t) \gamma(\x-\y),
\eeq
where $n$ is the density of color charges, and $\gamma(\x-\y)$ will be specified shortly. This form of correlator is somewhat analogous to that used in the McLerran-Venugopalan model \cite{McLerran:1993ni,McLerran:1993ka} that deals with cold nuclear matter of large nuclei.  It is also in line with  the Gyulassy-Wang model, in which the medium is modeled as a collection of static scattering centers \cite{Wang:1991xy,Gyulassy:1993hr,Wang:1994fx} . 

Assuming that  it can be reliably calculated from  perturbative QCD, the correlator $\gamma(\x)$ is given in leading order by
\beq\label{gamma}
\gamma(\x)=g^2\int_\q \frac{\rme^{i\q\cdot\x}}{\q^4},
\eeq
which is proportional to the Coulomb elastic cross-section, with  $\q$ the transverse momentum transfer. 
Here we introduced a shorthand notation for transverse momentum integrations, to be used throughout this review: $\int_\q \equiv \rmd^2\q/(2\pi)^2$. 
A somewhat more elaborate version of $\gamma$ for a QCD plasma in thermal equilibrium is obtained from the substitution \cite{Aurenche:2002pd} $\q^4 \to \q^2(\q^2+m_D^2)$.  

The picture that emerges form this discussion is one where the jet propagates through a random background field,  to be averaged over with the correlation function given by Eq.~(\ref{gamma}). Using another language, the high energy jet collides with the plasma constituents at a rate determined by the Coulomb cross-section. Note that this depends on the medium solely through the density of color charges, and the Debye mass. The motion of the jet in the transverse plane is a slow motion, and the interactions with the plasma constituents are effectively instantaneous. 

The treatment of collisions in terms of the propagation in a random field with an instantaneous correlation function implicitly assumes that the collisions are independent. This requirement is satisfied in a weakly coupled quark-gluon plasma: The Debye mass is $m_D\sim gT$, with  $T$  the temperature. The elastic mean free path for soft collisions (with momentum transfer of order $m_D$) can be estimated as
\beq
\lambda_\text{el} \sim\frac{1}{n\sigma }\sim \frac{1}{g^2 T}, 
\eeq
where we have used $\sigma\sim g^4/m_D^2$ and $n\sim T^3$. 
This is parametrically larger (by one power of the coupling constant) than the duration of the interactions,  that is $\lambda_\text{el}  \gg  m_D^{-1}$.

\subsection{Momentum broadening}\label{glu-prop}
The multitude of kicks that it receives from its collisions with the medium constituents affects the propagation of a high energy parton. In particular, it modifies its transverse momentum. This leads to the so-called \emph{ momentum broadening}, which we now discuss. 

The amplitude (\ref{Mone}) introduced above allows us to calculate  the transverse momentum spectrum of a gluon at time $t$, given the spectrum at time $t_0$:
\beq\label{2p-fct}
\frac{\rmd N}{\rmd \Omega_p}= \int_{\p_0 } \cP(\p,t|\p_0,t_0) \frac{\rmd N}{\rmd \Omega_{p_0}}, \qquad \rmd\Omega_{\p_0}\equiv\frac{\rmd E\rmd^2\p_0}{2E(2\pi)^3},
\eeq
where  ${\rmd N}/{\rmd \Omega_{p_0}}$ is the spectrum at time $t_0$, and $\cP(\p,t|\p_0,t_0)$  will be referred to as the  $p_\perp$-broadening probability. This is calculated  by squaring the amplitude,  averaging over the initial color and polarization, summing over final polarization and color, and finally averaging over the background field. In doing this calculation, it is convenient to keep the momentum $\p_0$ in the amplitude distinct from that, $\bar\p_0$, in the complex conjugate amplitude. We get then
\beq\label{defS2kk}
{\cal M}(\p|\p_0)\,{\cal M}^*(\p|\bar\p_0) &=&\frac{1}{N_c^2-1}\left<{\rm Tr}
(\p|\,{\cal G}(t_1,t_0)\,|\p_0)(\bar\p_0|\,{\cal G}^\dagger(t_0,t_1)\,|\p)\right>\nn
&=&(2\pi)^2 \delta(\bar\p_0-\bar\p) \, {\cal P}(\p,t_1|\p_0,t_0)\,,
\eeq
where the trace is over color indices and in the last line we have factored out the momentum conserving delta function that reflects translational invariance. The angular brackets in the  first line denote medium average.

In Eq.~(\ref{defS2kk}), appears the Fourier transform of a fundamental quantity, which we refer to as a 2-ppoint function, $(X_1|\Sii(t,t_0)|X_0)$ (see Eq.~(\ref{2-pt-def})). This is one of a collection of $n$-point functions that play an important role in the discussion. This 2-point function, is not a usual field theory 2-point function. It is built from the product of a propagator ${\cal G}$ of an amplitude and the corresponding propagator ${\cal G}^\dagger $ in the complex conjugate amplitude. This is given, quite generally, by  
\beq\label{2-pt-def}
(X_1|\Sii(t,t_0)|X_0)=\frac{1}{N_c^2-1}\langle \text{Tr}~  (\x_1|{\cal G}(t_1,t_0)|\x_0)\,(\bar\x_0|{\cal G}^\dag(t_0,t_1)|\bar\x_1)\rangle,
\eeq
where $X_i=(\x_i;\bar\x_i)$.
This two point function     describes the evolution from coordinates $|X_0)\equiv |\x_0,\bar\x_0)$ to $|X_1)\equiv |\x_1,\bar\x_1)$ \footnote{Some explanation about the notation is needed here. We regard $S^{(2)}(t_1,t_0)$ as a matrix in transverse space, $(Y|S^{(2)}(t_1,t_0)|X)$, with $X=(\x;\bar\x)$, $Y=(\y;\bar\y)$ with the coordinates $\x,\bar\x,\y,\bar\y$ attached to ${\cal G}$ and ${\cal G}^\dagger $ as indicated in Eq.~(\ref{2-pt-def})  (the time variables are not considered as matrix indices, and $S^{(2)}(t_1,t_0)$ is proportional to $\theta(t_1-t_0$)).},  of a (fictitious) dipole made of two  gluons in a color singlet state, one in the amplitude and the other in the complex conjugate amplitude. From the equation satisfied by the propagator ${\cal G}$, Eq.~(\ref{SchroG}), one easily deduces the equation satisfied by $S^{(2)}$, or equivalently that satisfied by  the broadening probability ($\cP(\p,t|\p_0,t_0)\equiv\cP(\p,t) $)\cite{Blaizot:2012fh}
\beq\label{ME-P}
\frac{\del}{\del t} \cP(\p,t) = -\frac{N_c}{2}n \int_\q \sigma(\q)\, \cP(\p-\q,t), 
\eeq
where $\sigma(\q)$ is the Fourier transform of the so-called dipole cross-section
\beq\label{dipole-cs}
\sigma(\x)=2g^2\left[\gamma(\0)-\gamma(\x)\right],
\eeq
with $\x$ corresponding to the transverse size of the dipole.  Eq.~(\ref{ME-P}) can be solved by Fourier transforming to coordinate space. The r.h.s of Eq.~(\ref{ME-P}) becomes local in $\x$, and the solution exponentiate. Moving back to Fourier space, it is straightforward to get (for $t\equiv L$)
\beq\label{GM-P}
\cP(\p,L) = \int \rmd^2 \x\, \exp\left[-\frac{N_c}{2} n\sigma(\x) L-i\p\cdot\x \right],
\eeq
which relates the gluon scattering cross-section at high energy with the S-matrix for a dipole passing trough the medium, in the independent multiple scattering approximation\cite{Mueller:2012bn}. 

In the diffusion limit, namely when the final momentum $\p$ is acquired by a large number of small momentum transfers, i.e., $\q\ll \p$, Eq.~(\ref{ME-P})  takes the form of a Fokker-Planck equation
\beq\label{diff-P}
\frac{\del}{\del t} \cP(\p,t) = \frac{1}{4}\left(\frac{\del}{\del \p}\right)^2 \left[\hat q (\p^2) \cP(\p,t)\right],
\eeq
where the (momentum dependent) diffusion coefficient 
\beq\label{qhat}
\hat q (\p^2) =- \frac{N_c}{2}n\int_\q \q^2 \sigma(\q)\simeq \frac{g^4N_cn}{4\pi }\ln \frac{\p^2}{m_D^2},
\eeq
stands for the jet-quenching parameter \cite{Baier:1996sk}, estimated here to logarithmic accuracy: the lower cut-off of the logarithmic transverse momentum integration is given by the  Debye screening mass $m_D$.
We shall see later that this coefficient receives potentially large radiative corrections.  Another article in this volume discusses an independent set of corrections to the leading-order expression (\ref{qhat}) (see also Ref.~\cite{CaronHuot:2008ni}). 

Finally, assuming $\hat q $ to be roughly constant, one obtains from the diffusion equation (\ref{diff-P}) that the  transverse momentum squared acquired by a high energy gluon traversing a length $L$ of the medium is 
\beq\label{kt2-typ}
\lg p^2_\perp\rg_\text{typ}= \hat q L.
\eeq
 It can also be shown that, in the harmonic approximation, the dipole cross-section (\ref{dipole-cs}) relates to the quenching parameter as follows
\beq\label{sigma-H}
N_c n\sigma(\x) \simeq \frac{1}{2}\x^2 \,\hat q.
\eeq

\subsection{Quasi-local gluon branching}\label{br-prob}

We turn now to inelastic processes that are responsible for the formation of the gluon cascade. This will be handled in first order in the Hamiltonian $\delta \cH_2$. We follow here the derivation presented in Ref.~\cite{Blaizot:2012fh}.

\begin{figure}[htbp]\label{fig:gluon-br-amp}
\centering
\includegraphics[width=10cm]{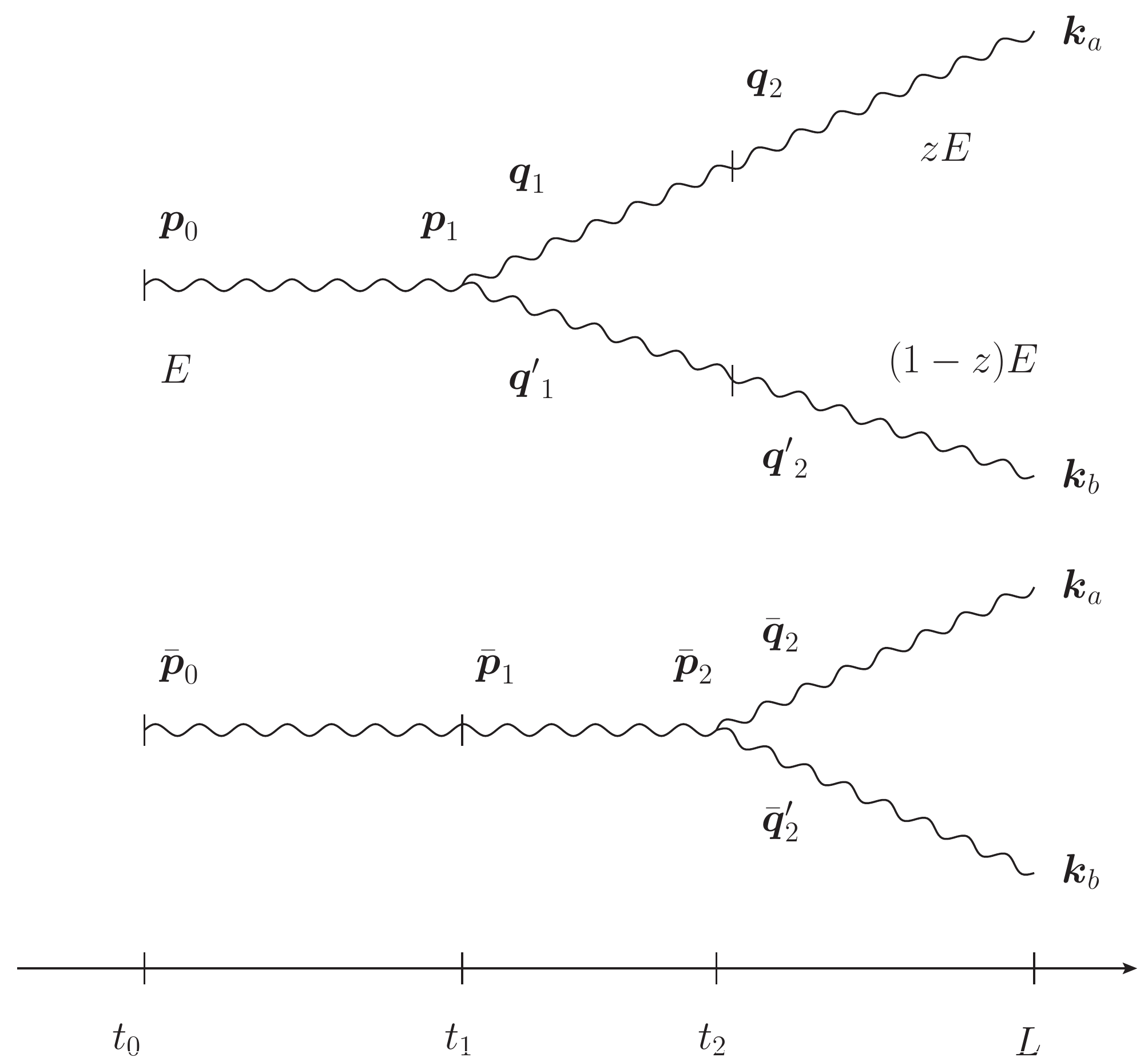} 
\caption{A diagrammatic representation of the time evolution of the amplitude for a gluon branching given by Eq.~(\ref{br-amp}) (upper diagram) and the complex conjugate amplitude (lower diagram) involved in the calculation of the branching probability Eq.~(\ref{g-br-1}), form $t_0$ to $L$. The branching takes place at $t_1$ in the amplitude and $t_2$ in the complex conjugate amplitude (see text for further explanations). }\end{figure}

We aim to extract the inelastic transition probability $\cP_2(\k_a,\k_b,L| \p_0,t_0)$, that relates the 2-gluon spectrum to the  single gluon spectrum  
\beq\label{2-gluon-cs}
\frac{\rmd N}{d\Omega_{k_a}d\Omega_{k_b}}= \int_{\p_0} \cP_2(\k_a,\k_b,L| \p_0,t_0) \frac{\rmd N}{d\Omega_{p_0}}.
\eeq
This is obtained from the amplitude for a gluon of momentum $p_0\equiv (E,\p_0)$ at time $t_0$ to propagate until the time $t_1$ where it has acquired the transverse momentum $\p_1$ and branches into two gluons with momenta $q_1\equiv (zE,\q_1)$ and $q_1'\equiv ((1-z)E, \q_1')$ respectively. The offspring gluons propagate then until they escape the medium at $L$ with momenta $k_a\equiv(zE,\k_a)$ and $k_b\equiv((1-z)E,\k_b)$. 
This amplitude is given by (we omit the color indices to simplify  the notations),  
\beq\label{br-amp}
 \frac{g}{2E}\int \rmd t_1\int_{\q_1\q'_1\p_1}\,(\k_a|\cG(L,t_1)|\q_1) \, (\k_b|\cG(L,t_1)|\q_1') \, (\q_1;\q'_1|  V|\p_1)(\p_1|\cG(t_1,t_0)|\p_0).\nn
\eeq 
It is represented in the upper diagram of Fig.~\ref{fig:gluon-br-amp}.
The 3-gluon vertex is given by 
\beq\label{vertex}
(\q a;\q' b | V |\p c)=\delta(\q-\q'-\p)\, 2 f^{abc}\epsilon^i (\q) \epsilon^j(\q')\epsilon^k(\p)
 \Gamma^{ijk}(\q-z\p,z),\nn
\eeq
with 
\beq
\Gamma^{ijk}(\Q,z) = \frac{1}{1-z} Q^k\delta^{ij}- Q^i\delta^{jk}+\frac{1}{z}Q^j\delta^{ik}.
\eeq 
and $z=\omega_a/E$ is the energy fraction of the parent gluon carried by gluon $a$. 
The particular combination of momenta $\Q= \q-z\p$, reflects a Gallilean invariance of the effective 2-D dynamics. 
 A detailed derivation of this formalism from Feynman diagrams in light-cone gauge, $A^+=0$, is given in Appendix~A of Ref.~ \cite{Blaizot:2012fh}.
Eq.~(\ref{br-amp}) provides the order $g$ correction to the evolution operator $\cG$.  

By squaring the amplitude (\ref{br-amp}) in order to obtain the 2-gluon production distribution,
we are left with the average over the gauge field. Because the correlations of two $A$ fields is instantaneous, this operation is conveniently visualized by plotting the amplitude and its complex conjugate as in  Fig.~\ref{fig:gluon-br-amp}).
By performing the average over the field configurations with the help of Eq.~(\ref{field-cor}), we identify then three factors corresponding to the three different time intervals, involving two, three and four propagators respectively. The 2-point function, $\Sii$ of the first interval $[t_0,t_1]$ is related to the broadening probability, cf. Eq~(\ref{2p-fct}). The 3-point function, $\Siii$ of the second interval,  $[t_1,t_2]$, will determine the branching rate as will shall see shortly. Finally, the last interval, involves a  4-point function, $\Siv$ which factorizes into two 2-point function in the limit of a short formation time \cite{Blaizot:2012fh}, 
\beq
&&(\k_a\k_b;\k_a\k_b|\Siv (L,t_2)|\bar\q_2 \bar\q'_2 ; \q_2 \q'_2)\simeq\nn
&&\qquad\qquad\delta(\q_2-\q'_2)\, \delta(\bar\q_2-\bar\q'_2)\, \cP(\k_a,L| \q_2,t_2) \cP(\k_b,t|\q'_2,t_2),\nn
\eeq
up to terms of the order of $\tau_\br/L$ that we neglect. One can indeed show that  the color correlations between the offspring gluons are exponentially suppressed after a time of the order of the branching time $\tau_\br$, as they move apart. 
This property is closely related to that of color decoherence that will be discussed in more detail in Section~\ref{decoherence}. 

By integrating over the intermediate states, summing over the final polarization vectors and averaging over the initial ones using the completeness relation $\sum_{\lambda=1,2} \epsilon_\lambda^i(\k)\epsilon^{\ast j}_\lambda(\k)=\delta^{ij}$, one finds that the amplitude squared depends on two quantities, the broadening probability and the reduced 3-point function as follows  
\beq\label{g-br-1}
&& \cP_2(\k_a,\k_b,L| \p_0,t_0) =\frac{g^2}{2E^2} 2 \rmR \int_0^L \rmd t_1 \int_0^{t_1}   \rmd t_2 \int_{\Q_1, \Q_2, \l}  \Gamma^{ijk}(\Q_1) \Gamma^{ijk}(\Q_2)\nn 
&& \cP(\k_a,t| \q_2,t_2)\cP(\k_b,t| \q'_2,t_2)  \, \tSiii(\Q_2,\Q_1\l;t_1,t_2) \, \cP(\p_1,t| \p_0,t_2), \nn
\eeq  
where the 3-point function $\tSiii$ reads (cf. Ref.~\cite{Blaizot:2013vha}) 
\beq\label{S3harm}
&&\tSiii(\Q_1,\Q_2,\l;t_1,t_2) = \int  \rmd \u_1 \rmd \u_1 \rmd \v\, \rme^{i\u_1\cdot\Q_1-i\u_2\cdot\Q_2-i\v\cdot\l}\int_{\u_1}^{\u_2}\cD \u\nn
&& \times  \exp\left \{\frac{iz(1-z)E}{2}\int_{t_1}^{t_2}\rmd t \dot\u^2-\frac{N_cn}{4}\int_{t_1}^{t_2}\rmd t \left[\sigma(\u)+\sigma(\v-z\u)+\sigma(\v+(1-z)\u)\right]\right\},\nn
\eeq
with $\Q_1=\q_1-z\p_1$, $\Q_2=\q_2-z\p_2$ and $\l=\bar\p_2-\bar\p_1$.  Momentum conservation at each of the two vertices implies that $\p_1=\q_1+\q_1'$ and $\bar\p_2=\bar\q_2+\bar\q_2'$.
The function $\tSiii$ is an important quantity that encodes the evolution, from $t_1$ to $t_2$, of the the system of two gluons with energies $zE$ and $(1-z)E$ respectively, and one gluon with energy $E$.  The transverse vector $\u$ corresponds to the transverse distance between the offspring gluons a and b, while $\v$ corresponds to the coordinate of the center of mass of the system. 

Furthermore, the sum of gluon momenta in the amplitude and the complex conjugate amplitude vanishes at any given time, $\q_2+\q_2'=\bar\q_2+\bar\q_2'$, $\p_1=\bar\p_2$ and at each vertex we have $\q_1'=\p_1-\q_1$ and $\bar\q_2'=\bar\p_2-\bar \q_2$. The contraction of  the three gluon vertices in the amplitude and complex conjugate amplitude yields
\beq
\frac{1}{2} \Gamma^{ijk}(\Q_1)\Gamma^{ijk}(\Q_2)=\frac{2}{z(1-z)}P(z) (\Q_1\cdot\Q_2),
\eeq
where 
\beq
P(z)=N_c\left[\frac{z}{1-z}+ \frac{1-z}{z}+z(1-z)\right], 
\eeq
 is the gluon-gluon Altarelli-Parisi splitting function \cite{Altarelli:1977zs}. 
 Eq.~(\ref{g-br-1}) can be further simplified in the short formation time approximation, i.e, $\tau_\br\ll L$. 
 
 The difference between $t_1$ in the amplitude and $t_2$ in the complex conjugate amplitude, $\tau=\Delta t =t_2-t_1$ is at most of the order of $\tau_\br$. The 3-point function decays exponentially for $\tau > \tau_\br$. It follows that the upper bound of the integral over $\tau$ can be send to infinity, 
\beq\label{t-integrals}
\int_0^L \rmd t_2 \int_0^{t_2} \rmd t_1=  \int_0^L \rmd t_2 \int_0^{t_2} \rmd \tau\approx  \int_0^L \rmd t_2 \int_0^{\infty} \rmd \tau.  
\eeq
Similarly, the relative transverse momenta generated in the branching, $\Q_1$, $\Q_2$, $\l$ are typically of the order of $k_\br\sim \sqrt{\hat q \tau_\br}$  and thus, can be neglected compared to the external momenta. This is the collinear branchings approximation. Hence, one can integrate over these internal transverse momenta and $\tau$, and this allows us to define the transverse momentum integrated kernel \cite{Zakharov:1996fv,Baier:1998kq}, \footnote{The vacuum part of the 3-point point function, $\tSiii_0$,  is implicitly subtracted.}
\beq\label{kernel}
{\cal K}(z,E)&= & \frac{P(z)}{[z(1-z)E]^2} \rmR\int_0^\infty \rmd \tau \int_{\Q_1\Q_2\l} (\Q_1\cdot\Q_2) \tSiii(\Q_1,\Q_2,\l,t_1,\tau), \nn
\eeq
In the harmonic approximation, we can ignore the logarithmic dependence of $\hat q $ that appears in the dipole cross-section (\ref{sigma-H}),   and the reduced 3-point function (\ref{S3harm}) can be explicitly calculated. This yields for the kernel (\ref{kernel}),
\beq\label{Kdef}
{\cal K}(z,E)=\frac{[1-z+z^2]^{5/2}}{[z(1-z)]^{3/2}}\,\sqrt{\frac{\hat q }{E}}.\,
\eeq
Finally, the transition probability (\ref{g-br-1}) takes the compact form\cite{Blaizot:2012fh} (letting $t_1\equiv t$) 
\beq\label{P2}
 &&\cP_2(\k_a,\k_b,z; L| \p_0,t_0) \nn
 &&=  2g^2z(1-z)\int_0^L \rmd t \, \cP(\k_a,L| z\p,t)\cP(\k_b,L| (1-z)\p,t)  \, {\cal K}(z,E)\, \cP(\p,t| \p_0,t_0).\nn
\eeq  
When the branching time is comparable to the size of the system, $\tau_\br\sim L$, finite size effects are no longer negligible but in this regime only a single branching occurs \cite{Blaizot:2012fh,Apolinario:2014csa}. The corresponding branching probability integrated over transverse momentum generalizes Eq.~(\ref{BDMPS-Z -L}) including finite energy corrections, that is $z \sim 1-z$, can be found in Refs.~\cite{Zakharov:1996fv,Baier:1998kq,Arnold:2008iy}. Also, the generalization to arbitrary parton species is straightforward and can be recovered from the aforementioned references. 

This is a leading order result that describes the instantaneous  branching  of a hard  gluon inside a medium of size $L$. The probability distributions (\ref{GM-P}) and (\ref{P2}) can serve as building blocks for the in-medium gluon cascade that we shall present in Section \ref{multi-br-approx}. 

\subsection{Independent multiple branchings approximation}\label{multi-br-approx}

The probability distribution (\ref{P2}) is parametrically of the form
\beq
\cP_2\sim \frac{ \bar\alpha}{\tau_\br}L,
\eeq
where $\bar\alpha/\tau_\br$ is the  rate of (quasi-instantaneous) branching, and the factor $L$ reflects the fact that the branching can occur anywhere along the medium, i.e., $0 < t< L$.
Consider now the branching probability $\cP_3$, where two successive branchings produce three gluons  in the final state. The two branchings occur successively at times  $t_1$ and $t_2$ in the amplitude and at times $t'_1$ and $t'_2$ in the complex conjugate amplitude. When the time intervals $[t_1,t'_1]$ and $[t_2,t'_2]$ do not overlap, that is,  $t_1<t'_1<t_2<t'_2$. the resulting successive integrations over the different times yields $L^2\, \tau^2_{\br}$,
provided $\tau_\br \sim \tau_{1\br}\sim \tau_{2\br}\ll L$. In this case, the branchings are independent from one another and the 3-gluon probability factorizes as 
\beq\label{fact-br}
\cP^{\text{indep}}_3\sim (\cP_2)^2 \sim 
\left(\bar\alpha \frac{L}{\tau_\br} \right)^2. 
\eeq
On the other hand, interferences between the subsequent branchings may occur when the time intervals $[t_1,t'_1]$ and $[t_2,t'_2]$ overlap. For  example this may happen when  $t_1<t_2<t'_1<t'_2$ with  $\tau_1 , \tau_2 <  \tau_\br \ll L $. The corresponding time integrals yield a factor $ L\, \tau^3_{\br} $. In that case,  the contribution to $\cP_3$ from the interferences,  $\cP_3^\text{inter}$, is suppressed by a factor $\tau_\br/L$ with respect to the  factorized form (\ref{fact-br}), and it is obvious to generalize the argument to all possible time configurations.

For consistency, and to complete the picture of independent multiple branching approximation, it is of course necessary that the time between successive branching be large compare to the branching time. But this is guaranteed for inclusive quantities at weak coupling since $t_\ast(\omega)=\tau_\br(\omega)/\bar\alpha\gg \tau_\br(\omega)$.

It follows that, as long as the length $L$ of the system is much larger that the branching time $\tau_\br$, successive branchings of a gluons in a large medium can be treated in first approximation as a probabilistic cascade. The properties of this cascade can be deduced from a set of $n$-point functions for which a generating functional was constructed in \cite{Blaizot:2013vha}.
 
The two building blocks needed in this construction are the elastic and collinear branching rates, or equivalently the  probabilities $P_1$ and $P_2$ given by 
\beq
P_1(\bk|\bp)&=&2(2\pi)E\delta(\omega-E)\cP(\k,t|\p,t),\nn
P_2(\bk_1,\bk_2|\bp )&=&2(2\pi)E\delta(\omega_1+\omega_2-E) \delta(\p-\k_1-\k_2)\delta(\k_1-z\p)  \cP_2(\k_1,\k_2,z;t| \p,t_0) .\nn
\eeq

\begin{figure}[htbp]
\centering
  \centerline{
\includegraphics[width=10.cm]{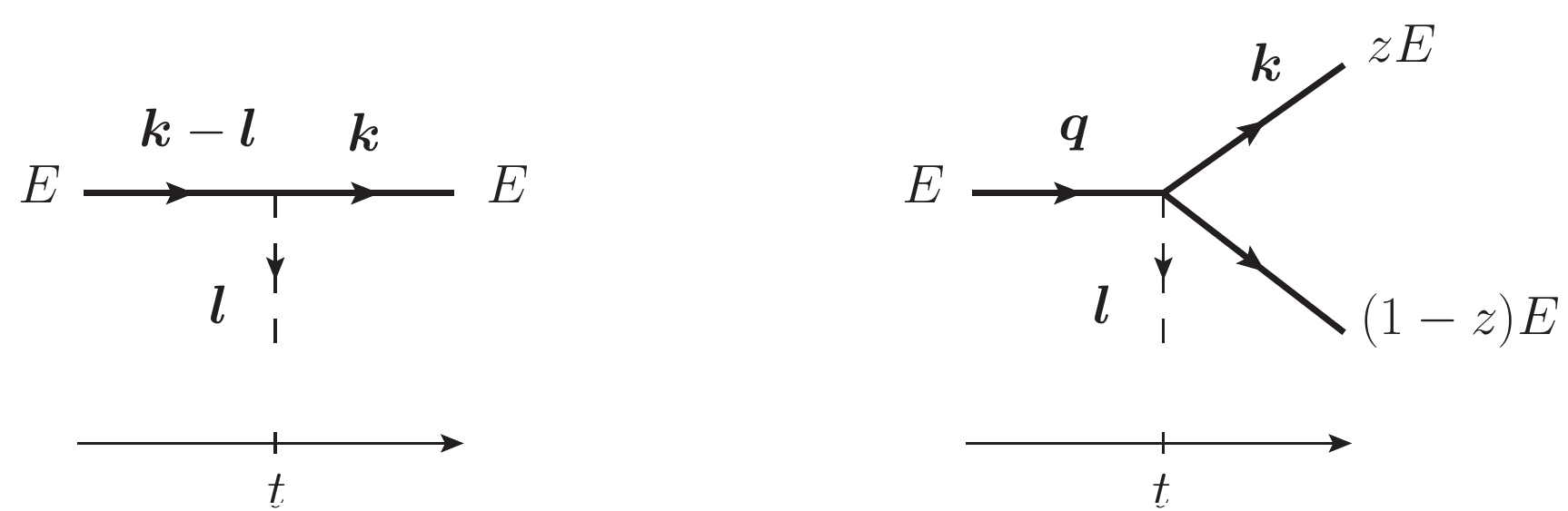}	\label{fig:inel-rate} 
		}
\caption{Graphical representation of the elastic (left panel)  and inelastic (right panel) rates. For collinear branching, $\q=\k/z$ and $\l=\0$. }\label{fig:el-inel-rate}
\end{figure}

In this review, we shall focus solely on the lowest of the aforementioned $n$-point functions, namely the single-inclusive gluon distribution  
\beq\label{D-def}
D(\omega,\k)=\omega \frac{\rmd N }{\rmd\omega \rmd^2\k}&=& \frac{1}{2(2\pi)^3}  \sum_{n=1}^\infty\frac{1}{n!}\int \prod_{i=1}^{n} \,\rmd\Omega_i\, n\, P_n(\bk, \bk_1,\cdots,\bk_{n-1}).
\eeq
It obeys the evolution equation 
\begin{align}\label{Dk}
\frac{\partial}{\partial t}D(x,\k,t)&=\frac{N_cn}{2}\int_{\q}\,
\sigma(\l,t)D\left(x,\k-\l,t\right)\nn
&+\alpha_s\int_0^1\rmd
z \bigg[\frac{2}{z^2}{\cal K}\left(z,\frac{x}{z}E\right)
D\left(\frac{x}{z},\frac{\k}{z},t\right)-{\cal K}\left(z,xE\right)D\left(x,\k,t\right)\bigg].
\end{align}
The first and second terms in the r.h.s. correspond, respectively, to  the elastic and the branching rates represented in Fig.~\ref{fig:el-inel-rate}, with in the latter case $\l=0$.

\subsection{The gluon distribution beyond the collinear approximation}\label{gluon-dist}

In constructing the probabilistic picture in the previous section, we have explicitly assumed that the dominant contribution to the transverse momentum broadening along the cascade comes from  elastic scatterings. 
Some increase in transverse momentum can also arise from the branching processes. 

Let us then estimate the contribution of multiple radiations to transverse momentum broadening. Recall that the typical transverse momentum generated by one radiation of frequency $\omega$ is $k_\br(\omega)\sim (\omega \hat q)^{1/4}$. Combining this with the radiative spectrum, $\omega \rmd I/\rmd\omega\sim L/t_\ast(\omega)$ (cf. Eq.~(\ref{BDMPS-Z })), we obtain the following parametric estimate for the total squared transverse momentum generated in splittings  
\beq\label{kt2-rad-estimate}
\lg k^2_\perp \rg_\text{typ,rad}\sim \int^{\omega_c}_{\omega_\BH}\frac{\rmd \omega}{\omega} k^2_\br(\omega) \frac{L}{t_\ast(\omega)}\sim \bar\alpha \hat q L \ln\frac{L}{\lambda_\el}.
\eeq
where we have integrated over the relevant range of gluon energies from $\omega_\BH$  to $\omega_c$, and we have used the relation $\omega_\BH=\hat q \lambda_{\rm el}$.
Compared to the typical transverse momentum acquired by elastic scattering, the latter estimate is suppressed by a factor $\bar\alpha$, yet enhanced by a potentially large logarithm $\ln L/\lambda_\el$. We will see that this not the dominant contribution. A more detailed analysis reveals  a double logarithmic enhancement  \cite{Liou:2013qya} that is missed by the estimate we just did. We shall return in the next section to a detailed discussion of the origin of this correction. Here, we just indicate how it can be simply obtained by taking into account the momentum broadening taking place during the branching process. This forces us to go beyond the collinear approximation, and hence perform a calculation whose validity is not immediately obvious, but will be justified later.

We note that the 3-point function defined earlier contains the information about the  transverse momentum broadening accompanying a single branching. Hence, instead of integrating over $\l$ and $\Q$, as we did for Eq.~(\ref{kernel}), we define the unintegrated kernel 
\beq\label{kernel-unint}
{\cal K}(\Q,\l,z,E)&= & \frac{P(z)}{[z(1-z)E]^2} \rmR\int_0^\infty \rmd \tau \int_{\Q} (\Q_1\cdot\Q) \tSiii(\Q_1,\Q,\l,t_1,\tau).\nn
\eeq
One can thus generalize Eq.~(\ref{Dkt}),
 \begin{align}\label{Dkt}
\frac{\partial}{\partial t}D(x,\k,t)&=\alpha_s\int_0^1\rmd
z\int_{\Q,\l}\,\bigg[\frac{2}{z^2}\,
{\cal K}\left(\Q,\l,z,\frac{x}{z}E\right)
D\left(\frac{x}{z},\q,t\right)\nn
&-{\cal K}\left(\Q,\l,z,xE\right)D\left(x,\k-\l,t\right)\bigg]-\frac{N_cn}{2}\int_{\l}\,\sigma(\l)D\left(x,\k-\l,t\right)\,.
\end{align}
where $\Q\equiv\k - z(\q+\l)$.  We shall now follow the strategy that we adopted in order to reduce Eq.~(\ref{ME-P})  to the  diffusion equation (\ref{diff-P}). This involves an expansion around the large momentum $\k$ of the followed gluon. The momenta $\Q$ and $\l$, which are at most of the order of $k_\br \equiv \sqrt{z(1-z)\hat q E}$, are small compared to $\k$ which is typically of the order of $\hat q L$. The expansion of the gluon distributions around $\k$ yields for the first term of Eq.~(\ref{Dkt})
\beq\label{D-exp1}
&&D\left(\frac{x}{z},\frac{\k-\delta \k}{z}\right)\nn
&&=  D\left(\frac{x}{z},\frac{\k}{z}\right)-\delta \k\cdot\frac{\del}{\del \k}
D\left(\frac{x}{z},\frac{\k}{z}\right)
+\frac{1}{2!}\, \delta k^i \delta k^j\frac{\del}{\del k_i}\frac{\del}{\del k_j}D\left(\frac{x}{z},\frac{\k}{z}\right)+\cdots\nn
 \eeq 
where we have set $\delta \k\equiv \Q +z\l$. One expands similarly $D\left(x, \k-\l\right)$. 
It is easy to see that the leading terms reproduce Eq.~(\ref{Dk}). The linear terms vanish upon angular integration. Remain the quadratic terms, whose contribution can be cast in the form of the diffusion term, thereby exhibiting a correction $\Delta\hat q$ to the jet quenching parameter.  
For consistency, we shall also simplify the collision term by using the diffusion approximation.
At this point, we anticipate that evaluation of the correction $\Delta\hat q$ meets with logarithmic divergences. These arise from the region $z\sim 1$. To the leading-logarithmic accuracy, we can set $z=1$ everywhere, except in the dominant singularity. The dominant contribution to $\Delta\hat q$ can be then written as
 \begin{align}\label{qhat-rad2}
\Delta\hat q (\k^2)\,=\, 2\alpha_s \int_{x}^1 \rmd z \int_{\Q,\l}\,
 \big[(\Q+\l)^2 - \l^2\big]\,
{\cal K}\left(\Q,\l,z,xE\right),
 \end{align}
where the $\k^2$ dependence arises from the integration boundary $\Q^2,\l^2\ll \k^2$. 
The complete calculation of the integral (\ref{qhat-rad2}) is presented in the next section. One gets\cite{Blaizot:2013vha}
 \begin{align}\label{qhat-right}
\Delta\hat q (\k^2)=\frac{\alpha_s \,N_c }{2\pi}\, \hat q \,
\ln^2\frac{\k^2}{\hat q \tau_{0}}, 
\end{align}
where $\tau_{0}^{-1}$ is the maximum energy that can be extracted from the medium in a single scattering (e.g. $\tau_{0} = 1/T$ for a weakly coupled plasma with temperature $T$). 
We return to this result in the next section. Finally, Eq.~(\ref{Dkt}) can be recast in a form that is similar to that of Eq.~(\ref{Dk}) (after performing the diffusion approximation)\cite{Blaizot:2013vha}, 
\begin{align}\label{Ddiff}
\frac{\partial}{\partial t}D(x,\k,t)&=\frac{1}{4}\left(\frac{\del}{\del \k}\right)^2 \left[(\hat q(\k^2) + \Delta \hat q (\k^2) )D\left(x,\k-\q,t\right)\right]\nn
&+\alpha_s\int_0^1\rmd
z \bigg[\frac{2}{z^2}{\cal K}\left(z,\frac{x}{z}E\right)
D\left(\frac{x}{z},\frac{\k}{z},t\right)-{\cal K}\left(z,xE\right)D\left(x,\k,t\right)\bigg],
\end{align}
where $p_\perp$-broadening due to soft gluon radiation is now taken into account effectively via a redefinition of the quenching parameter  $\hat q \to \hat q+\Delta \hat q$.

We are now ready to revisit the standard estimate of the typical momentum broadening Eq.~(\ref{kt2-typ}), making use of the redefinition of the quenching parameter above, substituting $\k^2 \sim \hat q L $ in Eq.~(\ref{qhat-right}),  we obtain 
\beq
\lg k^2_\perp\rg_\text{typ}  \simeq \hat q L \left[1+\frac{\alpha_s N_c}{\pi}\ln^2\frac{L}{\tau_0}\right].
\eeq
This result agrees with that obtained in Ref.~\cite{Liou:2013qya} using a different approach.\footnote{Note that NLO corrections to $p_\perp$-broadening have been also investigated in Deep Inelastic Scattering in the High-Twist approach \cite{Kang:2013raa}.}
Note that we obtain a double logarithm of the medium length, as opposed to the heuristic estimate (\ref{kt2-rad-estimate}), which involve a single logarithm. Although, the soft (short time) logarithmic enhancement is correctly captured in this estimate, the single scattering that contributes to the branching probability that in turn leads to the second logarithm, is missing in the BDMPS-Z spectrum that is used Eq.~(\ref{kt2-rad-estimate}).

\section{Renormalization of the jet quenching parameter}\label{RG-qhat}

The correction to the  jet quenching parameter discussed in the previous section points to the existence of potentially important radiative corrections. The calculation that we just presented focussed on a correction that is singular, and this is what allowed us to retain  in the branching kernel contributions that should not be kept a priori since, if they were not divergent,  they would be of the same order of magnitude as terms that  have been consistently neglected in arriving at Eq.~(\ref{Dk}). In this section we discuss a more systematic approach to the calculation of radiative corrections. Still, at present, the program of calculating the radiative corrections is incomplete\cite{Wu:2011kc,Liou:2013qya,Blaizot:2013vha,Blaizot:2014bha,Iancu:2014kga,Wu:2014nca}. As we shall see,  one is  able so far  to complete the calculation only for the leading singular part. Furthermore, it is only this singular part that can be interpreted as a correction to the jet quenching parameter; less singular corrections would presumably be non local, thereby spoiling their interpretation in terms of a correction to the jet quenching parameter.

\subsection{Radiative corrections to momentum broadening}
We start by revisiting the calculation of the momentum broadening. Recall that this is described by the 2-point function $\Sii (t_1,t_0)$, Eq.~(\ref{2-pt-def}), for which the following equation is easily established 
\beq\label{integralequation2}
\Sii (t_1,t_0) = \Siio (t_1,t_0) +\int \rmd t_3 \int \rmd t_2\Siio(t_1,t_3 ) \, \Sigma^{(2)}(t_3,t_2)\,  \Sii (t_2,t_0 ). 
\eeq 
In this equation, the instantaneous interaction kernel $\Sigma^{(2)}$ is related to the dipole cross-section (\ref{dipole-cs}), 
\beq\label{sigma-2}
(\x|\Sigma^{(2)}(t_3,t_2) |\y) =\frac{N_c n}{2}  \delta (\x-\y) \delta (t_3-t_2)\sigma(\x).
\eeq

\begin{figure}[htbp]
\begin{center}
\includegraphics[width=5cm]{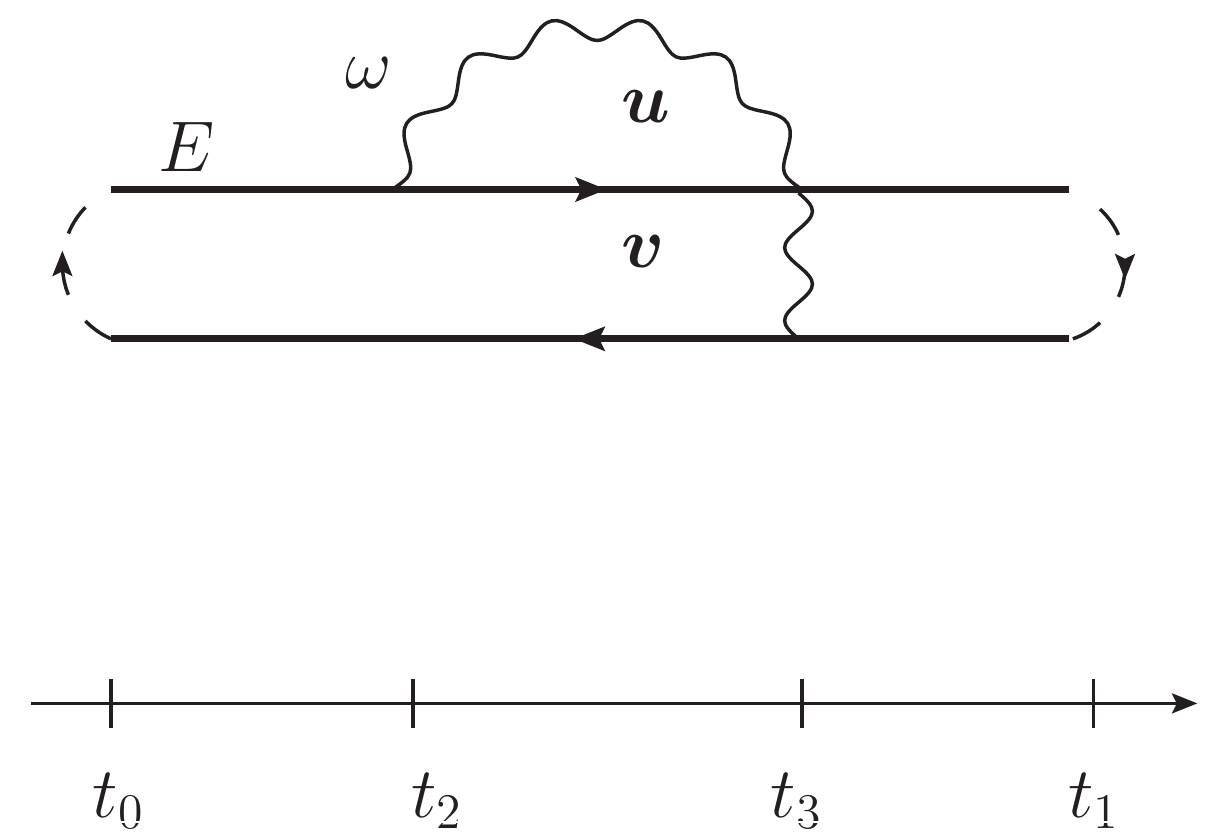}\qquad
 \includegraphics[width=5cm]{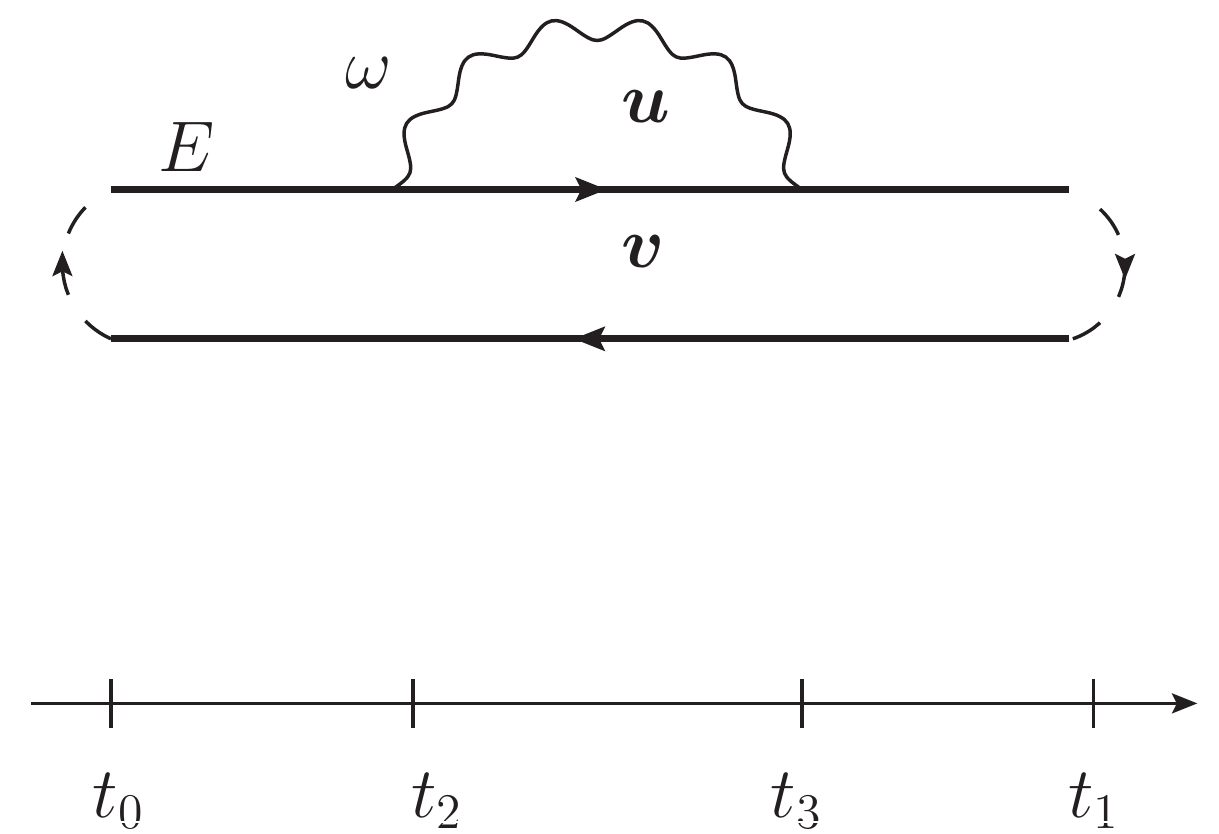}\\

\caption{Real (left) and virtual (right) contributions to the radiative correction to the 2-point function. There are other diagrams that are not shown, corresponding to a different time ordering of the emission and absorption vertices in the real term, and a diagram where the virtual correction is attached to the lower line. The transverse vector $\v$ stands for the transverse size of the frozen hard dipole (represented by the straight lines), whose constituents carry an energy $E$,  and  $\u$ corresponds to the time-dependent transverse distance between the radiated gluon (represented by a  wavy line) and its radiator whose dynamics is encoded in the 3-point function $\tSiii$. The integration over $\u \sim \q^{-1}$ in the region $ \u \gg \v$ yields the second logarithmic enhancement of the radiative correction to $\hat q$ (cf. Eq.~(\ref{qhat1})).   }
\label{fig5}
\end{center}
\end{figure}

We are interested in the radiative correction to $\Sigma^{(2)}$. The corresponding diagrams are displayed in Fig.~\ref{fig5}. The important par is that involving the three gluons evolving between the times $t_2$ and $t_3$, and to which we refer as a 3-point function. The dominant, double logarithmic contribution corresponds to the case where the radiated gluon (the wavy lines in Fig.~\ref{fig5}) is soft, i.e., its energy $\omega$ is  $\omega \ll E$. 
A simplification then occurs in the calculation of the diagrams since we may assume that the transverse coordinates of the hard gluons (in the amplitude and its complex conjugate) are frozen during the lifetime of the fluctuation, i.e., between $t_2$ (in the amplitude) and $t_3$ (in the complex conjugate amplitude). 
The calculation leads then, for the real correction, and using a symbolic matrix notation, 
\beq\label{DS2-2a}
&&\Delta \Sii (t_1,t_0)=  -\int _{t_0}^{t_1}d t_3 \int _{t_0}^{t_3}dt_2  \,  \Sii (t_1,t)\,\Delta \Sigma^{(2)}(t_3,t_2)\,\Sii (t_2,t_0)\,, 
\eeq
with 
\beq\label{sigmaSigma2}
&&(X_3|\Delta\Sigma^{(2)}(t_3,t_2)|X_2) \nn
&&=\delta(X_3-X_2)\frac{\alpha_sN_c}{2}\,2\text{Re}\int \frac{d\omega}{\omega^3}\,\bdel_{\u_2}\cdot\bdel_{\u_3}\tSiii(\u_2,\u_3,\v;\tau)|_{\u_2=0, \u_3=-\v}\,,\nn
\eeq
where $\tau\equiv t_3-t_2$. The 3-point function $\tSiii(\u_2,\u_3,\v;\tau)$  encodes the dynamics in the transverse plane of the system made of a frozen dipole and a soft gluon. It corresponds to the soft limit, i.e. $z=\omega/E \ll 1$, of Eq.~(\ref{S3harm}). Here, $\v$ stands for frozen transverse size of the hard dipole during the time interval $\tau$, while $\u_2=\0$ and  $\u_3=-\v$  stand for the transverse separation between the radiated gluon and its radiator (the hard gluon in the amplitude) at the times $t_2$ and $t_3$ respectively (see left panel of Fig.~(\ref{fig5}) for an illustration).
The virtual correction is given by the same formula in which $\u_3=\0$ since the gluon is emitted and reabsorbed by the same line.
Anticipating on an argument that will be provided later, we treat the correction $\Sigma^{(2)}(t_3,t_2)$ effectively as a local correction, i.e., set formally $\Sigma^{(2)}(t_3,t_2)\propto \delta(t_3-t_2)$.
 After replacing in Eq.~(\ref{DS2-2a}) the factors $S^{(2)}$ external to the radiative correction by free 2-point functions (which are independent of time), and performing the $t_2$ integration, we get
\beq\label{DS2-3}
&&(X_1|\Delta \Sii (t_1,t_0)|X_0)= -(t_1-t_0)\delta(X_1-X_0)\frac{\alpha_sN_c}{2}\,\int \frac{d\omega}{\omega^3} \int  d\tau  K(\v,\tau),\nn
\eeq
with

\beq\label{K-x}
K(\v,\tau)&\equiv & 2\text{Re} \int_{\q_2, \q_3, \l}\left( e^{i(\l-\q_3)\cdot\v}-e^{i\l\cdot\v}\right)\,
(\q_2\cdot\q_3)\tSiii(\q_2,\q_3,\l;\tau), 
\eeq
where we have performed a Fourier transform of the 3-point function. Besides an integration over $\tau$, we recognize in Eq.~(\ref{K-x}) the soft limit, i.e. $z\ll 1$ of the kernel Eq.~(\ref{kernel-unint}). 
Note the vanishing of this expression when $\v\to 0$. 
It results from a cancellation between real and virtual terms, reflecting the property of color transparency of a dipole cross section. 

The identification of the correction to the dipole cross section proceeds  by comparison with the integral equation (\ref{integralequation2}) (more properly, its generalization, whose validity relies on the  locality assumption)  for fixing the relation between $\Delta \sigma$ and $\Delta\Sigma^{(2)}$. One then gets, leaving the bounds on the $\tau$ integration unspecified for the moment, 
\beq\label{dsig-1}
\frac{N_c n}{2} \Delta\sigma(\v)&=&\alpha_s \frac{N_c }{2} \int \frac{\rmd\omega}{\omega^3} \int d\tau\,K(\v,\tau)\approx \frac{1}{4}\v^2\Delta\hat q,\nn
\eeq
where the last equality stems from Eq.~(\ref{sigma-H}) expressing $\Delta\sigma(\v)$ at small $\v$ in the harmonic approximation, by expanding the phases in Eq.~(\ref{K-x}). We use this relation to interpret the correction $\Delta\sigma(\v)$ as a correction to the parameter $\hat q$.  Note that in this approximation the resulting integrations over the transverse momenta $\q_2$ and $\l$ are bounded from above by $\v^{-2}$.

The explicit calculation of $\Delta\hat q$, using the expression (\ref{S3harm}) of the reduced 3-point function, yields \cite{Blaizot:2013vha,Blaizot:2014bha}
\beq\label{dsig-0}
\Delta \hat q\,&=&
\frac{\alpha_s \,N_c}{\pi}\, 2\text{Re }
\int \rmd\omega \,\int \rmd \tau  \,\frac{i \Omega^3}{\sinh(\Omega \tau)}\left[1+\frac{4}{\sinh^2(\Omega  \tau)}\right]\,,\nn
&\simeq & \frac{\alpha_s \,N_c }{\pi}\, \,
\int  \frac{\rmd\omega}{\omega}\int^{\Omega^{-1}}
\frac{\rmd\tau}{\tau}\,\hat q\,,\nn
\eeq
which as anticipated exhibits a double logarithmic divergence, when $\tau\to 0$ and $\omega\to 0 $.   
As clear from Eq.~(\ref{dsig-0}) the $\tau$ integral is bounded at the upper end by  $|\Omega|^{-1}\sim\tau_{\rm br}(\omega)=\sqrt{\omega/\hat q}$ (cf. Eq.~(\ref{Omega})) corresponding to the onset of the multiple scattering regime and the LPM effect: the relevant gluon fluctuation experiences a single scattering with the medium constituents. 
In order to systematically account for the boundaries of the double integral,\footnote{For a detailed discussion on the boundaries of the logarithmic integrals see Ref.~\cite{Liou:2013qya,Iancu:2014kga}}  it is in fact more convenient to change variables, from $(\omega,\tau)$ to $ (\q, \tau)$,  
with $\tau\equiv \omega/\q^2$ the formation time of the radiated gluon, and $\q$ its transverse momentum which can run up to $\p^2\equiv \v^{-2}$ (the logarithmic transverse momentum integration can be rephrased in terms of the transverse coordinate of the radiated gluon $\q^{-1}\sim \u \gg \v$, see Fig.~\ref{fig5}). We obtain then
\beq\label{qhat1}
\Delta \hat q (\tau_\text{max}, \p^2)\equiv \frac{\alpha_sN_c}{\pi} \, \int^{\tau_\text{max}}_{\tau_0} \frac{\rmd\tau}{\tau}\int^{\p^2}_{\hat q\tau}\frac{\rmd\q^2}{\q^2}\, \hat q (\q^2)\,,
\eeq
where we have explicitly indicated the scale dependence of $\hat q$. The boundary corresponding to the region of multiple scattering now appears as the lower bound of the $\q$ integration,  $ \q^2 \gg  k^2_\text{br}\equiv \hat q\tau $. 
 Since medium-induced gluons forms inside the medium, the largest value for $\tau$ is the length of the medium $L\sim t-t_0$. As for the lowest value $\tau_0$, it can be interpreted as the inverse of the largest energy that can be extracted from the medium through a single scattering. 

For a constant $\hat q $ in the integral, one can easily perform the integrations, and by keeping the leading contributions, we recover the result first derived in \cite{Liou:2013qya,Blaizot:2013vha}, 
\bal \label{deltaqhat}
 \Delta \hat q \simeq  \frac{ \alpha_s C_A}{2\pi}\, \hat q \ln^2\left(\frac{L}{\tau_0}\right),
\end{align}
where we have used the fact that $\p^2\sim \hat q L$.

\subsection{Locality of radiative corrections}\label{locality-RC}

So far we have discussed a single radiative correction. The logarithmic phase space for radiative corrections extends from $\tau_0$ to $L$ which makes them explicitly non-local. One may therefore  question why we have argued that these  corrections could be absorbed into a redefinition of $\hat q$, considered to be a local transport coefficient. Indeed, the quenching parameter makes sense strictly speaking only in the approximation of independent multiple scatterings. This requires the duration of each interactions to be much smaller than the mean-free path. 

The key lies in the nature of the logarithmic integral whose boundaries do not affect the overall multiplicative constant. Thus, to the extent that one restricts oneself to a leading order calculation, one can proceed as if the lifetimes of the fluctuations involved in the radiative corrections were small, and treat them as local. The coefficient of the leading double logarithm is calculated correctly, and corrections coming from overlapping contributions will be subleading. In short, the leading corrections  can be treated as being effectively local in time and thus independent \cite{Blaizot:2014bha}. 
This argument is what allows us to generalize Eq.~(\ref{integralequation2}) to a $\Sigma(t,t')$ that includes the radiative corrections, to treat the double logarithmic correction to  $\Sigma(t,t')$ as   effectively local, and multiple radiative corrections  as independent. As we have seen, such corrections can be interpreted, in the coordinate space description, as a modification of the dipole cross-section $
\sigma(\v)\,\to\, \sigma(\v)+\Delta\sigma(\v,\tau_\text{max})$, 
where $\tau_\text{max}\equiv t_1-t_0$, or in momentum space as a modification of the equation for the momentum broadening probability  \cite{Blaizot:2014bha}
\beq\label{P-mom-FP-rad}
\frac{\del}{\del t}{\cal P}(\p,t) = \frac{1}{4}\,\left(\frac{\del}{\del \p}\right)^2\,\left[ \hat q(\p^2)+\Delta\hat q(\p^2)\right]\,{\cal P}(\p,t)\,,
\eeq
which generalizes Eq.~(\ref{diff-P}). In both cases, the radiative corrections are accounted for by a correction to the jet quenching parameter.  
\subsection{Radiative corrections to Energy Loss and Universality }\label{rad-E-loss}
We now address the question of whether the absorption of the dominant logarithmic contributions to the radiative corrections in a redefinition of $\hat q$ is universal, that is, whether it holds for other observables than the broadening probability. This has important implications as it would affect the branching rate in Eq.~(\ref{Ddiff}), that controls the number of final state gluons. In order to address this issue, we consider radiative corrections to the BDMPS-Z spectrum (\ref{BDMPS-Z }) that can be recovered in limit $\omega=zE \ll E $ of the branching rate (\ref{kernel}).

\begin{figure}[htbp]
\centering{
\includegraphics[width=8cm]{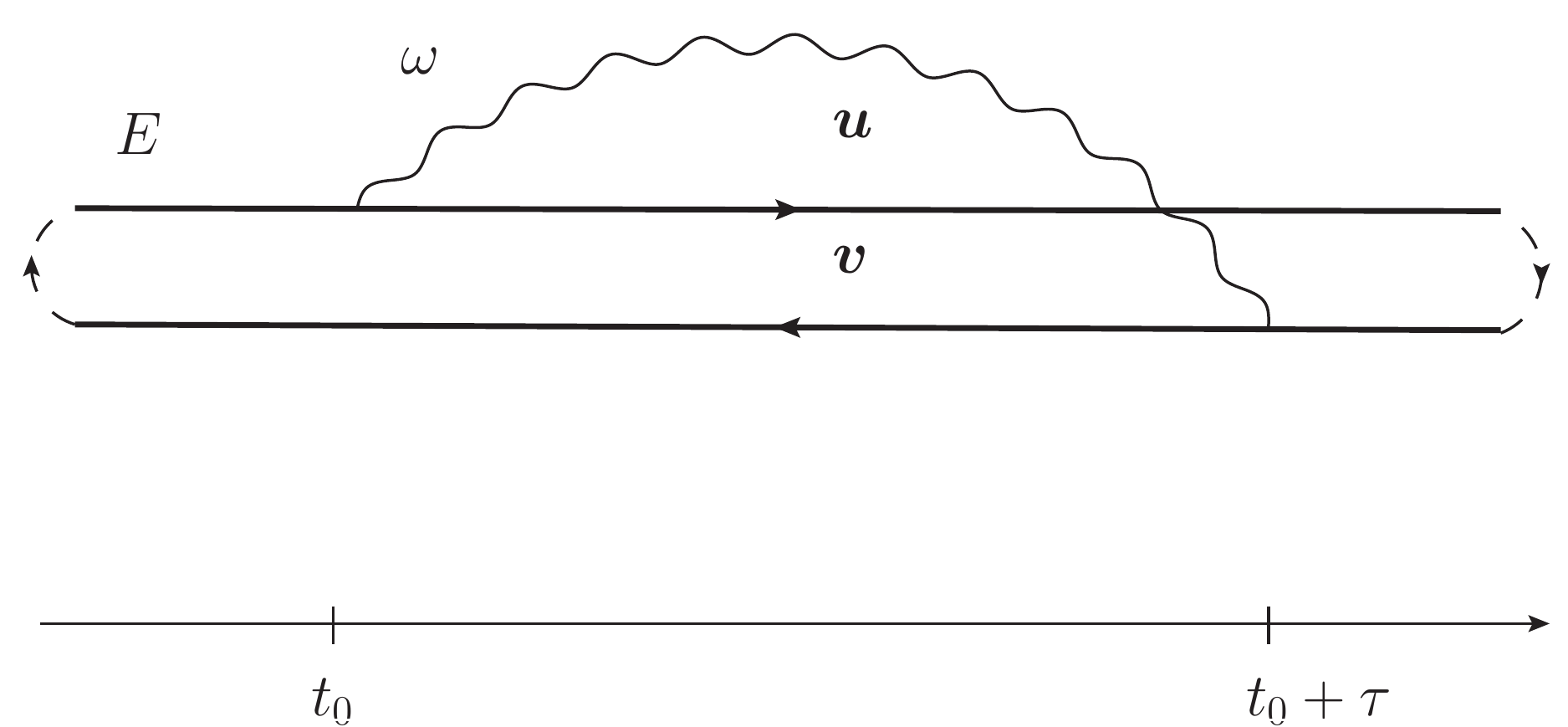} 
\caption{Diagrammatic illustration of the BDMPS-Z spectrum (\ref{kernel5}) that corresponds to the radiation of a soft gluon with energy $\omega$. We denote by $\u$ the transverse position of the radiated gluon with respect its radiator, the energetic dipole whose size is denoted by  $\v$.  }\label{fig:S3}}
\end{figure}

The BDMPS-Z spectrum (\ref{BDMPS-Z }) of radiated gluons with frequency  $\omega\ll E$, can be calculated from the reduced 3-point function (\ref{S3harm}) \cite{Zakharov:1996fv}, according to \footnote{The vacuum part of the 3-point point function, $\tSiii_0$,  is implicitly subtracted.}
  \beq\label{kernel5}
\frac{\rmd I}{\rmd\omega \rmd t}\equiv\, \frac{\alpha_sN_c}{\omega^2}\, 2\text{Re}
\int_{0}^{\infty} \rmd\tau\, \del_\u\cdot\del_{\u'}\, \left.\tilde S^{(3)}(\u,\u', \v;\tau)\right|_{\v=\u=\u'=\0}.
\eeq
Here the variable $t$ runs up to $\sim L$. This spectrum is valid in the large medium length limit where the gluon branching time $\tau_{\rm br}=\sqrt{\omega/\hat q}\ll L$ and the integration over the gluon formation time $\tau < L$ is suppressed exponentially beyond $\tau_{\rm br}$. The transverse coordinates $\u$ and $\u'$ correspond to initial and final coordinates of the radiated gluon, and  $\v$ corresponds to the size of the energetic dipole that radiates the gluon (see Fig.~\ref{fig:S3}).   This is why we can integrate $\tau$ up to infinity in Eq.~(\ref{kernel5}).

The 3-point function $\tilde S^{(3)}$ is given explicitly by Eq.~(\ref{S3harm}) in the harmonic approximation, and $\tau=t_1-t_0$, with $t_0$ and $t_1$ denoting the times of the emission in the amplitude and its complex conjugate, respectively (see Fig. \ref{fig:S3}). The BDMPS-Z spectrum of Eq.~(\ref{BDMPS-Z }) is easily recovered from this expression by performing the integrations over the transverse momenta and over $\tau$ \cite{Mehtar-Tani:2013pia,Blaizot:2012fh}. As is clear from Eq.~(\ref{S3harm}) the reduced 3-point function depends explicitly on $\hat q$. One can show that the leading radiative corrections do not modify $\tilde S^{(3)}$, except for a  change in the value of the parameter $\hat q$, the correction to $\hat q$ being,  besides,  the same as that calculated in the previous section for momentum broadening. 

\begin{figure}[htbp]
\centering{
\includegraphics[width=8cm]{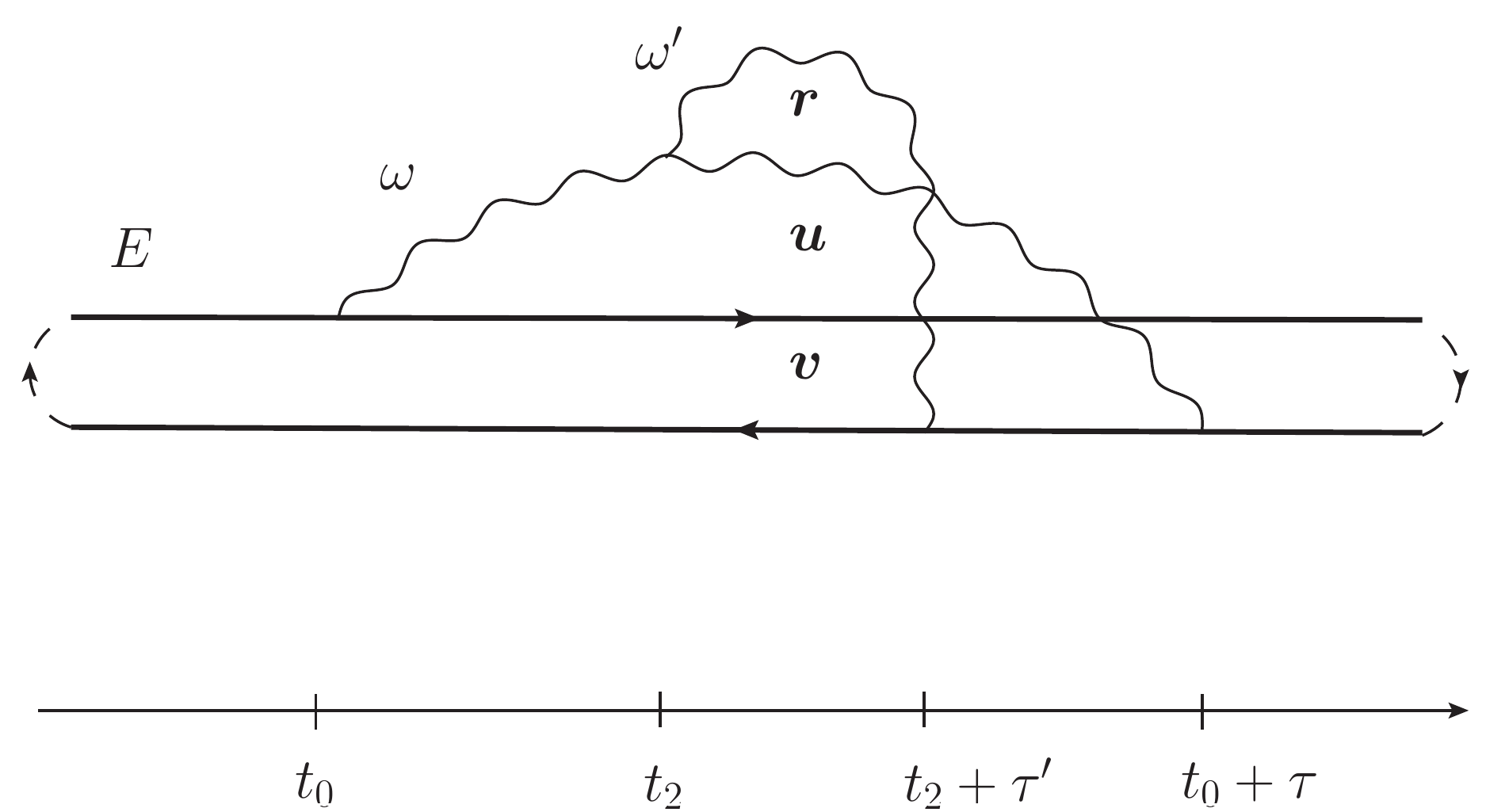} 
\caption{ Diagrammatic illustration of the radiative correction to the BDMPS-Z spectrum (\ref{kernel5}). The gluon $\omega'$ is integrated over and account for real and virtual contributions. Here one topology among 9 is displayed corresponding to the radiation of the gluon $\omega'$ from gluon $\omega$ and its absorption by the gluon $E$ in the complex conjugate amplitude. The labeling is as in the previous figure, with here, in addition, $\r$ denoting the transverse coordinate of the gluon fluctuation $\omega'$. The integration over $\r \sim \q^{-1}$ in the region $ \r \gg \u$ yields the second logarithmic enhancement of the radiative correction to $\hat q$ (cf. Eq.~(\ref{dsig-0})).  }\label{fig:S3-RC}}
\end{figure}

Quite generally, we are concerned with the radiative corrections of the 3-point function $S^{(3)}$, whose graphical interpretation is given in Fig.~\ref{fig:S3}.  The  diagram displayed there corresponds typically to a branching process where a gluon with initial energy $E$ (represented by the lower two thick lines in the amplitude and the conjugate amplitude, respectively), emits a  gluon at $t_0$ in the amplitude and reabsorbs it in the complex conjugate amplitude  at $t_1=t_0+\tau$, with energy $\omega=zE$.
The correction to the 3-point function (in the limit $\v\to\0$ as required by the integration over $\l$ in Eq.~(\ref{S3harm})) can now be written by analogy with Eq.~(\ref{DS2-2a}) as (see Fig.~\ref{fig:S3-RC} for an illustration) 
\beq\label{DS3-2}
\Delta \Siii  (t_1,t_0)\simeq - \int _{t_0}^{t_1}\rmd t_3 \int _{t_0}^{t_3}\rmd t_2   \Siii (t_1,t_2) \, \Delta\Sigma^{(3)}(t_3,t_2)\,\Siii (t_2,t_0)\,,
\eeq
where 
\beq
(X_3|\Delta\Sigma^{(3)}(t_3,t_2)|X_2)=\delta(X_3-X_2)\frac{\alpha_s N_c}{2}\int \frac{d\omega'}{\omega'^3}K(\u,\tau')\,.\nn
\eeq
At this point we proceed as for the 2-point function, and treat the correction $\Delta\Sigma^{(3)}(t_3,t_2)$ as a local correction. By comparing with Eq.~(\ref{S3harm}) in the limits $z\to 0$ and $\v\to 0$, one then gets
\beq\label{Deltasigma3}
\frac{N_c n}{2}\Delta\sigma(\u)=\frac{\alpha_s N_c}{2}\int^{\tau_\text{max} } \rmd \tau'\int \frac{\rmd\omega'}{\omega'^3}K(\u,\tau')\,,\nn
\eeq

Thus,  to double logarithmic accuracy, the radiative corrections to the reduced 3-point function are accounted for by correcting the dipole cross-section and thus  the jet quenching parameter.  This result applies to the particular 3-point function involved in the calculation of the gluon spectrum (\ref{kernel5}). In order to evaluate the corrected spectrum, we need to perform the integration over  $\tau$ in  Eq.~(\ref{kernel5}), replacing $\hat q$ by $\hat q +\Delta\hat q (\tau_\text{max})$. However, since the correction to $\hat q$ is computed to double logarithmic accuracy one can simply replace the variable $\tau_\text{max}$ in Eq.~(\ref{Deltasigma3}) (see also Eq.~(\ref{qhat1})) by its typical value in the radiation process, i.e., $\tau_\text{max}\simeq \tau_\text{br}\equiv \sqrt{\omega / \hat q}$. The integral over $\tau$ in the BDMPS-Z spectrum (\ref{kernel5}) can then be performed as for the case with no radiative correction. Doing so, we obtain \cite{Blaizot:2014bha} 
\beq\label{BDMPS-Z -rad}
\frac{\rmd I}{\rmd\omega \rmd t}\equiv \frac{\alpha_s N_c}{\pi } \sqrt{\frac{\hat q+\Delta \hat q}{\omega}},
\eeq
where for a constant $\hat q$ one gets, from Eq.~(\ref{qhat1}) letting $\p^2\simeq k^2_\text{br}(\omega)\equiv \sqrt{\omega\hat q} $,
\beq
\hat q+\Delta \hat q \approx \hat q\left[1+\frac{\alpha_sN_c}{2\pi}\ln^2\sqrt{\frac{\omega}{\hat q \tau^2_0}}\right]. 
\eeq
It can be shown that this result extends to the full kernel (\ref{kernel}) in the large $N_c$ limit \cite{Blaizot:2014bha}.  These results were recently confirmed by 
an alternative calculation of the radiative corrections to the mean-energy loss to double logarithmic accuracy\cite{Wu:2014nca}.
 
\subsection{Renormalization of the quenching parameter}

\begin{figure}[htbp]
\centering
\includegraphics[width=7.cm]{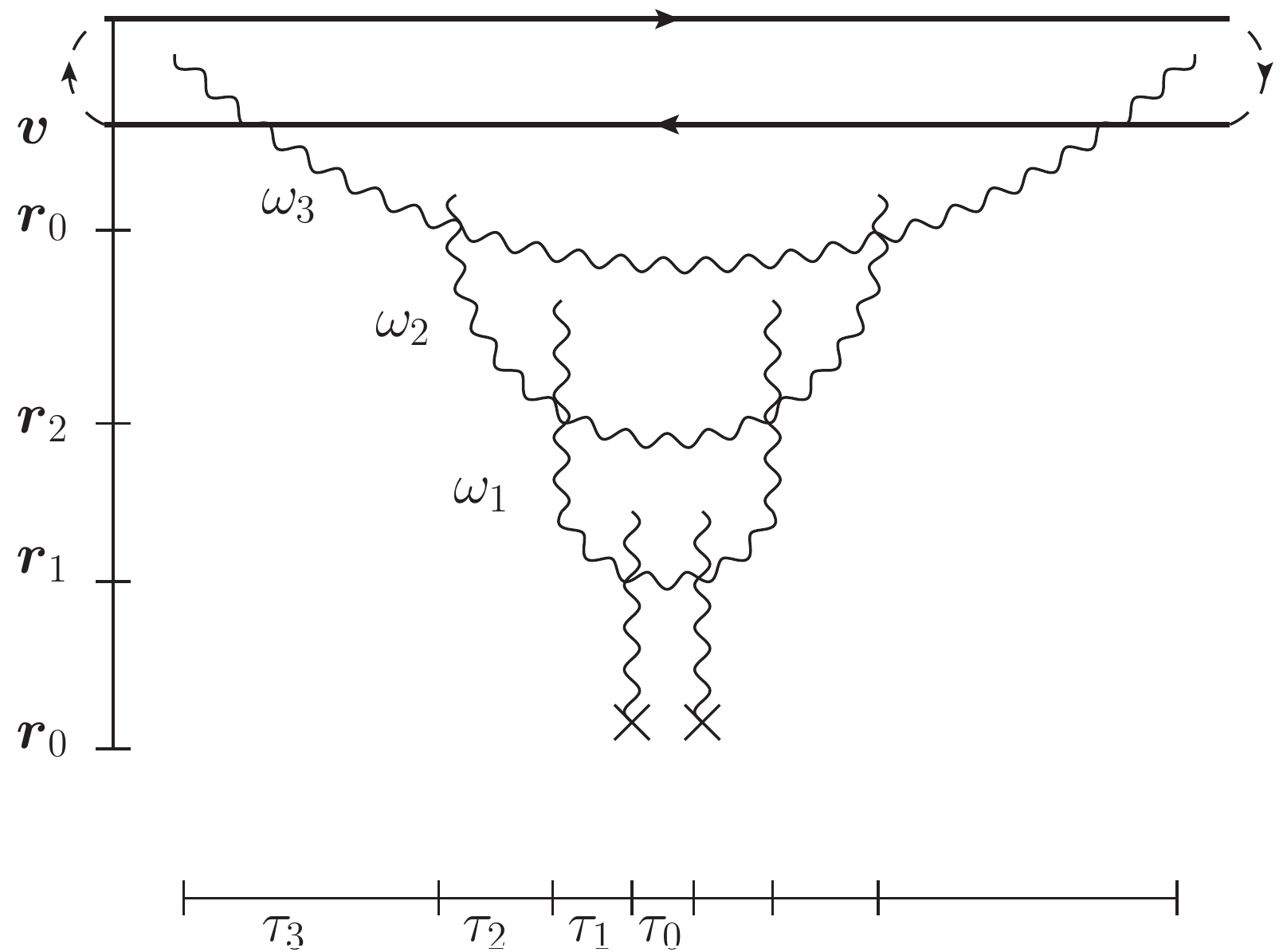}	\label{fig:RG-qhat} 
\caption{A diagrammatic representation of the resummation of double logarithmic contributions for a single scattering. Here three gluon emissions are depicted.  In the double logarithmic approximation successive radiations are strongly ordered in formation time, i.e. $\tau_0 \ll \tau_1\ll \tau_2\ll \tau_3\ll L$, and in transverse sizes, i.e., $1/m_D \sim \r_0 \gg \r_1\gg \r_2\gg \r_3\gg \v $. For the original dipole the subsequent radiations look {\it as if } they were instantaneous.  }		
\end{figure}

For large media, as soon as  $\bar\alpha\ln^2(L/\tau_0)\sim1$, one has to resum the double logarithmic power corrections. Unlike the previous resummation of independent multiple radiative corrections, this now involves radiative corrections that are correlated to each other. To understand how this resummation proceeds, we denote the standard leading order definition of the jet-quenching parameter by $\hat q_0$ and we note that the first correction to the jet-quenching parameter, $\hat q_1(\tau,\k^2)\equiv \Delta \hat q(\tau,\k^2)$ is proportional to the 3-point function, $S^{(3)}[\hat q_0]$ which is itself  a function of the leading order $\hat q_0$.  As we have shown, under radiative corrections,  the 3-point function gets renormalized by a simple modification of $\hat q$, that is,  $S^{(3)}[\hat q_0]\to S^{(3)}[\hat q_0+\hat q_1] $. This allows us to compute the second correction from Eq.~(\ref{qhat1}),
\beq
\hat q_2(\tau,\k^2)= \bar\alpha \, \int^{\tau }_{\tau_0} \frac{\rmd\tau'}{\tau'}\int^{1/\x^2}_{\hat q_0\tau'}\frac{\rmd\q^2}{\q^2}\,\hat q_1(\tau',\q^2).
\eeq
One sees  emerging a self-similarity that results from the separation of time scales involved in the computation of the leading logarithms.
The structure of the first double logarithmic corrections being set, the next corrections that yield double logarithms will follow the same systematics, with successive gluonic fluctuations ordered in formation time $\tau_0 \ll \tau _1\ll ...\ll\tau_n\equiv \tau_{\text{max}} $, or in transverse size $\r_0 \gg \r _1\gg ...\gg \r_n\equiv \r_{\text{max}} $, or in transverse momentum $m_D  \ll \q _1\ll ...\ll\q_n\equiv \k $. A diagrammatic illustration is given in Fig.~\ref{fig:RG-qhat}. The difference with the standard Double-Logarithmic Approximation (DLA) is the limits of logarithmic phase-space set by the LPM effect since, i.e., multiple-scatterings since in the DLA only a single scattering contributes, which imposes that the formation time of a fluctuation to be smaller than the BDMPS-Z formation time, or in terms of our transverse momentum variables,  $\q^2 \gg  \hat q_0\tau $ \footnote{The resummation of double logarithms in the quenching parameter was postulated earlier in Ref.~\cite{CasalderreySolana:2007sw} where the LPM suppression was not taken into account.}. The following equation resums the  double logarithmic corrections to all orders
\beq\label{qhat-evol}
\frac{\del \hat q (\tau,\k^2)}{\del \ln \tau} = \, \int_{\hat q  \tau }^{\k^2}\frac{d\q^2}{\q^2}\, \bar \alpha(\q)\, \hat q (\tau,\q^2).
\eeq
with some initial condition $ \hat q (\tau_0,\k)$. We have let the coupling running at the transverse scale $\q$.
The important feature of this equation is that it predicts the evolution of the jet-quenching parameter from an initial condition $\hat q_0$ (which can be computed e.g. on the lattice, or  to leading order in $\alpha_s$, which implies, $\hat q(\tau_0) \equiv\hat q_0  $ as given by the leading order result (\ref{qhat}). The $\tau_0$ cut-off that was introduced  to cut the logarithmic divergence in the radiative corrections, can be seen as a factorization scale.  
The solution to this equation was derived in \cite{Liou:2013qya} for the pt-broadening in the case where $\hat q_0=\hat q(\tau_0)$ is constant and for a final $\tau=L$ and $\k^2=\hat q_0L$, merging the 2 independent variables at the end of the evolution. The solution reads 
\beq\label{qhat-L}
\hat q(L)=  \frac{1}{\sqrt{\bar\alpha}}\,I_1\left(2\sqrt{\bar\alpha}\ln\frac{L}{\tau_0}\right)\,\hat q(\tau_0).
\eeq
A semi-analytic analysis of this equation including the running of the coupling has been performed in Ref.~\cite{Iancu:2014sha}.
For large $L$, the quenching parameter scales like $\hat q(L)\sim L^\gamma $, with the anomalous 
\beq\label{anomalous-d}
\gamma=2\sqrt{\bar\alpha}\,.
\eeq
Interestingly, the resummation of large double logarithms modifies the scaling of the energy loss with $L$,
 \beq
\Delta E \sim L^{2+\gamma},
\eeq
a correction that  seems to fall between the standard small coupling result, $\Delta E  \sim L^2 $ and the strong coupling result obtained with the help of the AdS-CFT correspondence in ${\cal N}=4$ SYM theory\cite{Hatta:2007cs,Chesler:2008uy}, $\Delta E\sim L^3 $.

Finally, let us make now a rough estimate of the renormalized quenching parameter. The standard perturbative estimate yields a value of about $\hat q _0\sim 1$ GeV$^2$/fm \cite{Baier:1996sk}.  For $\alpha_s\sim0.5$, $L\sim5$ fm and $T=\tau_0^{-1}\sim 500$ MeV, Eq.~(\ref{qhat-L}) yields a sizable increase by a factor 2, $\hat q \sim 2$ GeV$^2$/fm. This result is in the ballpark of the jet quenching parameter values extracted from the data \cite{Burke:2013yra}. 

\section{Energy flow and angular structure of the in-medium cascade}\label{energy-angle}

We return now to the medium induced cascade, and investigate some of its characteristic features as revealed by the analysis of the inclusive distribution function $D(x,\k)$. We shall discuss in particular the energy and angular distributions of the emitted gluons. As we shall see, these properties make the BDMPS-Z  cascade very different from the more common parton cascades in vacuum that are described by the DGLAP equation. 

\begin{figure}[htbp]
\centering
		\includegraphics[width=12.cm]{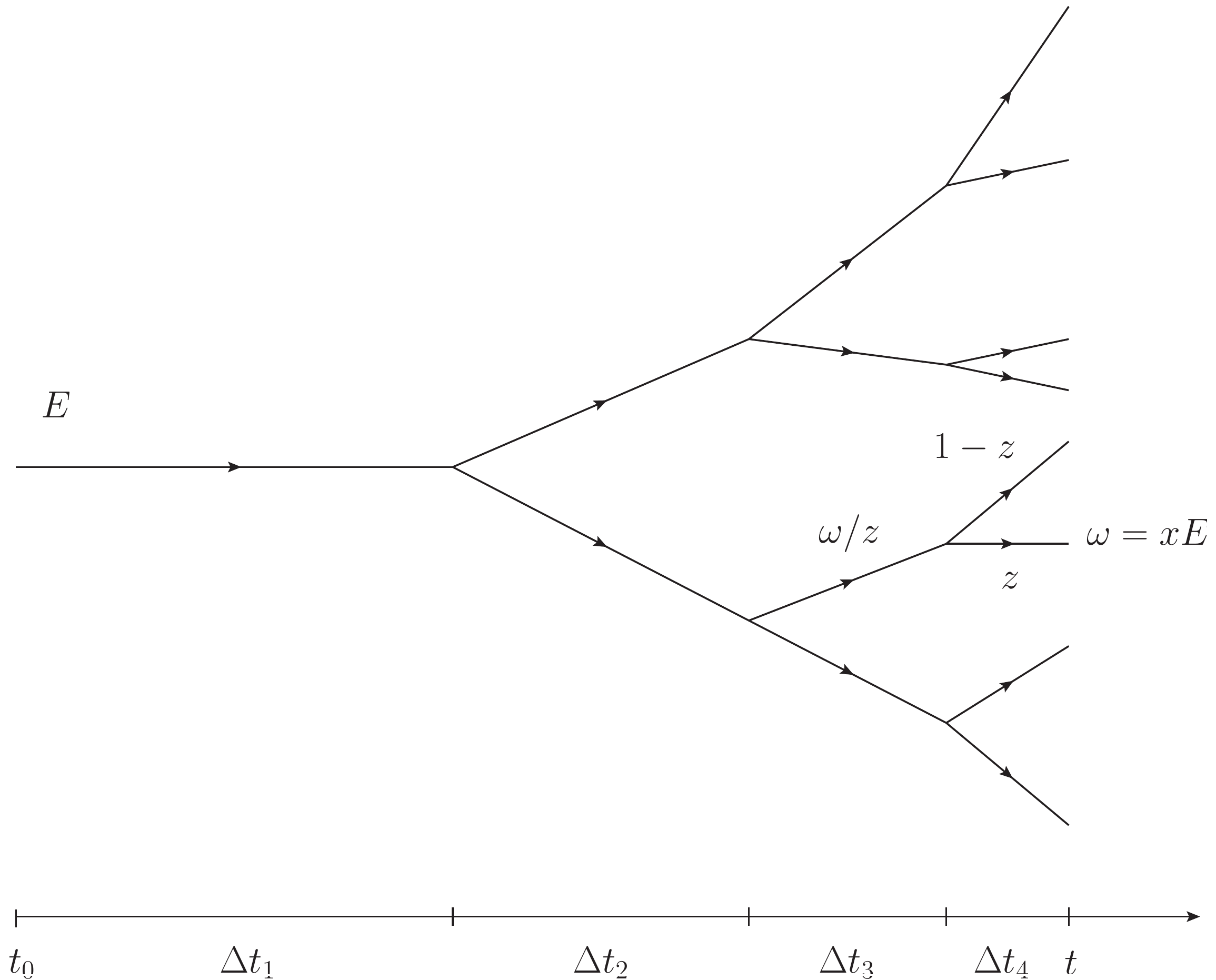}	\label{fig:shower2}
\caption{Illustration of a gluon cascade that is initiated by a gluon with energy $E$. Four generations are displayed. The branching time $\Delta t_i\sim t_\ast(x_i)$, that corresponds to the lifetime of generation $i$, decreases after each branching as in a BDMPS-Z  cascade. The inclusive distribution $D(x,\btheta)$ measures the probability to find in the cascade, at time $t$,  a gluon with energy $xE$.  The rate equation (\ref{evol-eq-ang0}) describes how this distribution evolves with time $t$. }

\end{figure}

Our starting point is Eq.~(\ref{Dk}) for the inclusive distribution. Since the transverse momentum $\k$ remains small compared to the longitudinal momentum, it  is actually convenient to use angle variables rather than transverse momenta. Accordingly, we set
\beq
D(x,\btheta)\equiv (2\pi)^2 x\frac{\rmd N}{\rmd x\rmd^2\btheta}\,,
\eeq
where 
\beq
x = \frac{\omega}{E}, \qquad \text{and}\qquad\btheta\equiv\frac{\k}{\omega}=\frac{\k}{xE}\,.
\eeq
Note that $\btheta$ is a 2-dimensional vector collinear to $\k$, whose (small) magnitude equals the polar angle of the emitted gluon with respect to the initial direction of the  leading particle. In this new variable, Eq. (\ref{Dkt}) reads 
\begin{eqnarray} 
\frac{\del}{\del t}D(x,\btheta,t)&=&\int \rmd z\, {\cal K}(z)\left[\frac{D\left(x/z,\btheta,t \right)}{t_\ast(x/z)}-z\frac{D\left(x,\btheta ,t\right)}{t_\ast(x)} \right]\nn
&&+\frac{N_cn}{2}\int \frac{\rmd^2\btheta'}{(2\pi)^2}{\cal \sigma}(\btheta',x) D(x,\btheta-\btheta',t)\,,
\label{evol-eq-ang0}
\end{eqnarray}
with (see Eq.~(\ref{dipole-cs})) 
\beq\label{calCx}
{\cal \sigma}(x,\btheta)=(xE)^2{\cal \sigma}(\q)= \frac{32\pi^2 \alpha^2_s}{(xE)^2} \left[\frac{1}{\btheta^4}- \delta(\btheta)\int \frac{\rmd^2\btheta' }{\btheta'^4} \right]\,.
\eeq
An illustration of the gluon cascade described by Eq.~(\ref{evol-eq-ang0}) is given Fig.~\ref{fig:shower2}. Note that the locality (in angle) of the splitting term reflects the effective collinearity of the splitting.

In addition to changing to angular variables, in Eq.~(\ref{evol-eq-ang0}) we have explicitly factorized the kernel (\ref{kernel}) into 
\beq
2\alpha_s\cK(z,xE)= \frac{\cK(z)}{t_\ast(x)},     
\eeq
where now
\beq\label{factorizedkernel}
\cK(z)= \frac{[1-z(1-z)]^{5/2}}{z^{3/2}(1-z)^{3/2}},\quad\text{and}\quad t_\ast(x)=\frac{1}{\bar\alpha}\sqrt{\frac{xE}{\hat q}}.
\eeq
We shall refer to $\cK(z)$ as the \emph{reduced kernel}. The quantity 
\beq\label{stop-time}
t_\ast(E)\equiv \frac{1}{\bar\alpha}\sqrt{\frac{E}{\hat q}}. 
\eeq
is a characteristic time scale of the BDMPS-Z  cascade. After a time of order $t_\ast(E)$, most of the initial energy has been radiated into soft gluons (provided $L>t_\ast(E)$, of course), hence the often used denomination of \emph{stopping time}, or \emph{stopping distance}\cite{Arnold:2009ik}. In the following, we often denote $t_\ast(E)$ simply by $t_\ast$, except when confusion may arise. Note that $t_\ast(x)=t_\ast\sqrt{x}$.

As such the  kernel (\ref{factorizedkernel}) does not account for finite size effects that become important when $t_\ast(x) >  L$, corresponding to $x_c> 1$ ($\omega_c > E$)  (cf. Eq.~(\ref{BDMPS-Z -L})). In the following discussion we neglect these corrections, we thus implicitly restrict jet energies to be $E\sim\omega_c$ or smaller (see Refs.~\cite{Blaizot:2014rla,Fister:2014zxa} for discussions on this matter). 

 In the next subsection, we analyze the energy distribution, obtained by integrating $D(x,\btheta,t)$ over the angle. That is, we define
\beq
D(x,t) =\int \frac{\rmd^2 \btheta}{(2\pi)^2}D(x,\btheta, t),
\eeq
which obeys the following equation \cite{Baier:2000sb,Arnold:2002zm,Jeon:2003gi,Blaizot:2013hx}
\begin{eqnarray}\label{evol-eq}
\frac{\del}{\del t}D(x,t)&=&\int \rmd z\, {\cal K}(z)\left[\frac{D\left(x/z,t \right)}{t_\ast(x/z)}-z\frac{D\left(x ,t\right)}{t_\ast(x)} \right]\equiv {\cal I}[D].
\end{eqnarray}
The angular distribution of the radiated gluons will be discussed next. 

\subsection{Energy distribution and wave turbulence}  \label{energy-dist-WT}

Since initially the total energy is carried by a single gluon, the initial condition for Eq.~(\ref{evol-eq}) is simply $D_0(x)= \delta(1-x)$. In the regime where the length of the medium is so short that at most a single branching can be induced,  $L\lesssim t_\ast(x(1-x))$,  Eq.~(\ref{evol-eq}) can be solved by iterations, with the first  one given by 
\beq
D_1(x,L)= \frac{L}{t_\ast} x\cK(x),
\eeq
which coincides with  the BDMPS-Z  spectrum (\ref{BDMPS-Z }) at small $x$. In the opposite regime where $L \gg t_\ast(x)$, which corresponds to gluon energies $\omega\ll \omega_s\equiv \bar\alpha^2 \,\hat q \,L^2$, multiple branchings are important. A non perturbative solution becomes mandatory.
 
Eq.~(\ref{evol-eq}) can be solved exactly for a simplified version of the reduced kernel (\ref{factorizedkernel}) in which one  neglects the $z$ dependence of the numerator, that is
\beq\label{reduced-kernel}
\cK(z)\approx \frac{1}{z^{3/2}(1-z)^{3/2}}. 
\eeq
This turns out to be an excellent approximation. In fact, we shall argue later that the exact form of the kernel plays a minor role in the determination of the general features of the cascade. 
The solution of Eq.~(\ref{evol-eq}) for the simplified kernel reads \cite{Blaizot:2013hx,Blaizot:2015jea} 
\beq\label{Gsol2Ap}
D(x,\tau)=\frac{\tau}{\sqrt{x}\,(1-x)^{3/2}}\, \exp\left(-\pi\frac{\tau^2}{1-x}\right)\,, \quad\text{with}\quad  \tau=\frac{L}{t_\ast}.
\eeq
This solution exhibits two remarkable features: a peak near $x=1$ associated with the leading particle, and a scaling behavior in $1/\sqrt{x}$ at small $x$ where the $x$ dependence factorizes from the time dependence, i.e. 
\beq\label{Gsol2-soft}
D(x,\tau)\approx \frac{\tau}{\sqrt{x}}\, \rme^{-\pi \tau^2 }.
\eeq

 An illustration of this solution is given in Fig.~\ref{fig2}, left panel. The energy of the leading particle, initially concentrated in the peak at $x\lesssim 1$, gradually disappears into radiated soft gluons,  and after a time $t\sim t_\ast$ (i.e. $\tau\sim 1/\sqrt{\pi}\approx 0.5$)  most of the energy is to be found in the form of radiated soft ($x\lesssim 0.1$) gluons. This is also the time at which the peak corresponding to the leading particle disappears (see Fig.~\ref{fig2}). At the same time the occupation of the small $x$ modes increases (linearly) with time, keeping the characteristic form of the scaling spectrum. When the peak has disappeared, the cascade continues to lower $x$, causing a uniform, shape conserving, decrease of the occupations of the modes, and a flow of energy towards small $x$. 
 
\begin{figure}[ht]

  \centerline{
		\subfigure[]{\includegraphics[width=6.cm]{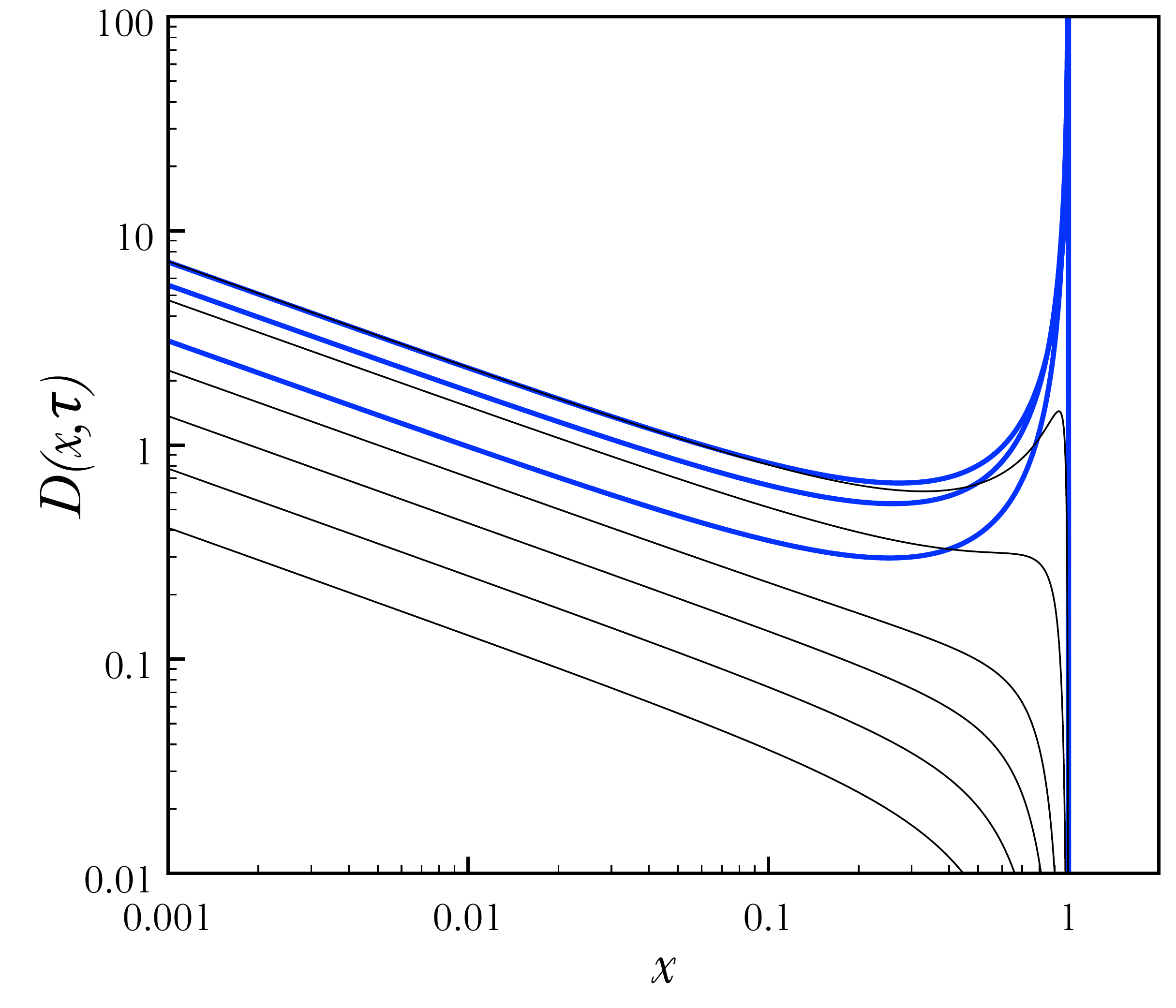}	\label{fig:BDMPS-Z } }
  	\subfigure[]{\includegraphics[width=6.cm]{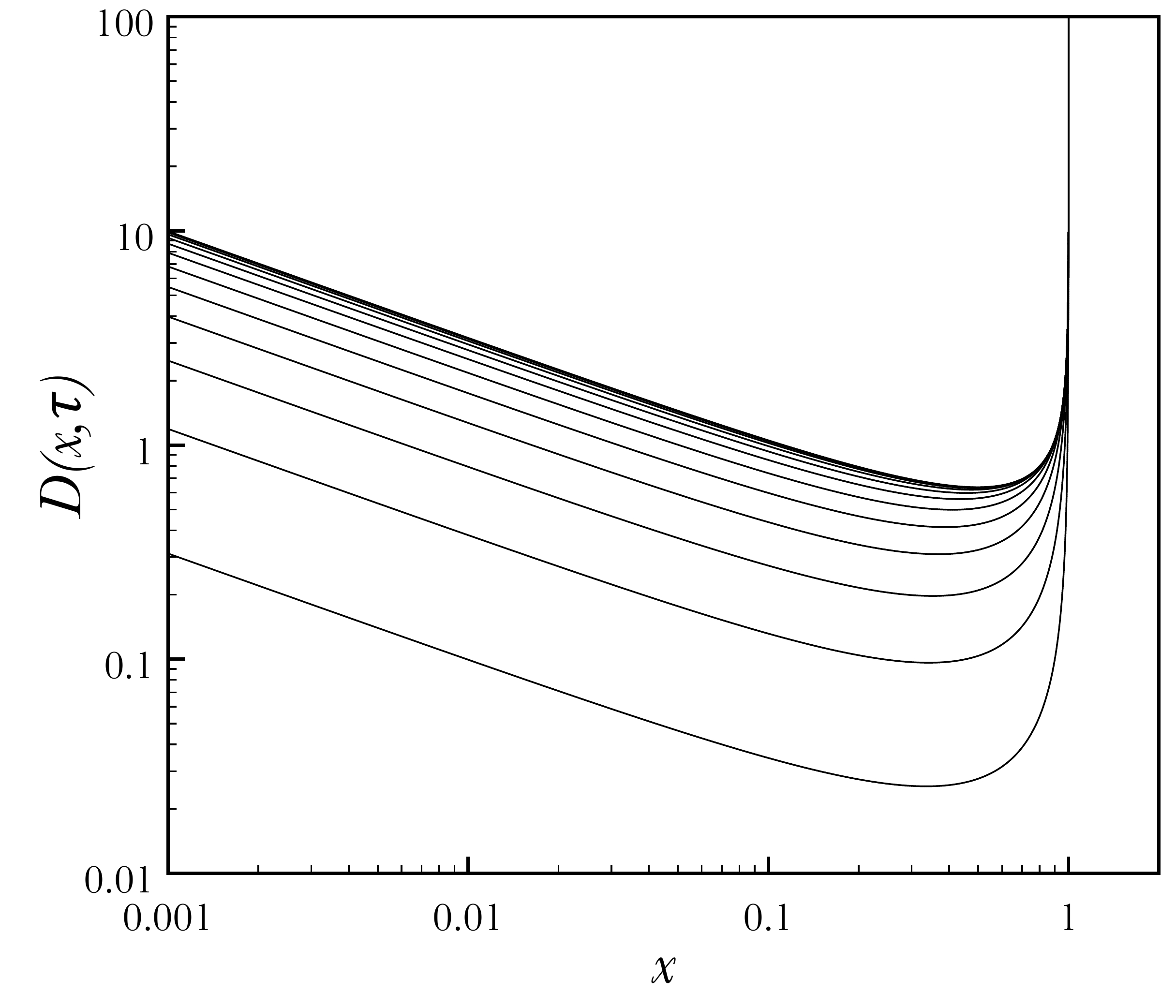}	\label{fig:BDMPS-Z _ST} }		
		}
		\caption{(Color online.) The function $D(x,\tau)$ (Eq. (\ref{Gsol2Ap}))  at
		various times. Left panel: the filling of the modes, which proceeds till the disappearance of the leading particle peak. 
		The values of $\tau$ are, for the thick (blue) curves, from bottom to top:  0.1, 0.2, 0.3 (during this stage the leading particle acts as a source for soft gluon radiation), and for the thin (black) curves, from top to bottom: 0.5, 0.7, 0.9,1.0,1.1,1.2  (the leading parton has exhausted its energy  and the peak has disappeared, while energy continues to flow to small $x$, the amount of energy in each mode decreasing exponentially fast). Right panel: energy is constantly injected into the system by a source located at $x=1$ (see Eq.~(\ref{DfinSource})). After a transitory regime, characterized by a uniform increase with time of the scaling spectrum, the system reaches a steady state. The values of $\tau$ are, from bottom to top: 0.1,0.2,0.3, 0.4, 0.5, 0.6,0.7,0.8,0.9,1.0. 
		 }
		\label{fig2}

\end{figure}


We define 
\beq\label{flowdef}
{\cal F}(x_0,\tau)=-\frac{\del {\cal E}(x_0,\tau)}{\del \tau}, \qquad {\cal E}(x_0,\tau)\equiv \int_{x_0}^1 \rmd x D(x,\tau), 
\eeq
where  ${\cal E}(x_0,\tau)$ is the amount of energy contained in the modes with $x>x_0$, and ${\cal F}(x_0,\tau)$ is the corresponding flux of energy,  counted positively for energy moving to values of $x$ smaller than $x_0$. 
These quantities can be calculated explicitly. We have for instance
\beq
{\cal E}(x_0,\tau)=\int_{x_0}^1 \rmd x\, D(x,\tau) ={\rm e}^{-\pi \tau^2}\,{\rm erfc}\left(  \sqrt{\frac{\pi x_0}{1-x_0}}\,\tau\right),
\eeq
with  ${\rm erfc}(x)$ the complementary error function. 
We note that the fraction of the total energy ``stored in the spectrum'', namely
\beq
\lim_{x_0\to 0} {\cal E}(x_0,\tau)={\rm e}^{-\pi \tau^2},
\eeq
 decreases with time, and accordingly there is a non vanishing flux of energy reaching $x=0$
\beq\label{flowexactwo}
{\cal F}(0,\tau)=2\pi \tau \, \rme^{-\pi\,\tau^2}\,.
\eeq
 It follows that the complete, energy conserving, solution involves a contribution 
 \beq\label{condensate}
 D_c(x)=n_c(\tau)\delta(x)\quad \text{with}  \quad n_c(\tau)=1-{\rm e}^{-\pi\tau^2},
 \eeq
somewhat analogous to a condensate where the radiated energy accumulates. Note that when $\tau\sim 1/\sqrt{\pi}$, corresponding to the disappearance of the leading particle into soft radiation,  about 60\% of the initial energy has flown into the condensate.

It is interesting to consider also the situation where the leading particle is replaced by a source that injects energy at a constant rate ${\cal A}$ at $x=1$. In this case, we are led to look for the solution of the following equation
\beq\label{DfinSource}
\frac{\partial}{\partial t}D(x,\tau)={\cal A}\delta(1-x) + {\cal I}[D],
\eeq
where ${\cal I}[D]$ denotes the r.h.s. of Eq.~(\ref{evol-eq}). 
The exact solution of  Eq.~(\ref{DfinSource})  with initial condition $ D(x,\tau=0)=0$ reads
  \beq\label{Dtbex}
  D_\text{st}(x,\tau)\,=\,\frac{{\cal A}}{2\pi \sqrt{x(1-x)}}\ 
  \left(1-\rme^{-\pi\frac{\tau^2}
  {1-x}}\right),
  \eeq
and  is plotted in Fig.~\ref{fig2}, right panel. As time goes on, this  solution converges to the stationary solution 
$({\cal A}/2\pi) /\sqrt{x(1-x)}$, keeping the
shape of the small $x$ spectrum, with just an overall time--dependent scaling. Remarkably, the small $x$ scaling form $D(x,t)={f(t)}/{\sqrt{x}}$ is reached well before the stationary state is achieved.
When the steady state is reached, all the energy provided by the source  flows throughout the entire system towards the condensate at $x=0$, that plays the role of a sink, while the population of the various modes stays unchanged. \\

Let us contrast these properties with those of the DGLAP cascade \cite{Altarelli:1977zs}. A simplified version of the corresponding evolution equation for the inclusive one particle distribution reads 
\beq\label{DGLAP}
\frac{\del }{\del t} D(x,t) = \bar\alpha \int_x^1 \frac{\rmd z}{z(1-z)} D(x/z,t) - \bar\alpha\int_0^1 \frac{\rmd z}{1-z}  D(x,t),
\eeq 
where here the time variable is related to the virtuality $Q^2$ of the emitting parton, $t\equiv \ln Q^2/Q^2_0$. 
This equation has the  form of Eq.~(\ref{evol-eq}) if one identifies  
\beq
\frac{1}{t_\ast(x)}=\bar\alpha,\qquad \text{and}\qquad {\cal K}(z)=\frac{1}{z(1-z)}.
\eeq
Thus the DGLAP equation differs from the BDMPS-Z  equation in two major aspects. First, the kernel ${\cal K}(z)$ is less singular near $z=0$ and $z=1$. Second, the rate of successive branchings is independent of the parent energy, i.e., it is constant along the cascade. We shall see that the latter property is what makes the major difference between the BDMPS-Z  and the DGLAP cascades. 

The solution of Eq.~(\ref{DGLAP}) is obtained using a Mellin transform \cite{Blaizot:2015jea}. The initial condition reads  $\tilde D(\nu,0)=1$, with $\tilde D(\nu,0)$ the Mellin transform of $D(x,0)=\delta(1-x)$. The solution can be expressed as the following integral 
\beq\label{DGLAP-sol}
D(x,t)=\int_{c-i\infty}^{c+i\infty} \frac{\rmd\nu}{2\pi i}  \exp\left[-(\psi(\nu)+\gamma)\,t+\nu\ln\frac{1}{x}\right],
\eeq
where $\phi(\nu)$ stands for the digamma function and $\gamma \simeq 0.58$ stands for the Euler constant. 

\begin{figure}[ht]
\centerline{
		\subfigure[]{\includegraphics[width=6.cm]{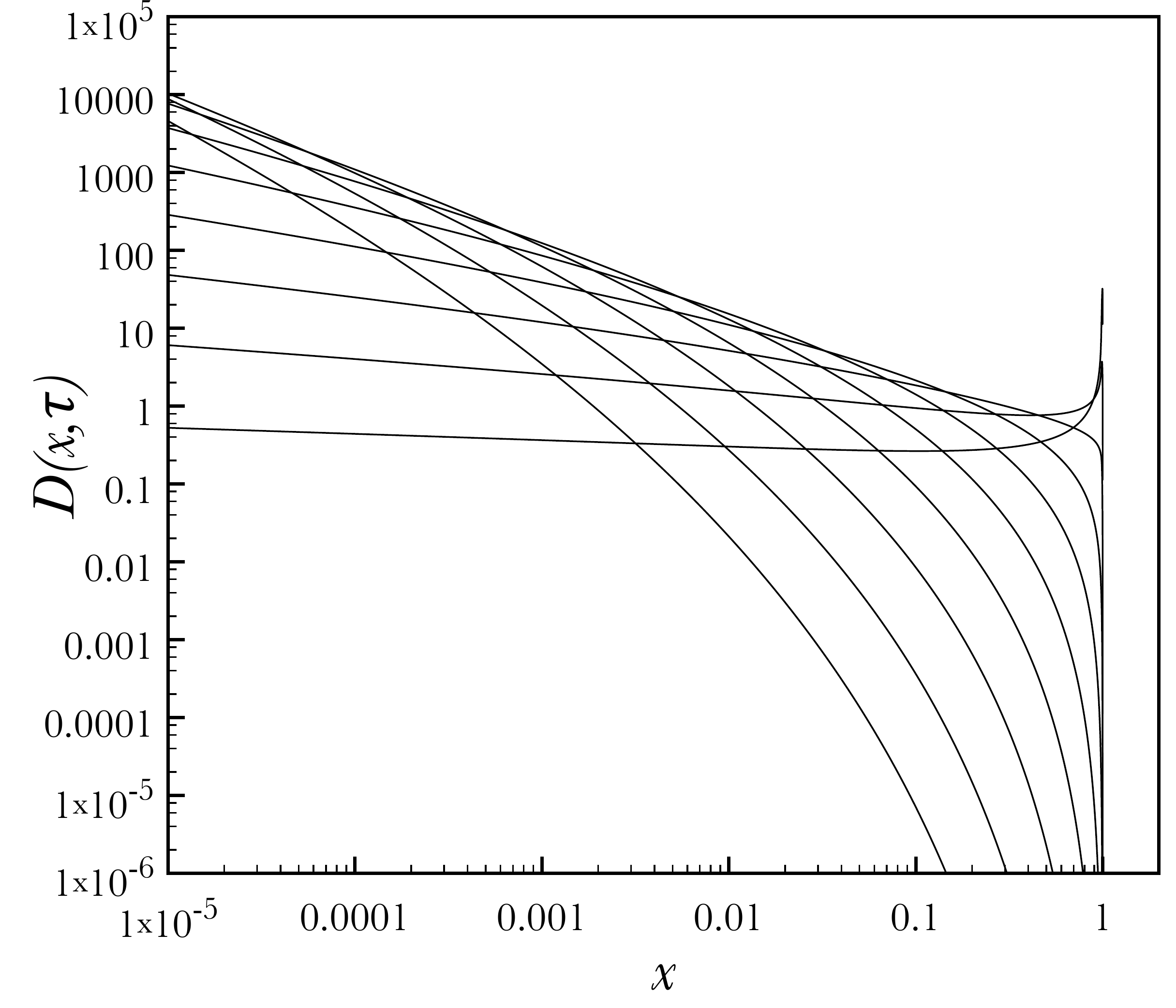}	 }
  	\subfigure[]{\includegraphics[width=6.cm]{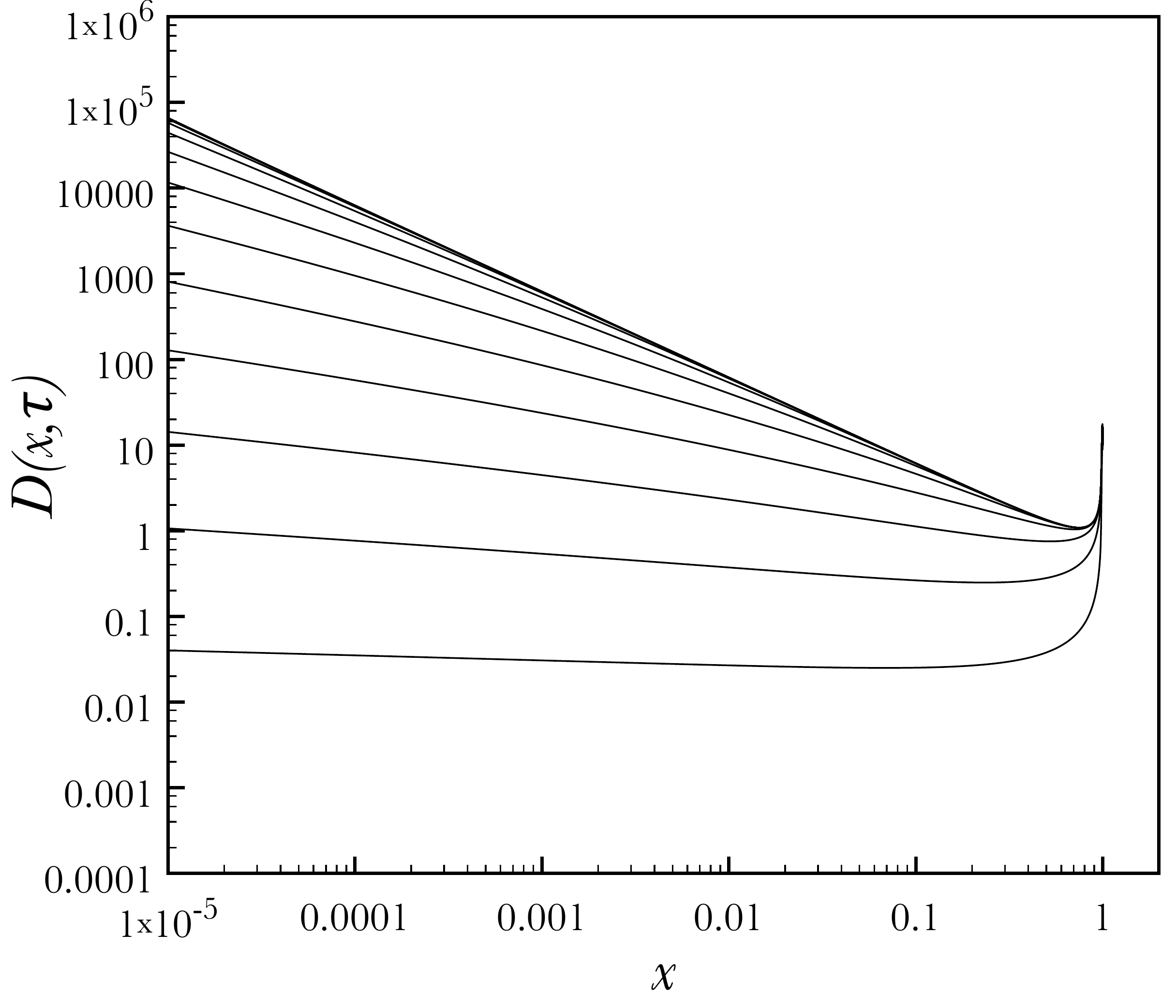}	 }		
		}
		\caption{Model of a DGLAP cascade according to Eq.~(\ref{DGLAP}) (left panel),  and with a source added at $x=1$ (right panel). The curves corresponds to $\tau=0.2,0.4,0.6,0.8,1.0,1.2,1.4,1.6,1.8,2.0$ from bottom to top. In the right panel, the emergence of the scaling solution $D(x)\sim 1/x$ is clearly visible, as well as the persistent deviation from it of the true solution at very small $x$.\label{fig:DGLAP}}

\end{figure}

This integral representation allows us to verify a few properties that are relevant for our discussion. First, 
it is easy to check that  the energy is conserved by the evolution: from Eq.~(\ref{DGLAP-sol}) we have 
$
{\cal E}(0,t)=\int_0^1 \rmd x D(x,t) = \tilde D(1,t) =1.
$
Furthermore, 
Eq.~(\ref{DGLAP-sol}) has following asymptotic behavior when $\ln 1/x \gg t$: 
\beq\label{DGLAP_as}
D(x,t)\approx \left(\frac{1}{x}\right)^{2\sqrt{\frac{t}{\ln 1/x}}}.
\eeq
Here, in contrast to what happens with the BDMPS-Z  cascade, the time dependence of the energy distribution does not factorize, and no obvious scaling behavior emerges at small $x$. 
 The growth of the spectrum at small $x$ is in fact tamed by the exponent in Eq.~(\ref{DGLAP_as}), making the spectrum integrable when $\ln 1/x >t $: all the energy remains in the spectrum, and no energy flows to $x=0$, unlike in the BDMPS-Z  cascade. 
 
These features are illustrated in Fig.~\ref{fig:DGLAP} where a 
 numerical calculation of the DGLAP cascade is presented. One recognizes in the left panel features analogous to that of the BDMPS-Z  cascade, in particular the disaperence of the leading particle peak, but no simple scaling pattern can be identified. In the right  panel, the the solution in the presence of a source is displayed. The solution in that case approaches the stationary solution $D(x)\sim {\cal A}/x$  at not too small $x$, but it takes a long time to populate the very small $x$ modes according to this $1/x$ law. In fact, for any time $t$, the true solution at very small $x$ deviates from the stationary solution in the same fashion as in Eq.~(\ref{DGLAP_as}).\\

The properties of the two cascades that we have discussed in this section can be understood in a broader context \cite{Blaizot:2015jea}.The two cascades are governed by the same general equation, Eq.~(\ref{evol-eq}) with, in each case, a specific kernel ${\cal K}(z)$ and  a specific  time scale $t_\ast(x)$.  The kernel ${\cal K}(z)$ controls how, in a given splitting,  the energy is shared between the two offsprings. The time scale $t_\ast (x)$ controls the rate at which successive splittings occur. The characteristic features of the cascades result from two properties: energy conservation and the existence of approximate scaling solutions. The latter involves crucially $t_\ast(x)$. The role of the kernel ${\cal K}(z)$ is less important.

To better understand the role of $t_\ast(x)$, let us observe that Eq.~(\ref{evol-eq})  admits an approximate fixed point solution, which we refer to as a scaling solution, of the form
\beq\label{scal-spect}
D_{\rm sc}(x,t)= \frac{t_\ast(x)}{x}. 
\eeq
Indeed, in the region where $D(x,\tau)\sim D_{\rm sc}(x,\tau)$, there is complete cancellation between gain and loss terms.  Note that this compensation occurs only for $z \ge x$ in the integrals of Eq.~(\ref{evol-eq}), so that the scaling solution cannot be an exact fixed point of the equation. However the general solution is driven to 
 this approximate fixed point.

Now, since energy is conserved, and since it is continuously moving towards the lower values of $x$, the distribution will generically develop a divergent behavior at small $x$. 
 Depending on the behavior of the scaling spectrum (\ref{scal-spect}) when $x\to 0$, 
  the approximate fixed point solution may lead to a divergent expression for the total energy ${\cal E}(0,t)$  contained in the modes between 0 and $1$. In order for this not to happen, either the fixed point solution  is never reached at very small $x$, or  $t_\ast(x)$ is an increasing function of $x$. The former occurs in the DGLAP cascade for which $t_\ast(x)=\text{cst}$, while the latter is what occurs in the BDMPS-Z  cascade for which $t_\ast(x)\sim \sqrt{x}$. Another way to see the role of the $x$ dependence of $t_\ast(x)$ is to observe that in the DGLAP cascade, where $t_\ast$ is independent of $x$, it takes an infinite time to propagate a finite amount of energy from $x=1$ to $x=0$. In contrast, in the BDMPS-Z  cascade, because the branching rate accelerates along the cascade (since $t_\ast(x)$ decreases with decreasing $x$ (see Fig.~\ref{fig:shower2} for an illustration)), the corresponding time is finite, and in fact of the order of $t_\ast$. This is the reason why there is a finite flow of energy all the way down to $x=0$ in the BDMPS-Z  cascade while the energy remains stored in the spectrum in the DGLAP cascade. For a more detail discussion see Ref.~\cite{Blaizot:2015jea}.
    
 At this point, it is useful to  comment briefly on a peculiarity of the BDMPS-Z  cascade, that makes the transition between the dilute, single-branching regime, and the the multiple branching regime completely smooth, with no sign of a change of regime as  $x$ crosses the value $x_s$. Indeed, the perturbative solution, valid at $\tau\ll 1$), is proportional to the BDMPS-Z  spectrum 
\beq
D(x,\tau)\simeq \frac{\tau}{\sqrt{x}},
\eeq  
and this is already in the scaling form. We shall see that the scale $x_s$ becomes visible in the angular distribution, to which we now turn. 

The properties of the cascades depend also on the splitting kernel, that is, on the way the energy is distributed between the offsprings during a splitting. However, this turns out to have a minor effect on the main characteristics of the cascade, as compared to that of the transport time scale just mentioned, at least as long as the splitting kernel is not too singular. In that case the cascades develop as if the branching were completely democratic, with the two offsprings taking each half the energy of the parent gluon. The interactions responsible for the splittings can then be considered as local (in energy space). As already mentioned, these properties of the BDMPS-Z  cascade that we have briefly listed, are typical of wave turbulence \cite{Nazarenko}.

\subsection{Angular distribution}\label{angle-dist}

 We turn now the angular distribution.  Our starting point is Eq.~(\ref{evol-eq-ang0}), that we write in Fourier space as 
\beq\label{evol-eq-ang}
\frac{\del}{\del t}D(x,\u,t)&&=\frac{1}{t_\ast}\int \rmd z\, {\cal K}(z)\left[\sqrt{\frac{z}{x}} D\left(\frac{x}{z},\u,t\right)-\frac{z}{\sqrt{x}} D\left(x,\u,t\right)\right]\nn && \quad +\frac{N_cn}{2}\sigma(x,\u) D(x,\u,t),
\eeq
with the Fourier transform given by
\beq\label{FT}
D(x,\u,t)=\int \frac{\rmd^2\btheta}{(2\pi)^2} D(x,\btheta,t)\,\rme^{-i\btheta\cdot\u}. 
\eeq
and similarly for $\sigma (x,\u)$. We shall also consider the  diffusion approximation  ($\btheta' \ll \btheta $) of Eq.~(\ref{evol-eq-ang0})
\begin{eqnarray}\label{evol-eq-diff0}
\frac{\del}{\del t}D(x,\btheta,t)&=&\int \rmd z\, {\cal K}(z)\left[\frac{D\left(x/z,\btheta,t \right)}{t_\ast(x/z)}-z\frac{D\left(x,\btheta ,t\right)}{t_\ast(x)} \right] \nn
&&+\frac{1}{4(xE)^2} \,  \left(\frac{\partial}{\partial \btheta}\right)^2 \, \left[\hat q \, D(x,\btheta,t)\right]\,,
\end{eqnarray}
with $\hat q\simeq  4\pi\alpha_s^2 N_c n\ln ({\btheta^2}/{\btheta_D^2})$ and 
 $\btheta_D\equiv m_D/\omega=m_D/(xE)$. \\
 
\begin{figure}[!ht]
\begin{center}
\includegraphics[width=12cm]{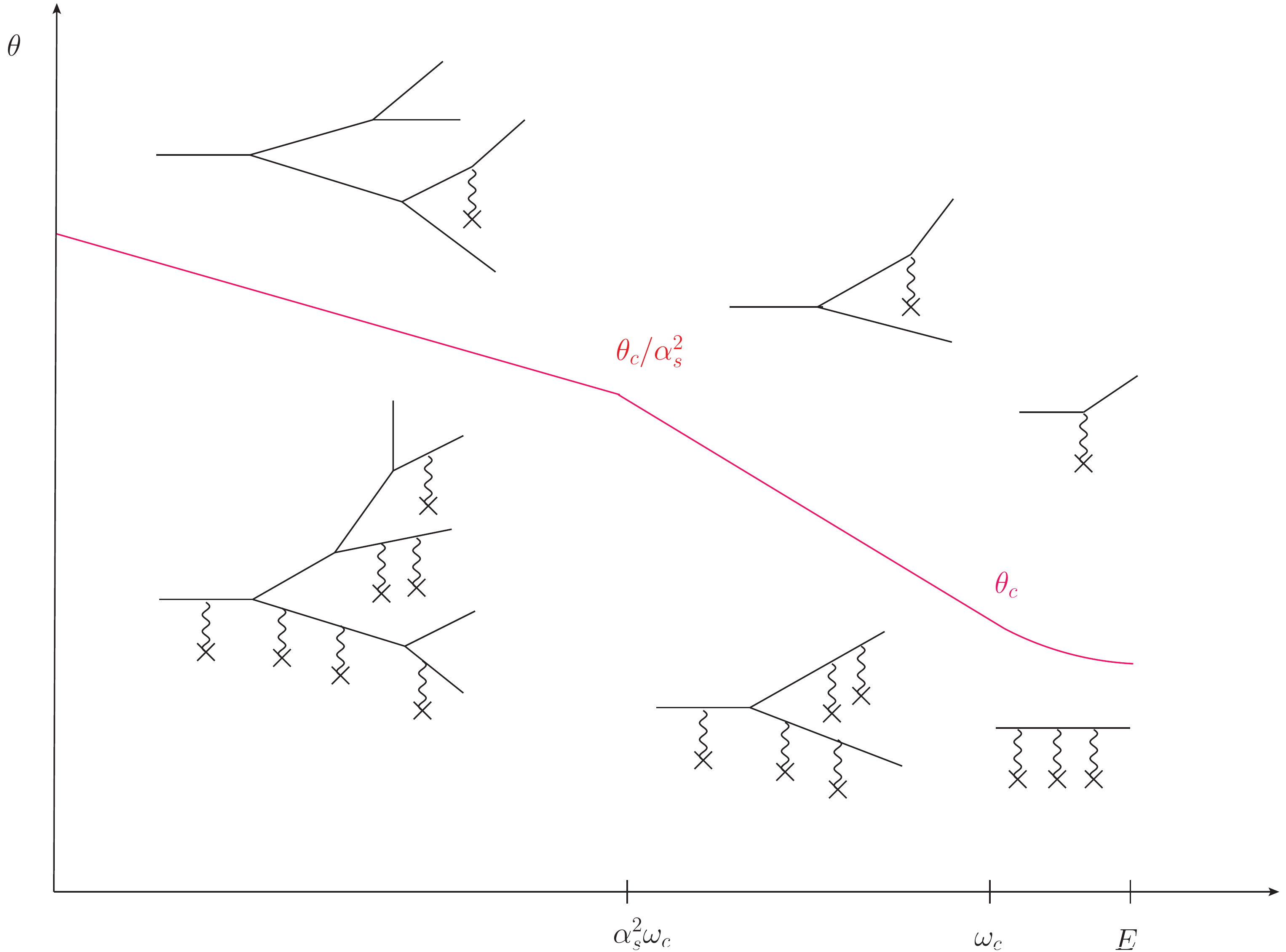}
\caption{(Color online.) The various regimes of the physical processes (branching and scattering) that accompany the propagation of a fast parton with energy $E\gtrsim \omega_c$ in a dense medium. The red curve stands for the root of $\lg \theta^2\rg$. It separates the regime of single hard, large angle, scattering, from that of soft, small angle, multiple scattering. Accordingly, the soft component of the angular distribution lies below the red line, the hard component above. The line $\alpha_s^2 \omega_c\sim \omega_s$ separates the region dominated by single branching from that of multiple branchings. $\omega \sim E$ indicates the region where the leading particle propagates without splitting, and merely suffers  momentum broadening due to its collisions with the constituents of the medium.  \label{fig1}}
\end{center}
\end{figure}

Solving Eq.~(\ref{evol-eq-ang}) exactly is difficult. 
Part of the difficulty  comes from the fact that  the angular distribution has two distinct components: a hard component, corresponding to large angles produced by single hard scatterings, and a soft component that can be obtained as the solution of the diffusion equation (\ref{evol-eq-diff0}). 
Each of these two components is strongly modified by gluon branching. The soft component admits moments, which is not the case for the hard component. The characteristic angle that marks the boundary between the soft and the hard components depends on $x$, i.e., on the amount of branching. It can be estimated by calculating the mean squared angle of the soft component. This provides a first orientation into the various regimes that characterize the general solution of Eq.~(\ref{evol-eq-ang0}), and that are illustrated in Fig.~\ref{fig1}.

We define the typical squared angle as follows 
\beq\label{theta2}
\langle \theta ^2 \rangle = \frac{1}{D(x,L)}\int \frac{\rmd^2\btheta}{(2\pi)^2}\, \btheta^2 D(x,\btheta,L)\,,
\eeq
where $D(x,\btheta,L)$ is here the solution of Eq.~(\ref{evol-eq-diff0}). This can be determined explicitly in the three regimes characterizing  the following ranges of $x$ values: i) $x\lesssim 1$, corresponding to the leading particle; ii) $x_s\ll x\ll 1$ where single emission dominates; iii) $x\ll x_s$, the regime of multiple emissions. We refer to Ref.~\cite{Blaizot:2014rla} for the details of the calculation, and discuss now the main results. 

When $x\lesssim 1$, 
\beq\label{ang-sol-2}
\langle \theta ^2 \rangle =\frac{ \hat qL}{E^2}= \theta_s^2(1,L),\qquad \theta_s^2(x,L)\equiv \frac{ \hat qL}{\omega^2}=\frac{ \hat qL}{x^2E^2}.
\eeq
This angle reflects directly the  momentum broadening of the leading particle propagating in the medium over a distance $L$. 
 
In the regime $x_s\ll x\ll 1 $, where typically one emission occurs, the dominant contribution comes from momentum broadening of the radiated gluon from the time of its emission. The other contribution, corresponding to  the momentum broadening of the leading particle before the splitting, is suppressed by a factor $x^2$. 
Keeping the dominant contribution, we get
\beq\label{ang-sol-3}
\langle \theta ^2 \rangle =\frac{1}{2}\theta_s^2(x,L)\,.
\eeq
The factor $1/2$ originates from the average over the emission time.  

Note that for realistic situations $\omega_c < E$, the maximal frequency is thus $x_c\equiv \omega_c/E < 1$. In this case, $x_s\ll x\ll x_c$, and hence, 
these rare radiations are on average confined to the angular corona, 
\beq\label{BDMPS-Z -angle}
\theta_c\, \ll\, \theta\, \ll \, \frac{1}{\alpha_s^2}\theta_c.
\eeq
where the critical angle 
\beq\label{theta_c}
\theta_c = (\hat q L^3)^{-1/4},
\eeq
corresponds to the smallest BDMPS-Z   radiation angle. 
The third regime, $x\ll x_s$, is dominated by multiple branchings at parametrically large angles, 
\beq \label{multi-br-angle}
\theta \gg \frac{1}{\alpha_s^2}\theta_c.
\eeq
It requires a non perturbative treatment and the result is (see Refs.~\cite{Blaizot:2014ula,Blaizot:2014rla} for details and Ref.~\cite{Kurkela:2014tla} for a similar discussion)

\beq\label{angle2-Multi-br}
\langle \theta ^2 \rangle = \frac{1}{4\bar \alpha}\left [\frac{\hat q}{(xE)^3}\right]^{1/2}\equiv \theta^2_\ast(x).
\eeq
As was the case for the previous regime,  this angle  can be identified with that corresponding to the momentum broadening of the observed gluon since the time of its emission.  The major difference with the previous regime is that now $\theta_\ast(x)$ does not depend on the size of the medium. This is because, in the multiple branching regime, only gluons that are emitted off the leading particle at a distance, that is shorter than their characteristic stopping time $t_\ast(x)$,  from the medium boundary $L$, escape the medium. Accordingly, their measured squared angle is typically $\theta^2_\ast(x)\sim \hat q t_\ast(x)/(xE)^2$, which recalling Eq.~(\ref{stop-time}) agrees with Eq.~(\ref{angle2-Multi-br}). 

The complete distribution in the three regimes can also be determined explicitly. Again, we refer to Ref.~\cite{Blaizot:2014rla} for details. 

When $x\lesssim 1$, the distribution takes the factorized form
\beq\label{dist-x0}
D(x,\btheta,L)\simeq {\cal P} (\btheta,L) \, D(x,L)\,,
\eeq 
where $D(x,L)$ is the energy distribution given by Eq.~(\ref{Gsol2Ap}) (for $x\lesssim 1$). In the regime of multiple scatterings, i.e., for $\theta^2\lesssim \theta_s^2(L,x)$ 
\beq\label{Gaussdistr2}
{\cal P}(\btheta,L)=\frac{4\pi}{\theta_s^2(x,L)}\exp\left[-\frac{\btheta^2}{\theta_s^2(x,L)}\right]\,.
\eeq
In the opposite case, $\theta \gg \theta_s(x,L)$, a single hard scatterings dominates the distribution and we have 
\beq\label{1-scat-P}
{\cal P}(\btheta,L)\approx \frac{n N_c g^4 L }{(xE)^2\,\btheta^4}\,.
\eeq

In the second regime the energy distribution is given by the leading order BDMPS-Z  distribution, 
\beq
D(x,L)\simeq \frac{L}{t_\ast \sqrt{x}}\,,
\eeq 
and  we expect also  the angular distribution to be given by a single radiation that undergoes multiple scatterings. For sufficiently small $x$, as already emphasized, the angular deviation of the radiated gluon is larger than that of the leading parton. Accordingly one can neglect the momentum broadening before the emission. We get \cite{Wiedemann:2000za}, 
\beq\label{evol-1b-3}
D(x,\btheta,L)\simeq \int_0^L\frac{\rmd t}{t_\ast}\,   {\cal P} (x,\btheta,L-t)   \,x{\cal K}(x)\,. 
\eeq

Finally, in the fully non-perturbative regime, i.e., $x\ll x_s$, and in the soft region,  both multiple scatterings and multiple branchings must be resummed.
A solution can be obtained as a power series in the number of scatterings\cite{Blaizot:2014rla}. This is achieved by using  an approximation that is equivalent to the diffusion approximation, hence not accurate for the very   the tail of the distribution (see below), but it allows us to determine the distortion of the main peak.  
The result reads
\beq\label{angular-dist-multi-br}
D(x,\btheta,L)=\frac{4\pi}{\theta^2_\ast(x)}\,  \eta\left(\frac{\btheta^2}{\theta^2_\ast(x)}\right)\,D(x,L)\,,
\eeq 
where the entire dependence on $\btheta$ is in the scaling function $\eta$,  given by
\beq\label{scal-eta-2}
\eta(z) &=&  \int_0^\infty \rmd \alpha \, J_0(2\sqrt{z\alpha})  \, \sum_{n=0}^{\infty} c_n \left(-\alpha^2 \right)^n\,,
\eeq
where  $J_0$ is a Bessel function, and we have the  normalization property.
\beq\label{eq:eta_norm}
\int_0^\infty \rmd z\, \eta(z) =1\,.
\eeq
In Fig.~\ref{fig2} we have plotted the angular distribution $\eta(z)$ in the multiple branching regime.  For the numerical evaluation we have computed the first 500 terms in the multiple scattering series (\ref{scal-eta-2}) \cite{Blaizot:2014rla}.

Consider finally the single scattering limit achieved when $\theta\gg \theta_\ast(x)$. This is easily found to be
 \beq
 D(x,\btheta,L) \approx \frac{N_c n g^4 t_\ast(x)}{(2xE)^2\,\btheta^4}D(x,L), 
 \eeq
where $N_c n g^4 t_\ast(x)/(2xE)^2 \sim \theta_\ast^2(x)$ is the typical angle squared of gluons in the multiple branching regime.
 \begin{figure}[!ht]
\begin{center}
\includegraphics[width=8cm]{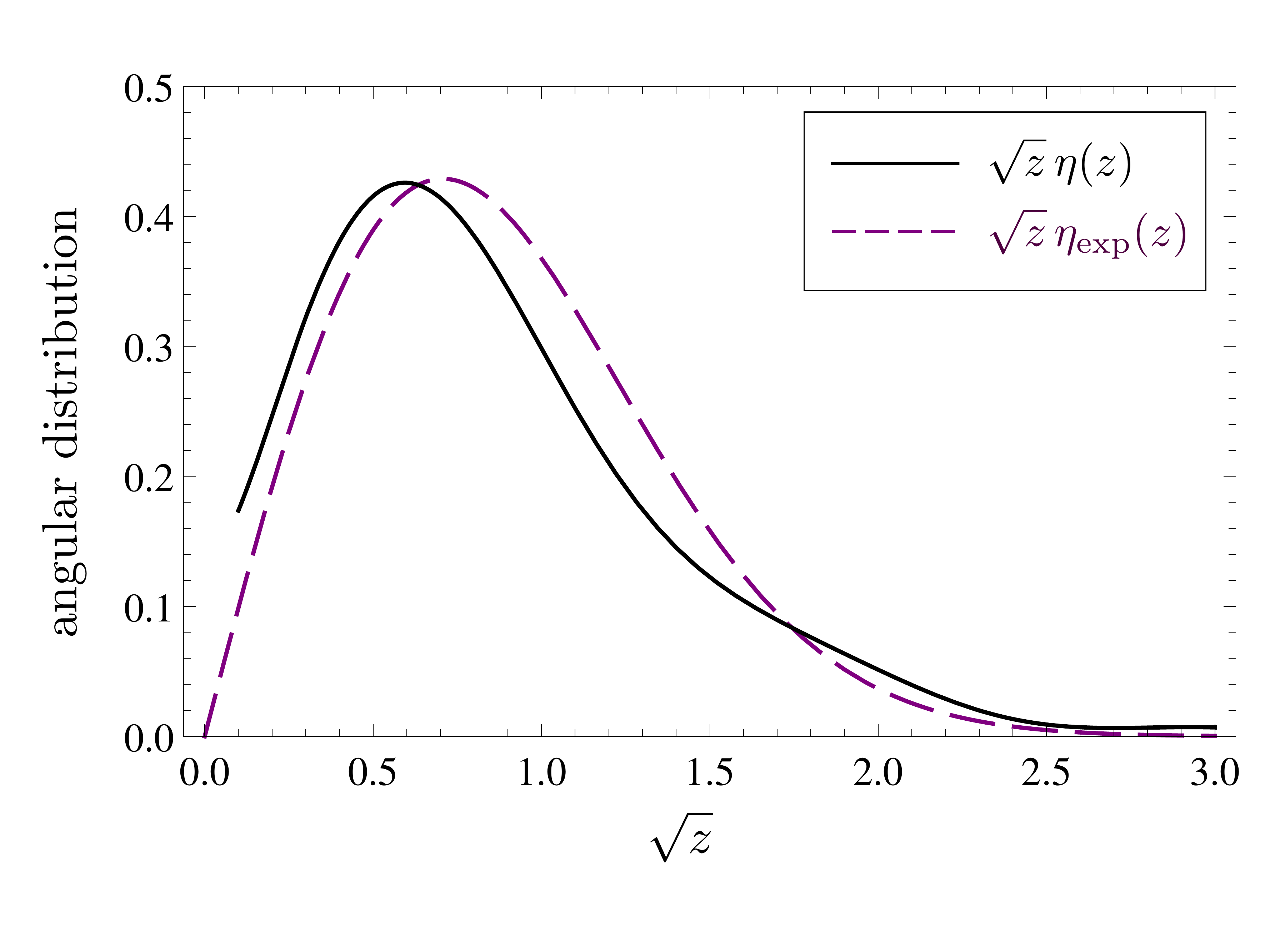}
\caption{(Color online.) The angular distribution  (\ref{scal-eta-2}) of a gluon in the multiple branching regime (solid black line)  compared to an exponential distribution with identical first two moments (dashed purple line).  Here $z\equiv \theta^2/\theta^2_\ast(x)$.}
\label{fig2}
\end{center}
\end{figure}

\subsection{Application: Energy imbalance in dijet events}

As we have argued in this section, a striking feature of the BDMPS cascade is that it provides an efficient mechanism to transport energy down to the lowest accessible gluon frequencies. Since soft gluons are typically emitted at large angles, the cascade naturally populates soft gluon modes at large angles with respect to the jet axis. We end this section with a brief analysis of the  phenomenological implications of this mechanism.  

Recently, the CMS collaborations investigated the energy imbalance in strongly asymmetric dijet events \cite{Chatrchyan:2011sx,CMS:2014uca}. As they are born back-to-back jets have roughly equal energy and momenta. The observed asymmetry is due to the fact that the leading jet (the most energetic of the two) traverses a shorter distance in the medium than the subleading one, and hence loses less energy. Surprisingly, the energy balance is recovered at quite large angles, $R> 0.8$, and most of the ``missing'' energy is carried by very soft particles \cite{Chatrchyan:2011sx}. \footnote{To our knowledge, this observable was only addressed in the context of the Q-PYTHIA Event Generator \cite{Apolinario:2012cg} which reproduces the qualitative features seen in the data. }

\begin{figure}[ht]
\centerline{
\includegraphics[width=7.cm]{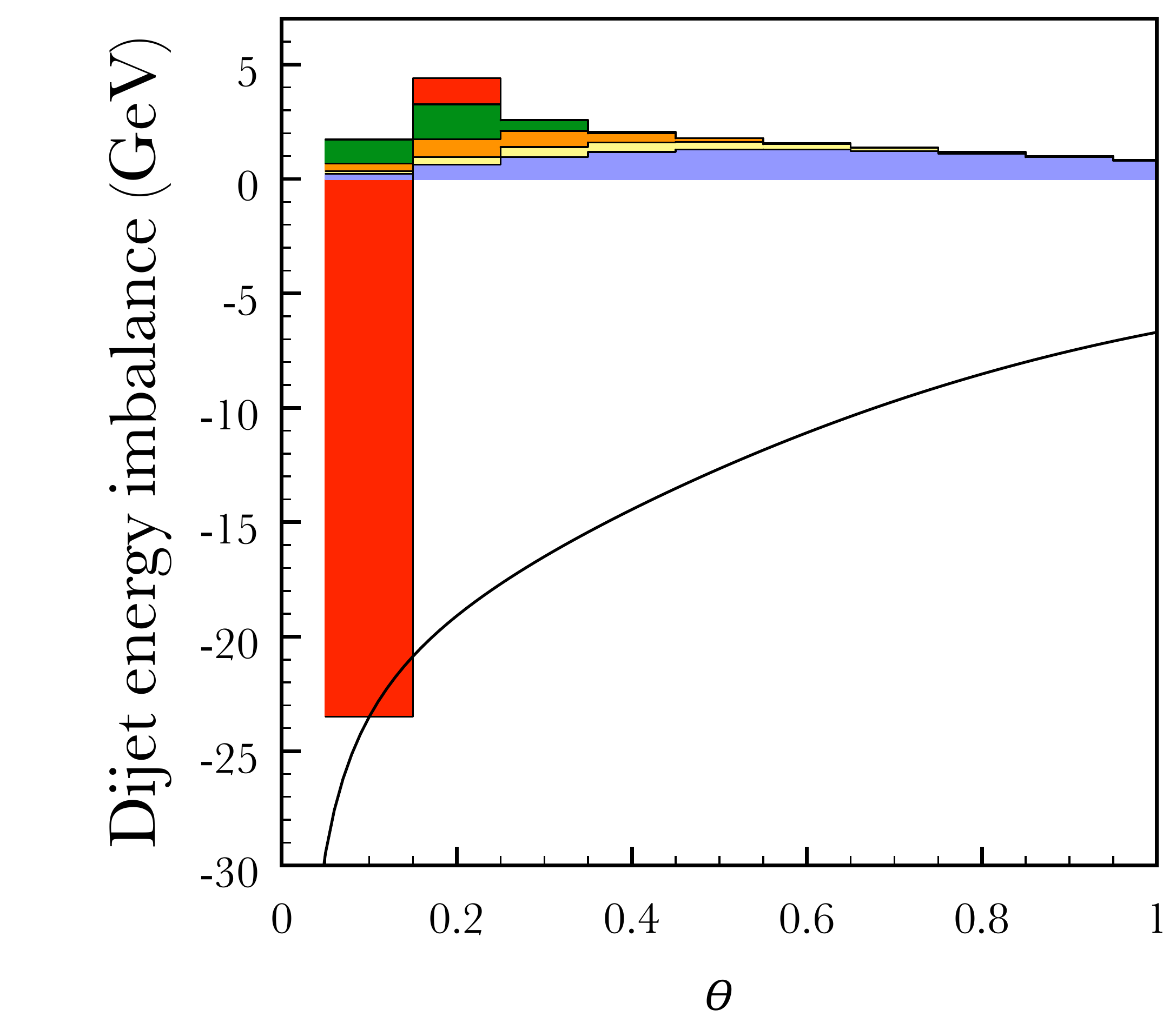}	\label{fig:In-cone-E}  
		}
		\caption{The  difference of the angular and energy distributions of sub-leading and leading jets for $\omega_\text{BH}=0.5$ GeV. The histograms account for the four binnings of energies: [0-1] GeV (grey),   [1-2] GeV (yellow), [2-4] GeV(orange), [4-8] GeV(green), [8-100] GeV(red). In the first angular bin we observe a large imbalance of energy in the hard particles. This energy is partly recovered at large angles by very soft particles.  The cumulative energy is given by the full line. }
\end{figure}

As a very rough model for an asymmetric dijet event, we consider a leading jet  and a subheading one that have traversed, respectively,  the path lengths $L=5$ fm and $L=1$ fm in opposite directions. We use the following set of parameters: $\hat q = 1$ GeV$^2$/fm, $\bar \alpha = 0.3$, and $E  \equiv E_1=E_2=100$ GeV. First, we find that the broadening of the subleading jet, $\theta_s\sim 0.02$ (cf. Eq.~(\ref{ang-sol-2})) is negligible compared to the opening angle of jet $\Tjet=0.3$, as observed in the data. 

We evaluate the energy imbalance by integrating the difference $D_2(x,\theta)-D_1(x,\theta)$, where $D_1$ and $D_2$ are the medium-induced gluon distributions of the leading and sub-leading jet respectively, in the different bins in energy and angle. The minus sign accounts for the fact that the constituents of the subleading jet propagate in opposite direction to those of the leading jet. The angular distribution is assumed to be Gaussian with a width determined by an interpolation between the mean squared angles in the three regimes discussed above. Finally, the evolution equation Eq.~(\ref{evol-eq}) is solved numerically where soft emissions with frequencies $\omega < \omega_\BH$ are cut-off to regulate the unphysical soft singularity in the branching kernel (\ref{Kdef}). \cite{Blaizot:2014ula}. 

In Fig.~\ref{fig:In-cone-E}, is plotted, in the form of histograms, the energy imbalance integrated on various angular and energy intervals \cite{Blaizot:2014ula}. At small angle $\theta \lesssim0.2 $ there is a deficit of hard particles with energies $\omega>8$ GeV (red negative histogram). As we increase the angle the energy imbalance turns in favor of the subleading jet and it is dominated by very soft particles with energies $\omega <1$ GeV (positive grey histograms).  The full line corresponds to the integrated energy imbalance for angles from $0$ to $\theta$. The imbalance is expected to be recovered at large $\theta$, when $\theta \to \pi/2$. The energy transported at large angles is clearly dominated by very soft gluons \footnote{These features are absent in the angular distribution of primary gluon radiations, as shown in the early work \cite{Salgado:2003rv}, confirming the important of multiple branchings to transport energy at large angles.}.  The jet structure that emerges from this plot is in semi-quantitative agreement with the recently released data by CMS  \cite{CMS:2014uca}.

\section{Decoherence of the jet core}\label{decoherence}

In this section we discuss briefly how the two cascades, vacuum and medium induced, can interfere in a jet produced in heavy ion collisions. The issue we want to address is to what extent color coherence, that constrains the pattern of successive parton branchings in vacuum, is altered when the jet propagates in a medium. As a paradigmatic example, we  consider the radiation  (in vacuum or in a QCD medium) of a quark-antiquark pair in the singlet state (the generalization to different color configurations is straightforward). This pair, which we refer to as an antenna,   can be thought of as being produced by the decay of the highly virtual primary parton \cite{MehtarTani:2010ma,MehtarTani:2011tz,MehtarTani:2011jw,CasalderreySolana:2011rz,MehtarTani:2011gf,MehtarTani:2012cy}.

In vacuum, the radiated spectrum off the quark reads\cite{Dokshitzer:1991wu} (see discussion in Section \ref{vacuum})
\beq\label{spect-vac-2} 
\rmd P_{gq}  =  \frac{\alpha_sC_F}{\pi} \Theta(\theta_{q\bar q }-\theta)\frac{\rmd \theta^2}{\theta^2} \frac{\rmd \omega }{\omega },
\eeq
and similarly for the antiquark, i.e., $\rmd P \equiv \rmd P_{gq}+\rmd P_{g\bar q}$. Here, the theta function accounts on average  for the suppression of large  angle radiation ($\theta \gg \theta_{q\bar q }$) caused by color coherence.   The total radiation spectrum exhibits two radiation cones aligned  on the quark and the antiquark momenta, respectively. Eq.~(\ref{spect-vac-2}) is at the basis of Eq.~(\ref{MLLA}).

The probability for the quark of the antenna to radiate soft gluons after traversing a QCD medium  is given 
by
 \cite{MehtarTani:2010ma,MehtarTani:2011tz},
\beq\label{decoh}
\rmd P_{g q } &=& \frac{\alpha_sC_F}{\pi} \Big[\Theta(\theta_{q\bar q}-\theta )+\Delta_\text{med}\, \Theta(\theta-\theta_{q\bar q})\Big]\frac{\rmd \theta^2}{\theta^2} \frac{\rmd \omega }{\omega }, \nn
\eeq
where
\beq\label{decoh-parameter}
\Delta_\text{med}\equiv \frac{1}{N_c^2-1}\lg \Tr \,\cU^\dag_{\bar\p}\cU_\p \rg \approx 1-\exp\left[ -  \frac{r_\perp^2}{l_\perp^2}  \right].
\eeq
 and ${\cal U}$ is a Wilson line in the adjoint representation, evaluated along the trajectory of the quark, and the harmonic approximation has been used to evaluate the correlator of the Wilson lines. 
Here  $r_\perp \equiv\theta_{q\bar q}L$  stands for the antenna transverse size and 
\beq \label{med-cor-length}
l_\perp\equiv \sqrt{\frac{12}{ \,\hat q L}},
\eeq
is a length characterizing the range of coherence in the transverse plane.  
The quantity $\Delta_\text{med}$ may indeed be  interpreted as a decoherence parameter: when  $r_\perp \ll l_\perp $, or equivalently $\theta_c \gg \Tjet$, where the BDMPS-Z critical angle $\theta_c$ is defined in Eq.~(\ref{theta_c}),  we have $\Delta_\text{med} \sim r_\perp^2/ l_\perp^2\ll 1$ and we recover Eq.~(\ref{spect-vac-2}). In this regime the inner structure of the antenna is not resolved by the medium and the antenna radiates coherently as in vacuum. 

\begin{figure}
\centering
 \includegraphics[width=0.5\textwidth]{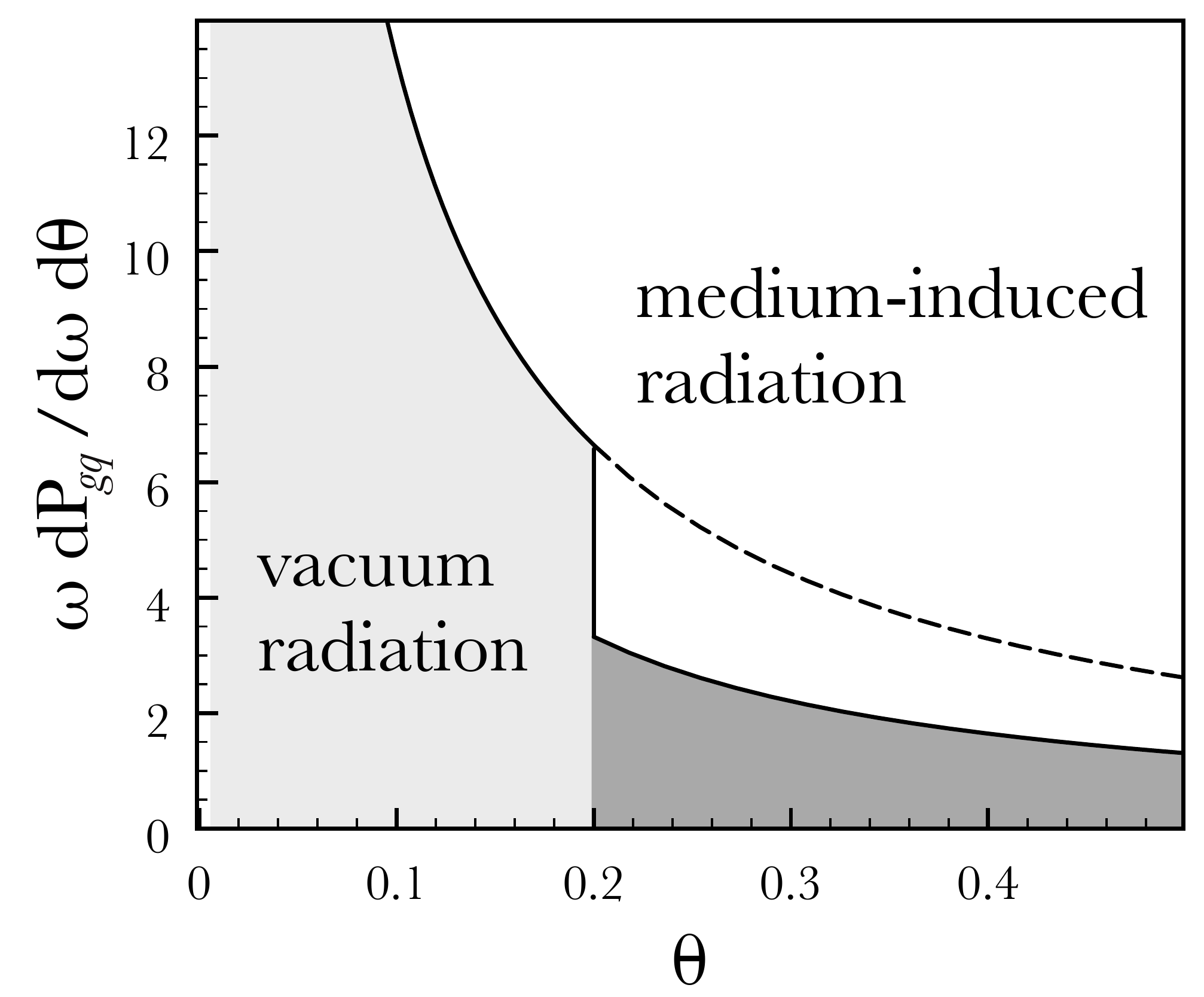}
\caption{The soft gluon emission spectrum off the quark constituent of a singlet antenna with opening angle $\theta_{q\bar q } = 0.2$, according to Eq.~(\ref{decoh}), in the presence of a medium with $\Delta_\text{med} = 0.5$ (solid line). Here $\equiv \alpha_s C_F/\pi=1$. On average vacuum radiation is confined within $\theta < \theta_{q\bar q }$, while the medium-induced radiation is radiated at $\theta > \theta_{q\bar q }$. The limit of opaque medium, given by $\Delta_\text{med} = 1$, is marked by the dashed line. Figure taken from Ref.~\cite{MehtarTani:2011tz}.}
\label{fig:DecoherenceSoft}
\end{figure}
%
On the  other hand, the case $r_\perp \gg  l_\perp$ ($\theta_c \ll \Tjet$), corresponds to full decoherence of the antenna, and we have $\Delta_\text{med}\simeq 1$. Here, the angular constraint disappears, and the quark and the antiquark radiate independently from one another \cite{MehtarTani:2011tz}\footnote{A analogous phenomena occurs between initial and final state radiation in a dense medium\cite{Armesto:2012qa,Armesto:2013fca}.},
\beq
\rmd P_{g q }  \approx  \frac{\alpha_sC_F}{\pi} \frac{\rmd \theta^2}{\theta^2} \frac{\rmd \omega }{\omega }.
\eeq
This discussion applies also to gluons radiated inside the medium \cite{MehtarTani:2011jw,CasalderreySolana:2011rz,MehtarTani:2011gf,MehtarTani:2012cy}. The decoherence parameter given by (\ref{decoh-parameter}) can be rewritten as $\Delta_\text{med}\sim 1-\exp(L/t_\text{decoh})$, from which we extract the decoherence time,
\beq
t_\text{decoh}\equiv \left( \frac{1}{\hat q \theta^2_{q\bar q}}\right)^{1/3}.
\eeq
The antenna spectrum (\ref{decoh}) is plotted in Fig.~(\ref{fig:DecoherenceSoft}) for $\Delta_\text{med}=0.5$. \\

The antenna setup serves as guideline to discuss the decoherence that a jet experiences  when passing through a dense medium. Consider  a jet with opening angle $\Tjet$. While inclusive jet observables in vacuum are characterized by two scales, $Q\equiv \Tjet E$, that correspond to the jet transverse mass typically of the order of 100 GeV, and $Q_0\sim \Lambda_\text{QCD}$, in the presence of a dense medium, an additional (hard) scale comes into play, $Q_\text{med} = \max \left(r_\perp^{-1}, \sqrt{\hat q L} \right)$, where here $r_\perp \equiv \Tjet L $. This medium-generated scale is of the order of a few GeV's, therefore, we typically have $Q_\text{med} \ll Q $. The medium affects the collimated parton shower, and thus causes its decoherence, only at scales of the order or smaller than $Q_\text{med}$. Hence, the medium alters the property of angular ordering by allowing for additional soft gluon radiations up to the limiting angle $Q_\text{med}/\omega$ \cite{MehtarTani:2011gf,MehtarTani:2012cy}. As a result of this aforementioned separation of scales a large portion of the jet is barely affected by its interactions with the medium.

 \begin{figure}[htbp]
\centering

		\includegraphics[width=8.cm]{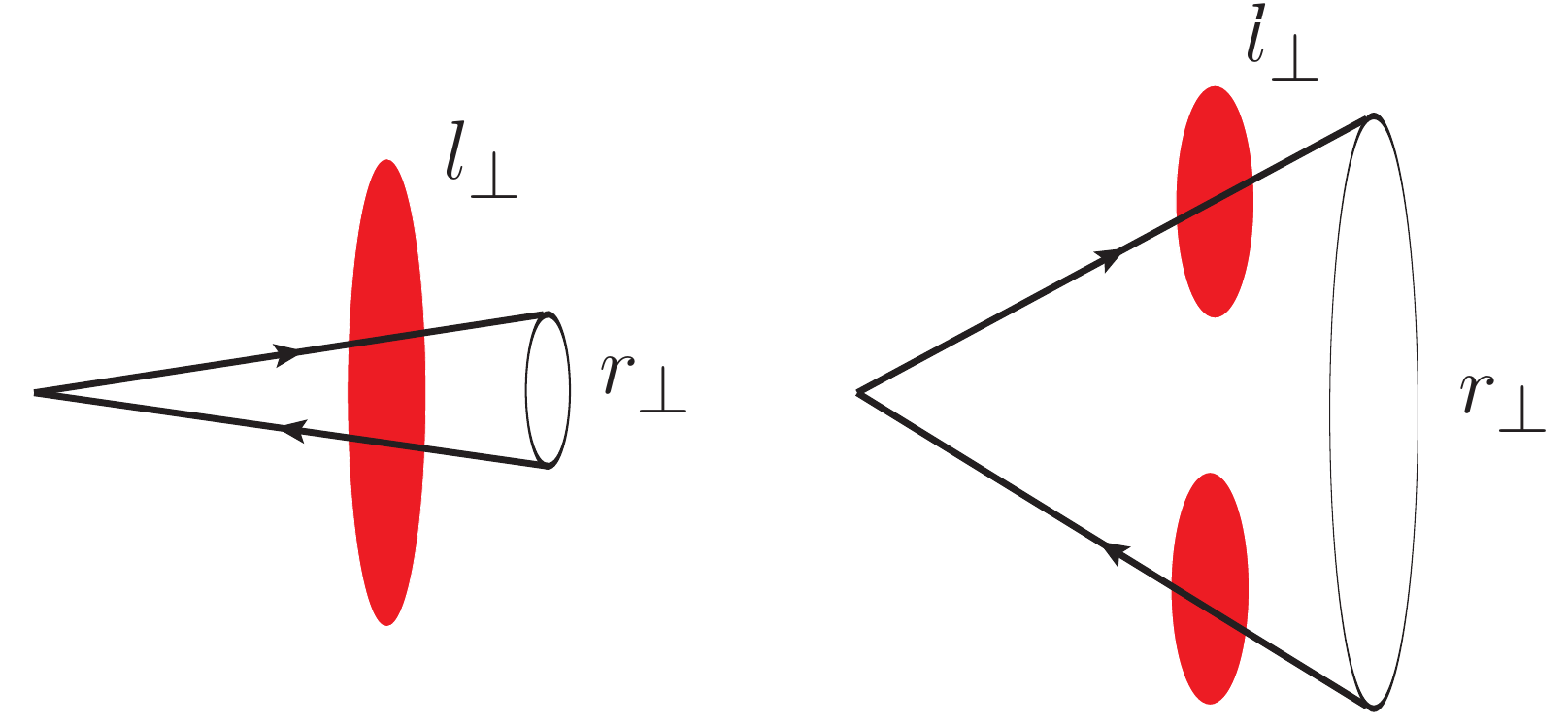}	\label{fig:antenna}
\caption{An illustration of a quark-antiquark antenna of typical transverse size $r_\perp=\Theta_{q\bar q}L$ in a medium of length $L$,  in the coherent regime (left) where the typical in-medium resolution length $l_\perp\equiv (\hat q L)^{-1/2} \gg r_\perp$. In this case, the medium doest not resolve the inner structure of the antenna,  and thus interacts with its total charge.  In the opposite case (right panel), $l_\perp\equiv  \ll r_\perp$, the medium resolves the quark and antiquark color charges, causing their decoherence. Then each of the antenna constituents radiates independently. 	}
\end{figure}

Comparing the two medium scales one is left with two regimes: 
\begin{enumerate}
\item  $r_\perp \ll l_\perp $ (or $\Tjet \ll \theta_c$): in this situation the jet is on average not resolved by the medium and remains color coherent.  Multiple interactions with the medium will induced gluon cascades off the total charge, typically at large angles (cf. Eq.~(\ref{multi-br-angle})),
\beq
\theta\,\gtrsim\, \frac{1}{\alpha_s^2}\, \theta_c\, \gg \, \Tjet. 
\eeq
The unresolved vacuum cascade looks as if it developed outside the medium. 

\item In the opposite case, $r_\perp \gg l_\perp$ (or $\Tjet \gg \theta_c$), the medium resolves more charges inside the jet, which results in an enhancement of the medium-induced radiation rate. In other words, the medium resolves a certain number of sub-structures characterized by the opening angle $\theta_c$, that evolve independently of each other.  On the other hand, the inner structure of these sub-jets remain unresolved by the medium\cite{CasalderreySolana:2012ef}. Consequently, the unresolved system loses energy {\it coherently} with a rate proportional to the total charge contained within $\theta_c$.

\end{enumerate}

\begin{figure}[ht]
\centerline{
\subfigure[]{\includegraphics[width=6.cm]{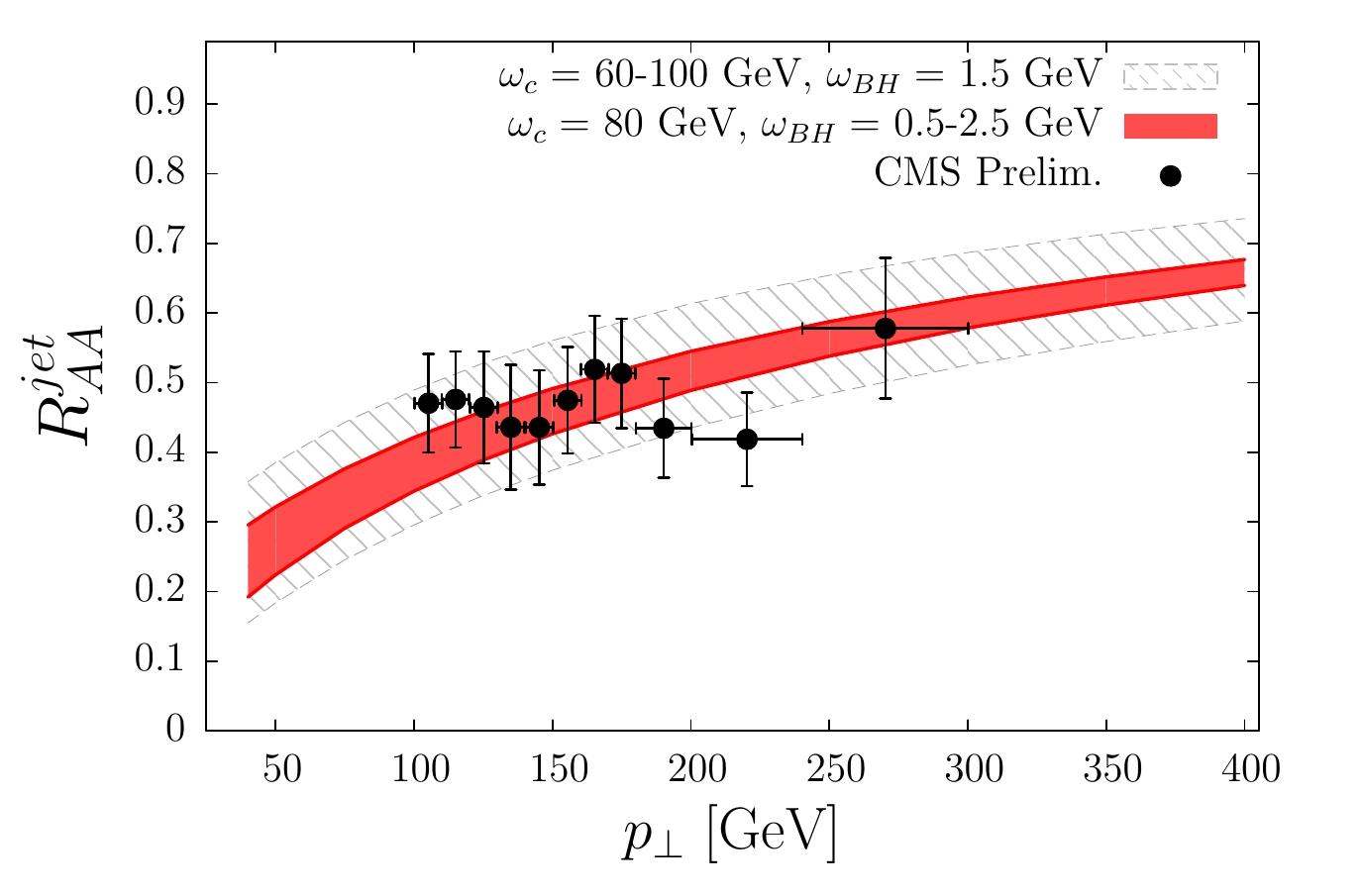}	\label{fig:RAA}}\quad  \subfigure[]{\includegraphics[width=6.cm]{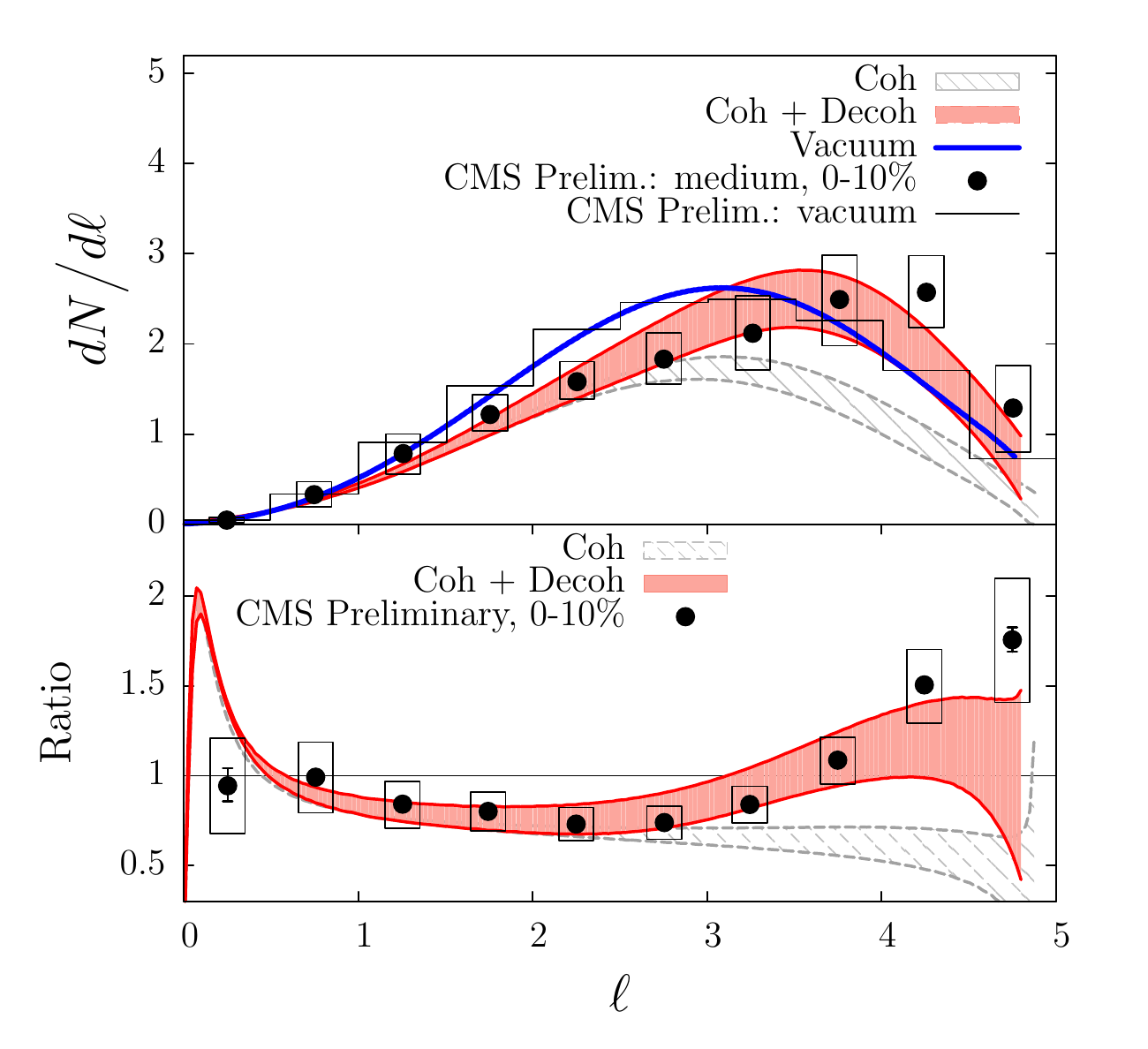}	\label{fig:FF}}  
		}
		\caption{(a) Calculation of the nuclear modification factor with $\omega_c$ = 80 GeV as a function of jet $p_\perp$ for central Pb-Pb collisions. The solid (red) band includes the variation of $\omega_\text{BH}$ around a central value of 1.5 GeV. The dashed (grey) band includes, in addition, a variation of $\omega_c \in$ [60, 100] GeV. The experimental data are taken from \cite{CMS:2012rba}. 
		(b) Upper panel: the longitudinal fragmentation function plotted as a function of $\ell = \ln 1\big/x$. Lower panel: the ratio of medium-modified and vacuum fragmentation functions. The experimental data are taken from \cite{CMS:2012vba}. See text for further details.}
\end{figure}

For realistic situations it was noted\cite{CasalderreySolana:2012ef} that a core containing most of the jet energy remains unresolved by the medium. 
Secondary substructures carry typically smaller energy fractions and are expected to be influenced significantly by medium effects. 
Under these circumstances, an unresolved jet lose energy as a single parton whose final distribution is then given by the rate equation  (\ref{evol-eq}). In this approximation, we can write the jet spectrum in nucleus-nucleus collision as a convolution of the hard parton spectrum due to incoherent binary collisions, and the probability distribution to find a parton with energy $p_\perp$ that originates from the nascent parton with momentum $p_\perp'=p_\perp/x$, 
\beq
\frac{\rmd N^{AA}_\text{jet}}{\rmd^2 p_\perp}\equiv N_\text{coll} \int_0^1 \frac{\rmd x}{x} D(x, p_\perp/x)  \frac{\rmd N^{pp}_\text{jet}}{\rmd^2 p_\perp} (p_\perp/x)
\eeq
where the distribution $D(x,p'_\perp)$, was discussed in Section \ref{energy-dist-WT}, identifying $p'_\perp\equiv E$ and $N_\text{coll}$ stands for the number of binary collisions. It is obtained by solving numerically Eq.~(\ref{evol-eq}), where finite size effects are taken into account (cf. Eq.~(\ref{BDMPS-Z -L})) . This procedure is analogous to that used to computed quenching of hadron spectra \cite{Baier:2001yt,Salgado:2003gb,Eskola:2004cr}, but proves to be best justified for high-$p_t$ jet spectra. 
The resulting nuclear modification factor, $R_{AA}\equiv  N^{-1}_\text{coll} \rmd N^{AA}_\text{jet}/ \rmd N^{pp}_\text{jet} $, is plotted in Fig.~\ref{fig:RAA}\cite{Mehtar-Tani:2014yea}, where the uncertainty bands shows the sensitivity of this calculation to variations of the medium length $L$ and the in-medium diffusion coefficient $\hat q$, via the variable $\omega_c=60-100$ GeV. In addition the sensitivity to the soft scale $\omega_\BH=0.5-1.5$ GeV is also shown, is implemented as a cut-off of the unphysical singularity in the branching kernel (\ref{Kdef}).
For $\hat q = 2$ GeV$^2$/fm and $L=4$ fm, we get $\omega_c=80$ GeV which corresponds to the average value. These values are consistent with the usual expectations \cite{Burke:2013yra}. 

In the CMS analysis the jet opening angle is chosen to be $\Tjet=0.3$, it follows that $r_\perp \simeq 1.5$ fm and $l_\perp \simeq 1.1$ fm. Obviously, the jet is neither in the fully coherent regime nor in the decoherent one. Hence, decoherent radiation off secondary substructures may be important. It was suggested in Ref.~\cite{Mehtar-Tani:2014yea} that a signature of partial decoherence may be found in intrajet structures, e.g., in the fragmentation functions, or equivalently the energy distribution of particles inside the jet-cone, $D(x)=x\rmd N/\rmd x$. 

The strategy to identify such effects consists in looking for departures from the fully coherent limit. In the coherence regime, the unresolved vacuum shower loses energy out of the jet-cone as a single parton. Hence, we can effectively factor out the in-medium distribution $D_\text{med} (z,p_\perp)$ described by Eq.~(\ref{evol-eq}) from the vacuum evolution (see Ref.~\cite{Mehtar-Tani:2014yea} for details): 
\beq\label{FF}
D^\text{coh}(x) \equiv \int_x^1 \frac{\rmd z}{z} D_\text{vac}(x/z,Q) D_\text{med} (z,p_\perp),
\eeq
where $D_\text{vac}(x/z,Q)$ is solution to the MLLA evolution equation (\ref{MLLA}) with $Q\equiv \Tjet p_\perp$. The result is plotted in Fig.~\ref{fig:FF} and is given by the grey band. Eq.~(\ref{FF}) properly matches the pp baseline with $D_\text{med} (z,p_\perp)=\delta(1-z)$ (see the blue curve in Fig.~\ref{fig:FF}).   The semi hard sector is well described and is consistent with the nuclear modification factor above. However, there is an excess of few soft particles that is not accounted for.   
We include now the decoherent contribution, $\Delta D^\text{decoh}$  which is computed to double logarithmic accuracy by convoluting the probability to form gluon-quark antenna in the medium which then subsequently radiates decoherent soft gluons according to  Eq.~(\ref{decoh}) (see Ref.~\cite{Mehtar-Tani:2014yea} for details). The total distribution,  $D^\text{tot}\simeq D^\text{coh}+\Delta D^\text{decoh}$ is displayed as red band in Fig.~\ref{decoh}. This in-cone decoherent radiation proves to be sufficient to account for the excess of soft gluons seen in the data.

\section{Conclusions and outlook}\label{outlook}

In this review, we have analyzed the dominant features of the medium-induced parton cascade, relying on the BDMPS-Z mechanism for medium induced gluon radiation. The medium induced cascade exhibits remarkable properties that originate on the one hand from the specific mechanism of medium induced radiation, on the other hand from the rapid loss of coherence among the radiated gluons caused by their multiple interactions with the medium constituents. These properties manifest themselves in a specific mechanism for the transport of energy along the cascade, with a characteristic angular structure. The transport of energy involves a constant  flow down to the lowest accessible frequencies. The angular structure is such that the softest quanta are emitted at large angles. We emphasize that these features, reminiscent of those revealed by the analysis of recent  CMS data on the angular distribution of the missing energy in inbalanced dijet events, are intrinsic properties of the BDMPS-Z cascade. They do not result from a specific coupling to the surrounding medium, nor are they influenced by the collective dynamics of this medium. In fact the only dependence on  the medium is hidden in the value of the jet quenching parameter $\hat q$.  

Much remains to be understood about the BDMPS cascade. In particular the treatment of what happens at the lower end of the cascade, where gluon frequencies become smaller than $\omega_\BH$ needs to be improved. In view also of the potential role of this cascade in the thermalization of the quark-gluon plasma \cite{Baier:2000sb,Kurkela:2014tea}, understanding how energy carried by the soft radiation  is dissipated in the medium is an important task. An important ingredient entering the description of the medium induced cascade is the  jet quenching parameters, $\hat q $. We have seen that this coefficient receives sizable radiative corrections. The calculations of these corrections is at present limited to the most singular contributions (double logarithmic). We have argued that it is the singular nature of the correction that allows us to absorb it in a redefinition of $\hat q$. It is unclear whether the simple description of energy loss and momentum broadening as local transport phenomena would hold in a more accurate treatment of these radiative corrections. 

Aside from the medium-induced cascade that we have just discussed, a jet in a medium is accompanied by another, vacuum-like, cascade. The two cascades are geometrically separated. The vacuum-like cascade is a collimated and coherent cascade, constituting the jet core,  characterized by collinear and angular ordered splittings triggered by the initial hard collision.  The medium-induced  cascade develops at parametrically large angles by successive branchings of the primary radiated gluons off the unresolved  total color charge of the jet core. Much remains to be understood about the interaction between the vacuum cascade and the vacuum induced  cascade. 
An interesting  phenomenon takes place when the medium charges resolve substructures in the inner jet core. The coherence of the resolved charges is broken, allowing  them to radiate independently in the soft sector.  This mechanism for medium-induced soft radiation is of a different nature than the BDMPS-Z  mechanism. It causes the decoherence of vacuum radiation which is not broadened by the medium and hence is expected to remain collimated. This mechanism may  provide a natural explanation for the excess of soft particles measured in the jet fragmentation region.  Finally, in addition to the effects on the intrajet structure, color decoherence is expected to be significant for processes at NLO, e.g., observables involving multi-jet events.

A complete description of parton cascades in a QCD medium  goes along with a refined understanding and modeling of the medium dynamics \cite{Polosa:2006hb,Dainese:2004te,Majumder:2014vpa,Majumder:2013re}. This, together with the details of the experimental setup, requires the development of event generators  \cite{Renk:2012ve,Majumder:2013re}, an effort undertaken by many groups \cite{Zapp:2008af,Zapp:2011ya,Armesto:2009fj,Schenke:2009gb,Young:2011ug,Majumder:2013re}.  Clearly, the microscopic physics that has been discussed in this review needs to be implemented in these generators. While this is the case, to some extent, of the  BDMPS-Z  cascade  \cite{Zapp:2008af,Schenke:2009gb,Young:2011ug}, as far as we know, the decoherence mechanism as well as the renormalization of the quenching parameter are not yet accounted for. \\

\noindent{\bf Acknowledgements}

We would like to thank N.~Armesto, J.~Casalderrey-Solana, F.~Dominguez, L.~Fister, E.~Iancu, M.~Martinez, J.~G.~Milhano, A.~H.~Mueller, C.~A.~Salgado, K. Tywoniuk, M.~A.~C.~Torres, B. Wu  for many discussions on  the issues discussed in this paper, as well as for direct collaboration on some of the works reported here. Y.~M.~-T is specially grateful to C.~A.~Salgado and K.~Tywoniuk for collaborating on Refs.~\cite{MehtarTani:2010ma,MehtarTani:2011tz}.  This research is supported by the European Research Council under the Advanced Investigator Grant ERC-AD-267258.

\bibliographystyle{ws-rv-van}


\printindex                         
\end{document}